\documentclass[a4paper,fleqn]{cas-dc}

\usepackage[numbers]{natbib}
\usepackage{graphicx}% Include figure files
\usepackage{dcolumn}% Align table columns on decimal point
\usepackage{multirow}
\usepackage{amsmath}
\usepackage{bm}% bold math
\usepackage{textcomp}
\usepackage{epsfig}
\usepackage{rotating}
\usepackage{xfrac}
\usepackage{threeparttable}
\usepackage{lipsum} % generate dummy text for estimating page length
\usepackage{longtable}
\usepackage{stfloats}
\usepackage[many]{tcolorbox}

\newtcolorbox{boxA}{
 sharp corners,
 colback = white,
 colbacklower=white,
 boxrule=0pt,%frame hidden    
 colframe = white % frame color
}
\newcommand*\Laplace{\!\!\vartriangle\!\!}
\RequirePackage{hyperref}
\hypersetup
{
bookmarks=true,%
unicode=false,%
pdftoolbar=true,
pdfmenubar=true,
pdffitwindow=false,
pdfstartview=FitH,
pdftitle={PRL},
pdfauthor={LAM Yi Hua},
pdfsubject={Nuclear Physics},
pdfcreator={LAM Yi Hua},
pdfproducer={LAM Yi Hua},
pdfkeywords={NucAstro},
bookmarksnumbered,%
colorlinks=true,%
linkcolor=blue,%
citecolor=magenta,%
filecolor=green,%
urlcolor=blue,
breaklinks, % To break long figure/table caption in the List of Figures / Tables
plainpages=false%
}
\usepackage{xcolor}
\definecolor{Mycolor}{HTML}{bfffbf}
\definecolor{blue}{HTML}{4477aa} % Color blind blue
\definecolor{red}{HTML}{ee6677} % Color blind red
\definecolor{magenta}{HTML}{ee3377} % Color blind magenta
\definecolor{green}{HTML}{228833} % Color blind green
\def\tsc#1{\csdef{#1}{\textsc{\lowercase{#1}}\xspace}}
\tsc{WGM}
\tsc{QE}
\tsc{EP}
\tsc{PMS}
\tsc{BEC}
\tsc{DE}
\begin{document}
\let\WriteBookmarks\relax
\def\floatpagepagefraction{1}
\def\textpagefraction{.001}
%\shorttitle{Nonlinear point-coupling interaction for the relativistic Hartree-Bogoliubov approach}
\shorttitle{The optimized point-coupling interaction for the relativistic energy density functional of Hartree-Bogoliubov approach}
\shortauthors{Liu \emph{et~al}.}
%The optimized point-coupling interaction for the relativistic energy density functional of Hartree-Bogoliubov approach  quantifying the nuclear bulk properties 
\title[mode = title]{The optimized point-coupling interaction for the relativistic energy density functional of Hartree-Bogoliubov approach quantifying the nuclear bulk properties}
%\title[mode = title]{Nonlinear point-coupling interaction for \\the relativistic Hartree-Bogoliubov approach}
%\tnotemark[1,2]
%\tnotetext[1]{This document is the results of the research project funded by the National Science Foundation.}
%\tnotetext[2]{The second title footnote which is a longer text matter to fill through the whole text width and overflow into another line in the footnotes area of the first page.}

\author[1,2,3]{Zi Xin Liu}[%
style=,
type=editor,
auid=000,
bioid=1,
prefix=,
suffix=,
role=,
orcid=0000-0001-5652-1516]
\cormark[1]
%\fnmark[1]
\ead{liuzixin1908@impcas.ac.cn}
%\ead[url]{www.cvr.cc, cvr@sayahna.org}
\credit{Conceptualization of this study, Fitting PC-L3R, Writing - Original draft preparation}
%\address[1]{CAS Key Laboratory of High Precision Nuclear Spectroscopy, Institute of Modern Physics, \\ Chinese Academy of Sciences, Lanzhou 730000, People's Republic of China}
\address[1]{Institute of Modern Physics, Chinese Academy of Sciences, Lanzhou 730000, People's Republic of China}
\address[2]{School of Nuclear Science and Technology, University of Chinese Academy of Sciences, Beijing 100049, People's Republic of China}
\address[3]{School of Physics Science and Technology, Lanzhou University, Lanzhou 730000, People's Republic of China}

\author[1,2]{Yi Hua Lam}[%
%\author[1,2]{Yi Hua Lam (\cntext{藍乙華})}[%
style=,
type=editor,
auid=000,
bioid=1,
prefix=,
suffix=,
role=,
orcid=0000-0001-6646-0745]
\cormark[1]
%\fnmark[1]
\ead{lamyihua@impcas.ac.cn}
%\ead[url]{www.cvr.cc, cvr@sayahna.org}
\credit{Conceptualization of this study, Fitting PC-L3R, Writing - Original draft preparation}
%\address[1]{CAS Key Laboratory of High Precision Nuclear Spectroscopy, Institute of Modern Physics, \\ Chinese Academy of Sciences, Lanzhou 730000, People's Republic of China}
%\address[2]{School of Nuclear Science and Technology, University of Chinese Academy of Sciences, Beijing 100049, People's Republic of China}

\author[1,2,4]{Ning Lu}[%
%\author[1,2,4]{Ning Lu (\cntext{盧寧})}[%
style=,
type=editor,
auid=000,
bioid=1,
prefix=,
suffix=,
role=,
orcid=0000-0002-3445-0451]
%\cormark[1]
%\fnmark[1]
%\ead{cvr_1@tug.org.in}
%\ead[url]{www.cvr.cc, cvr@sayahna.org}
\credit{Writing - Original draft preparation}
\address[4]{School of Nuclear Science and Technology, Lanzhou University, Lanzhou 730000, People's Republic of China}

\author[5]{Peter Ring}[%
style=,
type=editor,
auid=000,
bioid=1,
prefix=,
suffix=,
role=,
orcid=0000-0001-7129-2942]
%\cormark[1]
%\fnmark[1]
%\ead{cvr_1@tug.org.in}
%\ead[url]{www.cvr.cc, cvr@sayahna.org}
\credit{Important suggestions, Writing - Original draft preparation}
\address[5]{Fakult{\"a}t f{\"u}r Physik, Technische Universit{\"a}t München, D-85748 Garching, Germany}

\begin{abstract}
We propose a newly optimized nonlinear point-coupling parameterized interaction, PC-L3R, for the relativistic Hartree-Bogoliubov framework with a further optimized separable pairing force by fitting to observables, \textcolor{black}{i.e., the binding energies of 91 spherical nuclei, charge radii of 63 nuclei, and 12 sets of mean pairing gaps consisting of 54 nuclei in total.}
%of $91$ spherical nuclei, including the binding energies, charge radii, and mean pairing gaps. 
\textcolor{black}{The separable pairing force strengths of proton and neutron are optimized together with the point-coupling constants, and are justified in satisfactory reproducing the empirical pairing gaps.}
%This method improves the description of nuclear properties.
%\textcolor[rgb]{0.00,0.00,1.00}{The experimental values are obtained by AME2020.} 
%Overall, from comparing with the experimental data, the implementation of PC-L3R in relativistic Hartree-Bogoliubov successfully yields the lowest root-mean-square deviation, \textcolor[rgb]{1.00,0.00,0.00}{$1.350$~MeV}, among currently and commonly used point-coupling interactions. 
The comparison of experimental binding energies compiled in AME2020 for \textcolor{black}{91} nuclei with the ones generated from the present and other commonly used point-coupling interactions indicates that the implementation of PC-L3R in relativistic Hartree-Bogoliubov yields the lowest root-mean-square deviation. The charge radii satisfactory agree with experiment. 
Meanwhile, PC-L3R is capable of estimating the saturation properties of the symmetric nuclear matter and of appropriately predicting the isospin and mass dependence of binding energy. The experimental odd-even staggering of single nucleon separation energies is well reproduced.
%, e.g., the isotopes of $Z\!\!=\!20$, $50$, and $82$, and isotones of $N\!\!=\!20$, $50$, and $82$. 
The comparison of the estimated binding energies for $7,373$ nuclei based on the PC-L3R and other point-coupling interactions is also presented. 
%This template helps you to create a properly formatted \LaTeX\ manuscript.
%\noindent\texttt{\textbackslash begin{abstract}} \dots 
%\texttt{\textbackslash end{abstract}} and
%\verb+\begin{keyword}+ \verb+...+ \verb+\end{keyword}+ 
%which contain the abstract and keywords respectively. 
%\noindent Each keyword shall be separated by a \verb+\sep+ command.
\end{abstract}

%\begin{graphicalabstract}
%\includegraphics{grabs.pdf}
%\end{graphicalabstract}

%\begin{highlights}
%\item Research highlights item 1
%\item Research highlights item 2
%\item Research highlights item 3
%\end{highlights}

%https://ufn.ru/en/pacs/all/
\begin{keywords}
Nuclear Density Functional Theory \sep Relativistic Hartree-Bogoliubov \sep Point-coupling interactions \sep Binding energies and masses \sep Charge distribution \sep Saturation properties of nuclear matter 
%\sep Neutron-rich nuclei 
%\WGM \sep \BEC
\end{keywords}

\maketitle

%\begin{bibunit}
\section{Introduction}
With the recent advancements in radioactive ion beams and detectors, new experiments have revealed a wealth of structural phenomena in exotic nuclei with extreme isospin values, such as the halo phenomena~\cite{PPNP2013Tanihata}, island of inversion \cite{PPNP2021Nowacki}, neutron skin~\cite{PPNP2017Sakaguchi}, the disappearance of traditional magic numbers but with the occurrence of new ones~\cite{PRL2000Ozawa,nature2013Wienholtz}, masses and half lives of the neutron-rich nuclei important for understanding the synthesis of heavy elements in extreme astrophysical environments \cite{PRC2016Marketin}, and masses of proton-rich nuclei for Type-I X-ray bursts \cite{Lam2022}. Hence, theoretical nuclear models capable of providing a satisfactory description of the ground-state properties and collective excitations of atomic nuclei, ranging from the relatively light to superheavy nuclei, from the spherical to deformed nuclei, and from the valley of $\beta$-stability to the particle-drip lines \cite{PPNP1996Ring_37_193,RMP2003Bender,PR2005Vretenar_409_101,PPNP2006MengJ_57_470,PRL2009Goriely,PRC2010Kortelainen,PPNP2011Niksic,JPG2015MengJ_42_093101,PLB2018Afanasjev,PRC2019Sun,PRC2019Agbemava}, are crucial for us to understand the interactions and configurations of nucleons in atomic nuclei. %that are experimentally reachable or even unreachable nuclear regions.

The self-consistent relativistic mean-field (RMF) model with the Bardeen-Cooper-Schrieffer (BCS) method incorporating a nuclear density functional has been %successfully 
used to describe various nuclear properties \cite{PRC2010Zhao} and to obtain the relevant nuclear physics input for astrophysical calculations \cite{PRC2008Sun,PRC2009Niu,PRC2013Xu}. Such calculations were based on the RMF+BCS framework with the monopole pairing force, e.g., the $\delta$ pairing force \cite{PRC2010Zhao,PRC2002Burvenich,PRC2006Niksic_034308,PRC2006Niksic_064309}.
%, which is actually a phenomenologically zero-range pairing force. 
The BCS treatment on pairing force provides, however, a poor approximation for exotic nuclei far from the valley of $\beta$-stability \cite{PR2005Vretenar_409_101}. The common understanding is that a consistent description of neutron-rich nuclei requires a unified and self-consistent treatment of mean-field and pairing correlations, which plays a critical role in exotic nuclei. Therefore, the self-consistent relativistic Hartree-Bogoliubov (RHB) framework was developed to unify the treatment of pairing correlations and mean-field potentials, and to %successfully 
analyze the structural phenomena in exotic nuclei \cite{PPNP1996Ring_37_193,ZPA1991Kucharek_339_23,PRL1996Meng_77_3963,PRL1997Poschl_79_3841}. 

%In the applications of the RHB frameworks, 
The most widely used effective interactions serving as the mean-field potential of the RMF and RHB frameworks are the meson-exchange \cite{PRC1997Lalazissis,NPA1999Typel,PRC2002Niksic,PRC2012Abusara,PRC2014Dutra,JPG2017Lv,PRC2018Liu}. In recent years, the point-coupling model offers an alternative to the meson-exchange model as the exchange of heavy mesons associated with short-distance dynamics cannot be resolved at low energies \cite{PRC2010Zhao,PRC2002Burvenich,PRC1992Nikolaus,NPA1997Rusnak,PRC2008Niksic,PRC2017Zhao,PRC2019Yuksel,PLB2020Taninah,PRC2021Perera,PRC2021Vale,PRC2021Fan,ADNDT2022Zhang}. The presently and commonly used point-coupling parameterized interactions include 
PC-PK1~\cite{PRC2010Zhao}, 
PC-F1~\cite{PRC2002Burvenich}, 
PC-LA~\cite{PRC1992Nikolaus}, and 
DD-PC1~\cite{PRC2008Niksic}. 
%\textcolor{black}{These studies all minimizations of the functionals are performed within the RMF framework.}
The PC-LA parameterization does not consider a pairing force in the fitting procedure, whereas the PC-F1 and PC-PK1 interactions were fitted based on the RMF+BCS framework with the $\delta$ pairing force \cite{PRC2010Zhao,PRC2002Burvenich,PRC2006Niksic_034308,PRC2006Niksic_064309}. The DD-PCX~\cite{PRC2019Yuksel} and PC-X~\cite{PLB2020Taninah} interactions were recently proposed for the RHB model with a separable pairing force, however, its capability is yet to be assessed.
%However, a smaller amount of parameter is fitted within the point-coupling RHB framework. \textcolor[rgb]{0.00,0.50,0.75}{Recently, a nonlinear coupling parametrization PC-X was proposed~\cite{PLB2020Taninah} within the RHB framework with separable pairing. Although it reproduces .....}
Moreover, the proton and neutron pairing strengths in the separable pairing force of all presently available nonlinear point-coupling interactions are treated in an equal footing, and thus the separable pairing strength is not optimized.
Hence, \textcolor{black}{a robust and} fully optimized point-coupling interaction accustomed to the RHB framework with \textcolor{black}{justified pairing strengths} is highly desired by the community \textcolor{black}{to describe the experimentally determined finite nuclei and nuclear matter properties.}
%\textcolor{blue}{Hence, a fully optimized point-coupling interaction accustomed to the RHB framework with the capability of describing the finite nuclei and nuclear matter properties close to experimental findings is highly desired by the community}. \textcolor{black}{In order to obtain a high capability in describing a vast range of nuclear bulk properties, a fully optimizing the point coupling constant set together with the selected pairing interaction (pairing strengths) within RHB framework is very important.}

In this work, we propose a new set of fully optimized nonlinear point-coupling interaction for the RHB framework with a \textcolor{black}{justified} separable pairing force, named PC-L3R. We present the theoretical framework of the point-coupling interaction and relativistic Hartree-Bogoliubov model in Sec.~\ref{Sec:Theory}. Then, we discuss the assessments of the new point-coupling interaction in Sec.~\ref{Sec:checks}, and show the comparison of the nuclear binding energies resulted from various point-coupling interactions in Sec.~\ref{Sec:BindingEnergies}. Our summary is drawn in Sec~\ref{Sec:Summary}.
%of neutron-rich nuclei
%The Letter is organized as follows. 

%\section{THEORETICAL FRAMEWORK}
%\subsection{Point-coupling model}
\section{Point-coupling interaction for the relativistic Hartree-Bogoliubov model}
\label{Sec:Theory}
The effective covariant Lagrangian density of the point-coupling interaction for the relativistic mean field (RMF) model is expressed as~\cite{PR2005Vretenar_409_101,PPNP2006MengJ_57_470}
%,ANP1986Sert_16_1,RPP1989Reinhard_52_439
\begin{equation}
\label{eq:Lagrangian}
% \nonumber to remove numbering (before each equation)
   \mathcal{L} = \mathcal{L}^{\rm free}+\mathcal{L}^{\rm 4f}+\mathcal{L}^{\rm ho}+\mathcal{L}^{\delta}+\mathcal{L}^{\rm em}, 
\end{equation}
where $\mathcal{L}^{\rm free}$ is the free nucleons terms,
\begin{equation}
\label{eq:free}
\hspace{-5mm}
 \mathcal{L}^{\rm free} = \bar{\psi}(i\gamma_{\mu}\partial^{\mu}-M)\psi \,,
\end{equation}
$\mathcal{L}^{\rm 4f}$ is the four-fermion point-coupling terms,
\begin{eqnarray}
\label{eq:fermion}
\hspace{-5mm}
 \mathcal{L}^{\rm 4f} &=& -\frac{1}{2}\alpha_{S}(\bar{\psi}\psi)(\bar{\psi}\psi)-
 \frac{1}{2}\alpha_{V}(\bar{\psi}\gamma_{\mu}\psi)(\bar{\psi}\gamma^{\mu}\psi) \nonumber \\
 &-&\frac{1}{2}\alpha_{TV}(\bar{\psi}\vec{\tau}\gamma_{\mu}\psi)\cdot(\bar{\psi}\vec{\tau}\gamma^{\mu}\psi) \,,
\end{eqnarray}
$\mathcal{L}^{\rm ho}$ is the higher-order terms, which take into account of the effects of medium dependence,
\begin{eqnarray}
\label{eq:HO}
\hspace{-15mm}
 \mathcal{L}^{\rm ho} &=& -\frac{1}{3}\beta_{S}(\bar{\psi}\psi)^{3}-\frac{1}{4}\gamma_{V}[(\bar{\psi}\gamma_{\mu}\psi)
  (\bar{\psi}\gamma^{\mu}\psi)]^2 \nonumber\\
   &~&- \frac{1}{4}\gamma_{S}(\bar{\psi}\psi)^{4}\,,
\end{eqnarray}
$\mathcal{L}^{\delta}$ is the gradient terms,
\begin{eqnarray}
\label{eq:derivative}
 \mathcal{L}^{\delta} &=& -\frac{1}{2}\delta_{S}\partial_{\nu}(\bar{\psi}\psi)\partial^{\nu}
  (\bar{\psi}\psi)-\frac{1}{2}\delta_{V}\partial_{\nu}(\bar{\psi}\gamma_{\mu}\psi)\partial^{\nu}
  (\bar{\psi}\gamma^{\mu}\psi) \nonumber \\
  &~&-\frac{1}{2}\delta_{TV}\partial_{\nu}(\bar{\psi}\vec{\tau}\gamma_{\mu}\psi)\partial^{\nu}
  (\bar{\psi}\vec{\tau}\gamma_{\mu}\psi)\, ,
\end{eqnarray}
and $\mathcal{L}^{\rm em}$ is the electromagnetic interaction terms,
\begin{equation}
\label{eq:EM}
\hspace{-5mm}
 \mathcal{L}^{\rm em} = -\frac{1}{4}F^{\mu\nu}F_{\mu\nu} -e\frac{(1-\tau_{3})}{2} \bar{\psi}\gamma^{\mu}\psi A_{\mu} \, .
\end{equation}
%where $\mathcal{L}^{\rm free}$ is the free nucleons terms, $\mathcal{L}^{\rm 4f}$ is the four-fermion point-coupling terms, $\mathcal{L}^{\rm hot}$ is the higher-order terms which are responsible for the effects of medium dependence, $\mathcal{L}^{\rm der}$ is the gradient terms, and $\mathcal{L}^{\rm em}$ is the electronmagnetic interaction terms.
$M$ in Eq.~(\ref{eq:free}) is the nucleon mass. $\vec{\tau}$ in Eqs.~(\ref{eq:fermion}) and (\ref{eq:derivative}) is the isospin vector of the third component, $\tau_{3}$. In Eqs.~(\ref{eq:fermion}), (\ref{eq:HO}), and (\ref{eq:derivative}), $\alpha_{S}$, $\alpha_{V}$, $\alpha_{TV}$, $\beta_{S}$, $\gamma_{S}$, $\gamma_{V}$, $\delta_{S}$, $\delta_{S}$, $\delta_{V}$, and $\delta_{TV}$ are the respective coupling constants. The rotational symmetry of each coupling constant is indicated by the subscript. $S$ and $V$ refer to the scalar and vector nucleon fields, respectively. $TV$ refers to the isovector fields. $A_{\mu}$ and $F_{\mu\nu}$ are the four-vector potential and field strength tensor of the electromagnetic field, respectively.

With implementing the effective Lagrangian density, the relativistic Hartree-Bogoliubov (RHB) model that unifies the self-consistent mean field and a pairing field had been developed to describe the properties of open-shell nuclei \cite{PPNP1996Ring_37_193,ZPA1991Kucharek_339_23,PRL1996Meng_77_3963,PRL1997Poschl_79_3841}. Here, we use the point-coupling effective covariant Lagrangian density described above to be the self-consistent mean field of the RHB model, and take the separable form of the pairing force \cite{PRC2019Yuksel,PLB2020Taninah,PLB2009Tian_676_44,PRC2009Tian_064301,PRC2013Afanasjev} to be the pairing field.
The RHB energy density functional is given as
\begin{equation}\label{RHBE}
E_{\rm RHB}[\rho,\kappa]=E_{\rm RMF}[\rho]+E_{\rm pair}[\kappa],
\end{equation}
where $E_{\rm RMF}$ is the self-consistent mean field term, and the pairing term reads
\begin{equation}\label{Epair}
E_{\rm pair}[\kappa]=\frac{1}{4}\sum_{n_{1}n'_{1}}\sum_{n_{2}n'_{2}}
\kappa^{\ast}_{n_{1}n'_{1}}\langle n_{1}n'_{1} |V^{pp}| n_{2}n'_{2}\rangle \kappa_{n_{2}n'_{2}}\, ,
\end{equation}
where $n$ refers to the original basis, e.g., an oscillator basis, or the coordinates in space, spin and isospin $(\bm{r},s,t)$. The normal density, $\rho$, and the pairing tensor, $\kappa$ are given as
\begin{equation}\label{rho}
\rho_{nn'}=\langle\Phi_{0} |c^{\dag}_{n'}c_{n}| \Phi_{0}\rangle \,, {\rm and }\,\,
\kappa_{nn'}=\langle\Phi_{0} |c_{n'}c_{n}| \Phi_{0}\rangle,
\end{equation}
%\textcolor[rgb]{0.00,0.00,1.00}{To describe open-shell nuclei, it is necessary to include pairing correlations in the self-consistent relativistic framework. This has led to the formulation and development of the RHB model, which provides a unified description of the self-consistent mean field and a pairing field. For the self-consistent mean field part, the main contribution comes from point-coupling effective covariant Lagrangian density is provided. In general, the pairing correlations are described by the density-independent $\delta$ interaction~\cite{PRC2010Zhao,PRC2002Burvenich,PRC2006Niksic_034308,PRC2006Niksic_064309} or the finite range Gogny interaction~\cite{PLB2020Taninah,NPA1984Berger,PLB2009Tian_676_44,PRC2009Tian_064301}.}
%The relativistic Hartree-Bogoliubov (RHB) theory implementing the effective Lagrangian density as the density functional has been developed to describe the pairing effects in open-shell nuclei \cite{PPNP1996Ring_37_193,ZPA1991Kucharek_339_23,PRL1996Meng_77_3963,PRL1997Poschl_79_3841}. The RHB model provides a unified description of the self-consistent mean field and a pairing field. 
respectively. $|\Phi_{0}\rangle$ is the ground state for an even-even nucleus. $c^{\dag}_{n}$ and $c_{n}$ are the single-nucleon creation and annihilation operators, respectively. In the unitary Bogoliubov transformation, the single-nucleon operators is defined as the quasiparticle operators~\cite{ManybodyProb1980},
%The quasiparticle operators are defined by the unitary Bogoliubov transformation of the single-nucleon creation and annihilation operators~\cite{ManybodyProb1980}
%Starting from the effective Lagrangian density, 
\begin{equation}
\label{q-opterator}
  \beta^{\dag}_{k} = \sum_{n}U_{nk}c_{n}^{\dag}+V_{nk}c_{n},
\end{equation}
and $|\Phi_{0}\rangle$ is represented as a vacuum with respect to quasiparticles
\begin{equation}
\label{eq:gs-wave}
  |\Phi_{0}\rangle = \prod_{k}\beta_{k}|0\rangle \, ,
\end{equation}
and $|0\rangle$ is the bare vacuum.

The RHB equation for nucleons is derived by the variational procedure,
\begin{equation}\label{eq:RHB}
  \int d^3 \bm{r}' \left(
   \begin{array}{cc}
    h_{D}-\lambda_{\tau} &  \Delta \\
     -\Delta^{\ast} &  -h^{\ast}_{D}+\lambda_{\tau} \\
  \end{array}
\right)\left(
   \begin{array}{c}
    U_{k}\\
     V_{k}\\
  \end{array}
\right)=E_{k}\left(
   \begin{array}{c}
    U_{k}\\
     V_{k}\\
  \end{array}
\right),
\end{equation}
where $E_{k}$ is the quasiparticle energy, $\lambda_{\tau}$ ($\tau\!=\!\mathrm{n}, \mathrm{p}$) is chemical potential, $U_{k}$ and $V_{k}$ are quasiparticle wave functions and $h_{D}$ is the Dirac Hamiltonian,
\begin{equation}\label{Dirac-H}
  h_{D}(\bm r)=\bm{\alpha}\cdot\bm{p}+V(\bm r)+ \beta(M+S(\bm r)) \, ,
\end{equation}
where $\bm{\alpha}$ and $\beta$ are the Dirac matrices, $p$ is the momentum operator, and the scalar and vector potentials are
 \begin{eqnarray}
 % \nonumber to remove numbering (before each equation)
   S(\bm r) &=& \alpha_{S}\rho_{S}+\beta_{S}\rho^{2}_{S}+\gamma_{S}\rho^{3}_{S}+\delta_{S}\Laplace\rho_{S} \,\, {\rm and} \nonumber \\
   V(\bm r)&=&  \alpha_{V}\rho_{V}+\gamma_{V}\rho_{V}^{3}+\delta_{V}\Laplace\rho_{V}+eA_{0} \nonumber \\
   &~&+\alpha_{TV}\tau_{3}\rho_{TV}+\delta_{TV}\tau_{3}\Laplace\rho_{TV} \, , \nonumber
 \end{eqnarray}
respectively. The local densities are
\begin{eqnarray}
% \nonumber to remove numbering (before each equation)
  \rho_{S}(\bm r) &=& \sum_{k>0}\bar{V}_{k}(\bm r)V_{k}(\bm r) \,, \nonumber\\
  \rho_{V} (\bm r)&=&  \sum_{k>0} V^{\dag}_{k}(\bm r)V_{k}(\bm r) \,, {\rm and} \\
  \rho_{TV}(\bm r) &=& \sum_{k>0} V^{\dag}_{k}(\bm r)\tau_{3}V_{k}(\bm r) \,. \nonumber
\end{eqnarray}
The pairing field $\Delta$ in Eq.~(\ref{eq:RHB}) reads
\begin{equation}
\label{eq:Delta}
\Delta_{n_{1}n'_{1}}=\frac{1}{2}\sum_{n_{2}n'_{2}}\langle n_{1}n'_{1}|V^{pp}|n_{2}n'_{2} \rangle \kappa_{n_{2}n'_{2}} \, .
\end{equation}
The pairing force that we use in Eq.~(\ref{eq:Delta}) is the separable form of the pairing force, which is adjusted to model the pairing term of the finite range Gogny D1S force \cite{NPA1984Berger} in the particle-particle channel. The implementation of this separable form of the pairing force in the framework of RHB model produces the pairing properties of the nuclear matter and of finite nuclei in the same footing as the effect yielded from the corresponding pairing Gogny interaction \cite{PLB2009Tian_676_44}. The separable form of the pairing force is expressed as
%Applications in the framework of the relativistic Hartree–Bogoliubov approach show that the pairing properties are depicted on almost the same footing as by the original pairing interaction not only in nuclear matter, but also in finite nuclei.
%The finite range Gogny force that 
\begin{equation}
\label{eq:Vpp}
V^{pp}(\bm{r}_{1},\bm{r}_{2},\bm{r}'_{1},\bm{r}'_{2})=-G\delta(\bm{R}-\bm{R}')
P(\bm{r})P(\bm{r}')\frac{1}{2}(1-P^{\sigma}) \, ,
\end{equation}
with the center of mass, $\bm{R}=(\bm{r_{1}}+\bm{r_{2}})/2$, and the relative coordinates, $\bm{r} = \bm{r_{1}}-\bm{r_{2}}$. The form factor $P(\bm r)$ is of Gaussian shape
\begin{equation}
\label{eq:Gaussian}
P(\bm r)=\frac{1}{(4\pi a^{2})^{3/2}}e^{-r^{2}/4a^{2}} \, .
\end{equation}
According to Tian \emph{et al}.~\cite{PLB2009Tian_676_44}, the $G$ strength in Eq.~(\ref{eq:Vpp}) and the $a$ parameter in Eq.~(\ref{eq:Gaussian}) are defined as $-$728 $\rm MeV~fm^{3}$ and 0.644 fm, respectively. We can in fact follow the procedure of Y{\"u}ksel {\it et al.} \citep{PRC2019Yuksel} to further optimize these $G$ and $a$ factors with separating the $G$ strength to neutron ($G_\mathrm{n}$) and proton ($G_\mathrm{p}$) pairing strengths to improve the description of the bulk nuclear properties.

%\section{NUMERICAL DETAILS AND CHECKS}
\section{PC-L3R, the optimized coupling constants and pairing strengths for the RHB model}
\label{Sec:checks}
We perform a series of grid search up to $\approx\!\!10^9$ calculations with cross validation for  $91$ spherical nuclei to obtain a set of optimized coupling constants, and the $G_\mathrm{n}$, $G_\mathrm{p}$, and $a$ factors in the separable pairing force. 
%\textcolor{black}{The benefit of using $G_\mathrm{n}$ and $G_\mathrm{p}$ strengths in the PC-L3R interaction is manifested by a rather close reproduction of experimental findings.}
%\textcolor{black}{We find that the robustness of the point-coupling interaction in describing the finite nuclear properties were improved by using the different strengths of $G_\mathrm{n}$ and $G_\mathrm{p}$.}
The mean pairing gaps are calculated using the five-point formula \cite{EPJA2000Bender}.
The observables of 65 spherical nuclei selected in the previous work \cite{PRC2010Zhao} were also included for fitting the present point coupling interaction. The configuration space of harmonic oscillator wave functions with appropriate symmetry is employed to solve the RHB equation for nucleons. The densities are computed in the coordinate space. The fermionic shells, $N_{\!f}\!\! =\!\!26$, is adopted for these calculations. 
%The new parameter set is listed in Table~\ref{tab:PCL3M}. 

Meanwhile, we use the $\chi^{2}$ minimization to isolate and to confine the relativistic point-coupling and pairing-strength parameter space. The minimization is quantified as below, %most optimized
\begin{equation}
\label{eq:minimization}
\chi^{2}(\mathbf{P})=\sum_{i=1}^{N_i}\sum_{j=1}^{n_{i}} \left (\frac{O_{i,j}^{\rm {cal.}}(\mathbf{P})-O_{i,j}^{\rm {exp.}}}{\Delta O_{i}} \right ) ^{2}
\end{equation}
where $\mathbf{P}$ is the parameter set, $\mathbf{P} = (\rm {P_{1}, P_{2}, \ldots, P_{n}})$. $N_i$ is a set of physical observables, e.g., binding energies, charge radii, and empirical pairing gaps. Each type of observable with the label $i$ is associated with a set of $n_i$ data points.
For the theoretical weight, $\Delta O_i$, we refer to the previous works~\cite{PRC2010Zhao,PRC2008Niksic,PRC2019Yuksel,PLB2020Taninah} and assess several $\Delta O_\mathrm{BE}$, see Fig.~\ref{fig:X2-rms91}. With studying several sets of the optimized theoretical weights, 
$\Delta O_i$ for binding energies, charge radii, and empirical pairing gaps, are defined as $\Delta O_\mathrm{BE}\!=\!1$~MeV  (91 nuclei), $\Delta O_\mathrm{CR}\!=\!0.02$~fm (63 nuclei), and $\Delta O_\mathrm{PG}\!=\!0.1$~MeV (twelve sets of mean pairing gaps consisting of 54 nuclei in total), respectively.
Note that the consideration of the influence on the single neutron separation energies further help confines the pairing strengths.
%Using the observables, i.e., binding energies, we obtain
%charge radii, and empirical pairing gaps (single neutron separation energy),  
%The process of parameter selection we adopt the grid search with cross validation. 
This new set of point-coupling interaction is named PC-L3R (Table~\ref{tab:PCL3M}), which consists of the nine coupling constants in Eqs.~(\ref{eq:fermion}), (\ref{eq:HO}), and (\ref{eq:derivative}), the $G_\mathrm{n}$ and $G_\mathrm{p}$ pairing strengths and $a$ factor in Eqs.~(\ref{eq:Vpp}) and (\ref{eq:Gaussian}). The test of naturalness for nine coupling constants of PC-L3R indicates that these coupling constants are natural (Supplemental Material (SM) and Table~\ref{tab:PC-cln}).

We then assess the influence of the PC-L3R interaction on the ground state properties, i.e., binding energies and available charge radii of these $91$ finite nuclei and the saturation properties of symmetric nuclear matter as follows. 

\begin{table}[pos=t,width=\columnwidth]
\renewcommand\arraystretch{1.3}
\centering
\footnotesize
\caption{The PC-L3R point-coupling constants, pairing strengths, and the respective uncertainties.} %\textsc{PC-L3R}
\label{tab:PCL3M}
%\begin{tabular}{lcc}00
%\begin{tabular}[width=\linewidth]{@{\extracolsep{\fill}}lcc}
%\begin{tabular}{@{\extracolsep{\fill}}lr}
\begin{tabular}{@{}@{\hspace{2mm}}l@{\hspace{25mm}}r@{\hspace{2mm}}@{}}
\toprule
\midrule
%\hline
%\hlinehttps://www.overleaf.com/project/6078f9e923c7c00d2fd483b7
%$\rm {Coupling~~constant}$ & $\rm{Quantity}$ and $\rm {physical~unit(s)}$\\
Coupling & Quantity, uncertainty$^{a}$ and  \\
constant & physical unit(s) \\
\hline
% $\alpha_{S} $             & $-3.99289^{+0.00004}_{-0.00004}\times10^{-04}$  $\rm MeV^{-2}$   \\
% $\beta_{S}  $             & $8.65504^{+0.0007}_{-0.0007}\times10^{-11}$    $\rm MeV^{-5}$   \\
% $\gamma_{S} $             & $-3.83950^{+0.0009}_{-0.0009}\times10^{-17}$  $\rm MeV^{-8}$   \\
% $\delta_{S} $             & $-1.20749^{+0.0055}_{-0.0055}\times10^{-10}$  $\rm MeV^{-4}$   \\
% $\alpha_{V} $             & $2.71991^{+0.00004}_{-0.00004}\times10^{-04}$    $\rm MeV^{-2}$   \\
% $\gamma_{V} $             & $-3.72107^{+0.008}_{-0.008}\times10^{-18}$  $\rm MeV^{-8}$   \\
% $\delta_{V} $             & $-4.26653^{+0.005}_{-0.005}\times10^{-10}$  $\rm MeV^{-4}$   \\
% $\alpha_{TV}$             & $2.96934^{+0.016}_{-0.016}\times10^{-05}$   $\rm MeV^{-2}$   \\
% $\delta_{TV}$             & $-4.65682^{+0.25}_{-0.25}\times10^{-10}$  $\rm MeV^{-4}$   \\ 
%\hline
% $G_{n}~     $             & $749.552^{+3.5}_{-3.5}$                $\rm MeV~fm^{3} $ \\
% $G_{p}~     $             & $680.080^{+20}_{-20}$                $\rm MeV~fm^{3} $ \\ 
% $a~         $             & $0.62948^{+0.0015}_{-0.0015}$                   $\rm fm     $     \\
%%  \hline\hline
 $\alpha_{S} $             & $-3.99289\!\pm\!0.00004\!\times10^{-04}$ $\rm MeV^{-2}$   \\
 $\beta_{S}  $             & $8.65504\!\pm0.0007\!\times10^{-11}$     $\rm MeV^{-5}$   \\
 $\gamma_{S} $             & $-3.83950\!\pm0.0009\!\times10^{-17}$    $\rm MeV^{-8}$   \\
 $\delta_{S} $             & $-1.20749\!\pm0.0055\!\times10^{-10}$    $\rm MeV^{-4}$   \\
 $\alpha_{V} $             & $2.71991\!\pm0.00004\!\times10^{-04}$    $\rm MeV^{-2}$   \\
 $\gamma_{V} $             & $-3.72107\!\pm0.008\!\times10^{-18}$     $\rm MeV^{-8}$   \\
 $\delta_{V} $             & $-4.26653\!\pm0.005\!\times10^{-10}$     $\rm MeV^{-4}$   \\
 $\alpha_{TV}$             & $2.96688\!\pm0.016\!\times10^{-05}$      $\rm MeV^{-2}$   \\
 $\delta_{TV}$             & $-4.65682\!\pm0.25\!\times10^{-10}$      $\rm MeV^{-4}$   \\ 
\hline
 $G_\mathrm{n}$            & $-749.552\!\pm\!3.5$                    $\rm MeV~fm^{3} $ \\
 $G_\mathrm{p}$            & $-680.080\!\pm\!20$                     $\rm MeV~fm^{3} $ \\ 
 $a           $            & $0.62948\!\pm\!0.0015$                 $\rm fm     $     \\
\bottomrule
\end{tabular}
\begin{minipage}{\linewidth}
\footnotesize
\vskip5pt
{\sc Note}\\
%\textbf{Note.}\\ a_1, a_2, a_3, a_4, a_5, and
$^a$ The uncertainties of the coupling constants and pairing strengths are constrained by the close reproduction of the root-mean-square deviation (rms) value, 1.339~MeV (Table~\ref{tab:RMS}), of comparing the theoretical and experimental binding energies of $91$ spherical nuclei listed in Table~\ref{tab:BindingEnergy}. The close reproduction of rms is limited to $\lessapprox\!20$~kev different from the rms. 
%, and each calculated binding energy is $\lessapprox\!600$~kev ($0.85\%$) different from the respective value in Table~\ref{tab:BindingEnergy}. 
Meanwhile, the rms of the charge radii of $63$ nuclei is maintained in the range of $0.0185\!\leq\!\mathrm{rms}\!\leq\!0.0187$, and the change of $\chi^{2}$ in Eq.(\ref{eq:minimization}) is limited to $\Delta\!\chi^2(\mathbf{P})\!\lessapprox\!5$.
% Each individual binding energy and charge radii
\end{minipage}
\end{table}
\renewcommand\arraystretch{1.0}

\subparagraph{\label{Sec:Energies}\normalsize\textbf{\emph{Binding energies of the selected spherical nuclei}}}\!\!--The root mean square (rms) deviation values and the root of relative square (rrs) percentages of comparing the theoretical and experimental binding energies of $60$, $65$, and $91$  spherical nuclei are listed in Table~\ref{tab:RMS}.
%\textcolor{black}{The comparison of the rms values of binding energies of $60$, $65$, and $91$ spherical nuclei produced from all theoretical models in Table~\ref{tab:RMS} indicates} that 
The present optimized PC-L3R interaction used in the RHB model with the separable pairing force yields the lowest rms
%\footnote{\scriptsize${\rm rms }=\sqrt{\left(\sum_{i}^{N}(E_{i}^{\rm expt.}-E_{i}^{\rm calc.})^{2}\right)/N}$} 
values, i.e., $1.176$ MeV (rms$_{60}$), $1.245$ MeV (rms$_{65}$), and $1.339$ MeV (rms$_{91}$).
% among all theoretical models in Table~\ref{tab:BindingEnergy}. 
The PC-PK1, or PC-X, or DD-PC1, or DD-PCX, or PC-F1, or PC-LA interaction in the RHB model with the separable paring force produces a rather high rms deviations, $1.248$-$3.269$ MeV (rms$_{60}$), $1.312$-$3.206$ MeV (rms$_{65}$), and $1.453$-$2.925$ MeV (rms$_{91}$) %, compared with the experimental binding energies 
(models labeled with a superscript asterisk in Table~\ref{tab:RMS}).
The experimental and theoretical binding energies of the selected 91 spherical nuclei are listed in Table~\ref{tab:BindingEnergy}.
%\textcolor{blue}{Similar results are obtained of the rms$_{60}$ ($XXX$ MeV) and rms$_{65}$ ($XXX$ MeV).}
Hereafter, the labels of all models with the specific point-coupling interaction and pairing force are referred to the footnote of Table~\ref{tab:RMS}, except when otherwise specified.

%We remark that the RHB model with PC-PK1 interaction and separable pairing force produces an inverse order of the binding energies of $^{214}$Ra and $^{214}$Pb compared to experiment. By using the RHB model with PC-L3R interaction, the order is restored and comparable with experiment. The \textcolor{black}{yielded $^{214}$Ra and $^{214}$Pb energies} from PC-PK1, PC-PK1$^{\ddag}$, and DD-PC1 listed in Table~\ref{tab:BindingEnergy} also produce the same order as experiment but not the \textcolor{black}{yielded energies} from PC-PK$^\ddag$, DD-PC1$^*$, DD-PCX, PC-F1, PC-F1$^*$, PC-LA, and PC-LA$^*$.

%interaction of separable form of the Gogny force are used in DIRHB code~\cite{CPC2014}, the results of the binding energies get worse. Compared with the PC-L3R interaction, the RMS increase 220~keV.

%\textcolor{black}{Note that the density-independent $\delta$ interaction \cite{PRC2010Zhao,PRC2002Burvenich,PRC2006Niksic_034308,PRC2006Niksic_064309} is used as the pairing field for the RMF+BCS framework.} 

\begin{table*}[pos=ht!,width=\textwidth]
%\nonumber
%\renewcommand{\thetable}{2}
%\centering
\caption{\footnotesize  The root mean square (rms) deviation values and the root of relative square (rrs) percentages of comparing the theoretical and experimental binding energies (BE), and charge radii (CR) for 60, 65, and 91 spherical nuclei, using the PC-L3R, PC-PK1, PC-X, DD-PC1, DD-PCX, PC-F1, and PC-LA interactions. The numbers of the nuclei considered in the calculations are given in subscript. The detailed data is listed in Tables~\ref{tab:BindingEnergy} and \ref{tab:ChargeRadii} in the Supplemental Material.}
%$\scriptstyle {\rm RMS} = \scriptstyle \sqrt{\Sigma_{i}^{N}(E^{i}_{expt.}-E^{i}_{calc})^{2}/N}$
%The experimental values of the corresponding nuclei are used in the parametrization fitting.
\label{tab:RMS}
\scriptsize
%\tiny
%\resizebox{.90\columnwidth}{!}{
%\begin{tabular}{lcccccccccccc}
\begin{tabular}{@{}l@{\hspace{2mm}}l@{\hspace{2mm}}c@{\hspace{2mm}}c@{\hspace{2mm}}c@{\hspace{2mm}}c@{\hspace{2mm}}c@{\hspace{2mm}}c@{\hspace{2mm}}c@{\hspace{2mm}}c@{\hspace{2mm}}c@{\hspace{2mm}}c@{\hspace{2mm}}c@{\hspace{2mm}}c@{\hspace{2mm}}c@{}}
%\hline\hline
\toprule
\midrule
 &  &  PC-L3R$^a$ & PC-PK1$^b$ & PC-PK1$^{\ast c}$ & PC-PK1$^{\ddag d}$ & PC-X$^e$ & DD-PC1$^f$ & DD-PC1$^{\ast g}$ & DD-PCX$^{h}$ & PC-F1$^i$ & PC-F1$^{\ast j}$ & PC-LA$^k$ & PC-LA$^{\ast l}$  \\
%\hline
\midrule
BE & $\rm\mathbf{ rms_{60}}$$^m$ (MeV)   &     $ 1.176 $   &  $ 1.248  $  &  $ 1.417   $  &  $ 1.64   $ &  $ 1.378   $ &  $ 3.139  $  &  $ 3.269  $ &  $ 1.413  $ &  $ 2.589  $ &  $ 2.913  $ &  $ 2.581  $&  $ 2.890  $\\  

& $\rm\textcolor{blue}{\mathbf{rms_{65}}}$$^{n}$ (MeV)     &     $ 1.245 $   &  $ 1.312  $  &  $ 1.495   $  &  $ 1.69   $ &  $ 1.542   $ &  $ 3.088  $  &  $ 3.206  $ &  $ 1.461  $ &  $ 2.590  $ &  $ 2.895  $ &  $ 2.615  $&  $ 2.915  $\\                                                     
& $\rm rms_{91 }$$^{p}$ (MeV)  &     $ 1.339 $   &  $ 1.453  $  &  $ 1.796   $  &  $ 2.11   $ &  $ 2.168   $ &  $ 2.805  $  &  $ 2.925  $ &  $ 1.712  $ &  $ 2.493  $ &  $ 2.755  $ &  $ 2.555  $&  $ 2.714  $ \\ 
%\hline
\cline{2-14}
&$\rm\mathbf{rrs_{60}}$ ($\%$) &     $ 0.22 \% $   &  $ 0.18 \%  $  &  $ 0.22 \%   $  &  $ 0.39 \%   $ &  $ 0.20 \%   $ &  $ 0.47 \%  $  &  $ 0.50 \% $ &  $ 0.22 \%  $ &  $ 0.30 \%  $ &  $ 0.39 \%  $ &  $ 0.29 \%  $&  $ 0.34 \% $\\  

&$\rm\textcolor{blue}{\mathbf{rrs_{65}}}$  ($\%$)   &     $ 0.23 \% $   &  $ 0.19 \%  $  &  $ 0.23 \%   $  &  $ 0.39\%   $ &  $ 0.22 \%   $ &  $ 0.45 \%  $  &  $ 0.49 \%  $ &  $0.23 \% $ &  $ 0.32 \%  $ &  $ 0.40 \%  $ &  $ 0.30 \%  $&  $ 0.35 \%  $\\                                                     
&$\rm rrs_{91 }$ ($\%$) &     $ 0.24\% $   &  $ 0.23 \%  $  &  $ 0.27 \%   $  &  $ 0.39 \%   $ &  $ 0.27 \%   $ &  $ 0.44 \%  $  &  $ 0.48 \%  $ &  $ 0.26 \%  $ &  $ 0.35 \%  $ &  $ 0.40 \%  $ &  $ 0.32 \%  $&  $ 0.35 \%  $ \\ 
\hline
CR &$\rm \mathbf{rms_{23}}$$^{q}$ (fm) &     $ 0.0224 $   &  $ 0.0222 
$  &  $ 0.0220    $  &  $ 0.0217   $ &  $ 0.0222   $ &  $ 0.0197  $  &  $ 0.0197 $&  $ 0.0195 $ &  $ 0.0197 $  &  $ 0.0194 $ &  $ 0.0250  $ &  $ 0.0247  $\\  

&$\rm rms_{63}$$^{r}$  (fm)  &     $ 0.0187 $   &  $ ~  $  &  $ 0.0183   $  &  $ 0.0182   $ &  $ 0.0187   $ &  $ ~  $  &  $ 0.0157  $ &  $ 0.0153 $&  ~ &  $ 0.0153  $ &  $ ~  $ &  $ 0.0211  $ \\
%\hline
\cline{2-14}
&$\rm\mathbf{rrs_{23}}$ ($\%$)  &     $ 0.68 \% $   &  $ 0.68 \%  $  &  $ 0.67 \%   $  &  $ 0.67 \%   $ &  $ 0.69 \%   $ &  $ 0.58 \%  $  &  $ 0.59 \% $ &  $ 0.61 \% $&  $ 0.61 \%  $ &  $ 0.61\%  $ &  $ 0.67 \%  $ &  $ 0.66 \%  $\\  

&$\rm rrs_{63}$  ($\%$)   &     $ 0.51 \% $   &  $~  $  &  $ 0.50 \%   $  &  $ 0.50\%   $ &  $ 0.52 \%   $ &  $ ~  $  &  $ 0.44 \%  $ &  $ 0.43 \% $&  ~ &  $ 0.43\%  $ &  $ ~  $ &  $ 0.51\%  $ \\
%\hline\hline
\bottomrule
\end{tabular}
%}
%\begin{minipage}{\columnwidth}
\begin{minipage}{\textwidth}
%\vskip5pt
%{\sc Note}---\\
% \textbf{Note.}\\
\tiny
%The experimental binding energies compiled in AME2020 \cite{AME2020}. 
$^a$ PC-L3R: calculated from the RHB model with the PC-L3R interaction.
$^b$ PC-PK1: quoted from the RMF model with the Bardeen–Cooper–Schrieffer (BCS) theory using the PC-PK1 interaction \cite{PRC2010Zhao} and $\delta$ pairing force.
$^c$ PC-PK1$^*$: calculated from the RHB model using the PC-PK1 interaction \cite{PRC2010Zhao} and separable pairing force \cite{PLB2009Tian_676_44}.
$^d$ PC-PK1$^\ddag$: quoted from the relativistic continuum Hartree-Bogoliubov (RCHB) model \cite{ADNDT2018Xia} using the PC-PK1 interaction \cite{PRC2010Zhao} and $\delta$ pairing force.
$^e$ PC-X: calculated from the RHB model using the PC-X interaction \cite{PLB2020Taninah} and separable pairing force \cite{PLB2009Tian_676_44}.
$^f$ DD-PC1: quoted from the RMF model with the BCS theory using the DD-PC1 interaction \cite{PRC2008Niksic} and $\delta$ pairing force.
$^g$ DD-PC1$^*$: calculated from the RHB model using the DD-PC1 interaction \cite{PRC2008Niksic} and separable pairing force \cite{PLB2009Tian_676_44}.
$^h$ DD-PCX: calculated from the RHB model using the DD-PCX interaction \cite{PRC2019Yuksel} and separable pairing force \cite{PRC2019Yuksel}.
$^i$ PC-F1: quoted from the RMF model with the BCS theory using the PC-F1 interaction \cite{PRC2002Burvenich} and $\delta$ pairing force.
$^j$ PC-F1$^*$: calculated from the RHB model using the PC-F1 interaction \cite{PRC2002Burvenich} and separable pairing force \cite{PLB2009Tian_676_44}.
$^k$ PC-LA: quoted from the RMF model with the BCS theory using the PC-LA interaction \cite{PRC1992Nikolaus} and $\delta$ pairing force.
$^l$ PC-LA$^*$: calculated from the RHB model using the PC-LA interaction \cite{PRC1992Nikolaus} and separable pairing force \cite{PLB2009Tian_676_44}.
%of comparing the theoretical and experimental binding energies are obtained by equation
The rms values and rrs percentages are expressed as, 
%rms$=\!\!\sqrt{\left(\sum_{i}^{N}(E_{i}^{\rm expt.}-E_{i}^{\rm calc.})^{2}\right)/N}$ and 
%rrs$=\!\!\sqrt{\left(\sum_{i}^{N}(E_{i}^{\rm expt.}-E_{i}^{\rm calc.})^{2}/(E_{i}^{\rm expt.})^{2}\right)/N}$, 
rms$=\!\!\left[\left(\sum_{i}^{N}(E_{i}^{\rm expt.}-E_{i}^{\rm calc.})^{2}\right)/N\right]^{1/2}$ and 
rrs$=\!\!\left[\left(\sum_{i}^{N}(E_{i}^{\rm expt.}-E_{i}^{\rm calc.})^{2}/(E_{i}^{\rm expt.})^{2}\right)/N\right]^{1/2}$, 
respectively.
$^m$ rms$_{60}$ (bold text): obtained from comparing the experimental and theoretical binding energies of $60$ spherical nuclei, selected for fitting the PC-PK1~\cite{PRC2010Zhao}. 
$^n$ rms$_{65}$ (blue text): obtained from comparing the experimental and theoretical binding energies of $65$ spherical nuclei, referred by previous works in fitting point coupling interactions~\cite{PRC2010Zhao,PRC2019Yuksel,PLB2020Taninah}. 
$^o$ rms$_{91}$: obtained from comparing the experimental and theoretical binding energies of $91$ spherical nuclei, selected for fitting the PC-L3R. %in this work
$^p$ rms$_{23}$ (bold text): obtained from comparing the experimental and theoretical  charge radii of $23$ spherical nuclei, selected for fitting the PC-PK1~\cite{PRC2010Zhao}.
$^r$ rms$_{63}$: obtained from comparing the experimental and theoretical charge radii of $63$ spherical nuclei, selected for fitting the PC-L3R.
The rrs have a similar label as rms.
%\vspace{10mm}
\end{minipage}
\end{table*}

\subparagraph{\label{Sec:Radii}\normalsize\textbf{\emph{Charge radii of the selected spherical nuclei}}}\!\!--
The theoretical radii of the selected $63$ spherical nuclei are compared with the updated experimental data~\cite{ADNDT2013Angeli} %, which was recently compiled by Angeli \& Marinova \cite{ADNDT2013Angeli} 
(Table~\ref{tab:ChargeRadii}). Overall, the low rms deviations, 0.0153--0.0211~fm, of comparing the calculated radii from each theoretical model with experimental radii indicate that all theoretical radii are in good agreement with experiment, especially the DD-PCX and PC-F1 yields the lowest rms deviation, $0.0153$~fm. The rms value of the RHB model with PC-L3R interaction, 0.0224~fm (rms$_{23}$) and 0.0187~fm (rms$_{63}$), closely aligns with the rms values of other theoretical models (Table~\ref{tab:RMS}). 
In additions, the neutron skin thickness of $^{208}$Pb produced from PC-L3R is closely aligned with the results of RHB frameworks using other nonlinear point-coupling interactions, and agrees with the latest PREX-2 experiment \cite{PRLAdhikari2021} (Fig.~\ref{fig:rnp-Pb208}).
%~\cite{SupplementMaterial}
%We may presume that charge radii is not sensitive to the optimized point-coupling interaction PC-L3R.

%and the equation of state 
\subparagraph{\label{Sec:EOS}\normalsize\textbf{\emph{Saturation properties of the symmetric nuclear matter}}}\!\!-- 
%We study the saturation properties and equation of state for the symmetric nuclear matter based on the covariant density functional of the PC-L3R interaction. 
These properties include the saturation density, $\rho_{0}$, binding energy per nucleon, $E\!/\!A$, effective Dirac mass, $M^{\ast}$, incompressibility, $K_{0}$, symmetry energy, $E_{\rm sym}$, and slope of symmetry energy, $L_{0}$. 

Comparing the properties generated by PC-L3R with the ones resulted from 
%other point-coupling interactions, i.e., 
PC-PK1 \cite{PRC2010Zhao},
PC-X \cite{PLB2020Taninah},
DD-PC1 \cite{PRC2008Niksic},
DD-PCX \cite{PRC2019Yuksel},
PC-F1 \cite{PRC2002Burvenich}, and PC-LA \cite{PRC1992Nikolaus}, %overall, 
we find that all point-coupling interactions produce a set of rather similar $\rho_{0}$ with the average as $0.152^{+0.002}_{-0.004}$~fm$^3$, and $E\!/\!A$ with the average as $-16.11^{+0.08}_{-0.06}$~MeV (Table~\ref{Tab3}). This indicates that PC-L3R not only exhibits the capability of describing the saturation properties of nuclear matter conformed with other nonlinear effective point-coupling interactions, but also agrees with the empirical $\rho_{0}\!=\!0.166\!\pm\!0.018$~fm$^3$ and $E\!/\!A\!=\!-16\!\pm\!1$~MeV~\cite{PRC1990Brockmann}. Besides, we note that the nonlinear effective interactions PC-L3R, PC-PK1, PC-F1, PC-LA, and PC-X produce a set of $\rho_{0}$ and $E\!/\!A$ values somehow larger than the saturation properties produced by the density dependence point-coupling interactions, DD-PC1 and DD-PCX \textcolor{black}{(Fig.~\ref{fig:Eos_EA})}. 

%Similar results for saturation density $\rho_{0}\approx 0.152$~fm$^3$ and binding energy per nucleon $E/A \approx -16.1$~MeV are predicted by all the point-coupling interactions, which agree well with the empirical values $0.166\pm 0.018$~fm$^3$  and $-16\pm1$~MeV~\cite{PRC1965Brockmann}. Moreover, the nonlinear effective interactions PC-L3R, PC-PK1, PC-X, PC-F1 and PC-LA are always larger than the density dependence effective interactions DD-PC1 for the other saturation properties in Table~\ref{Tab3}.

\begin{table}[pos=t,width=\columnwidth]
\renewcommand{\thetable}{3}
%\vspace{-7mm}
%\scriptsize
\scriptsize
  \caption{\label{Tab3}\footnotesize The saturation properties of nuclear matter estimated by the nonlinear point-coupling interactions, i.e., PC-L3R, PC-PK1 \cite{PRC2010Zhao}, PC-X \cite{PLB2020Taninah}, PC-F1 \cite{PRC2002Burvenich}, and PC-LA \cite{PRC1992Nikolaus} and the density-dependent point-coupling interactions, DD-PC1 \cite{PRC2008Niksic} and DD-PCX \cite{PRC2019Yuksel}. 
  \textcolor{black}{The symmetric nuclear matter energy plot is shown in Fig.~\ref{fig:Eos_EA}.}
  }
  %\begin{tabular}{lcccccc}
  %\begin{tabular}{@{\hspace{0mm}\extracolsep{\fill}}lcccccc}
  \begin{tabular}{@{}l@{\hspace{1.5mm}}c@{\hspace{1.5mm}}c@{\hspace{1.5mm}}c@{\hspace{1.5mm}}c@{\hspace{1.5mm}}c@{\hspace{1.5mm}}c@{\hspace{1.5mm}}c@{}}  
   \toprule
   \midrule
%  \hline\hline
                          & PC-L3R & PC-PK1   & PC-X   & DD-PC1& DD-PCX & PC-F1 & PC-LA \\
   \hline
   $\rho_0$~(fm$^{-3}$)   &$0.153$ & $0.154$ &  $0.154$ &$0.152$&$0.152$ &$0.151$  & $0.148$  \\
   $E/A$~(MeV)            &$-16.12$&$-16.12$ &  $-16.14$&$-16.06$ &$-16.03$  &$-16.17$ & $-16.13$\\
   $M^{\ast}/M$           & $0.59$ &  $0.59$ &  $0.58$  &  $0.58$ &  $0.56$   &$0.61$   & $0.58$\\
   $K_{0}$~(MeV)          &$245$   & $238$   &   $240$  & $230$  & $213$         &$255$    & $264$  \\
   $E_{\rm sym}$~(MeV)    &$35.8$  & $35.6$  &   $35.2$ & $33$ & $31$  & $37.8$  &$37.2$ \\
   $L_{0}$ (MeV)          &$114   $& $113$   &   $112$  & $70$ & $46$  & $117$   & $108$   \\
%   $K_{\rm asy}$ (MeV)    &        &         &         &        &       &       \\
%  \hline\hline
  \bottomrule
  \end{tabular}
%\begin{tablenotes}
%%\begin{minipage}
%%\footnotesize
%\item[a] The reproduced results in this work.
%%\end{minipage}
%\end{tablenotes}
%\vspace{-10mm}
\end{table}

%Note that the inconsistent description at the high-density nuclear matter region, $\textcolor{black}{\rho_{\!B}\geq0.30}$, between the nonlinear point-coupling interactions and the DD-PC1 and \emph{ab initio} variational calculations does not affect the capability of point-coupling interactions in describing the low-energy bulk nuclear properties. \textcolor{blue}{(Note that the inconsistent description at the high-density nuclear matter region, $\textcolor{black}{\rho_{\!B}\geq0.30}$, which has little influence on the description of low-energy bulk nuclear properties.)}

\begin{figure}[pos=t]
\includegraphics[width=0.45\textwidth]{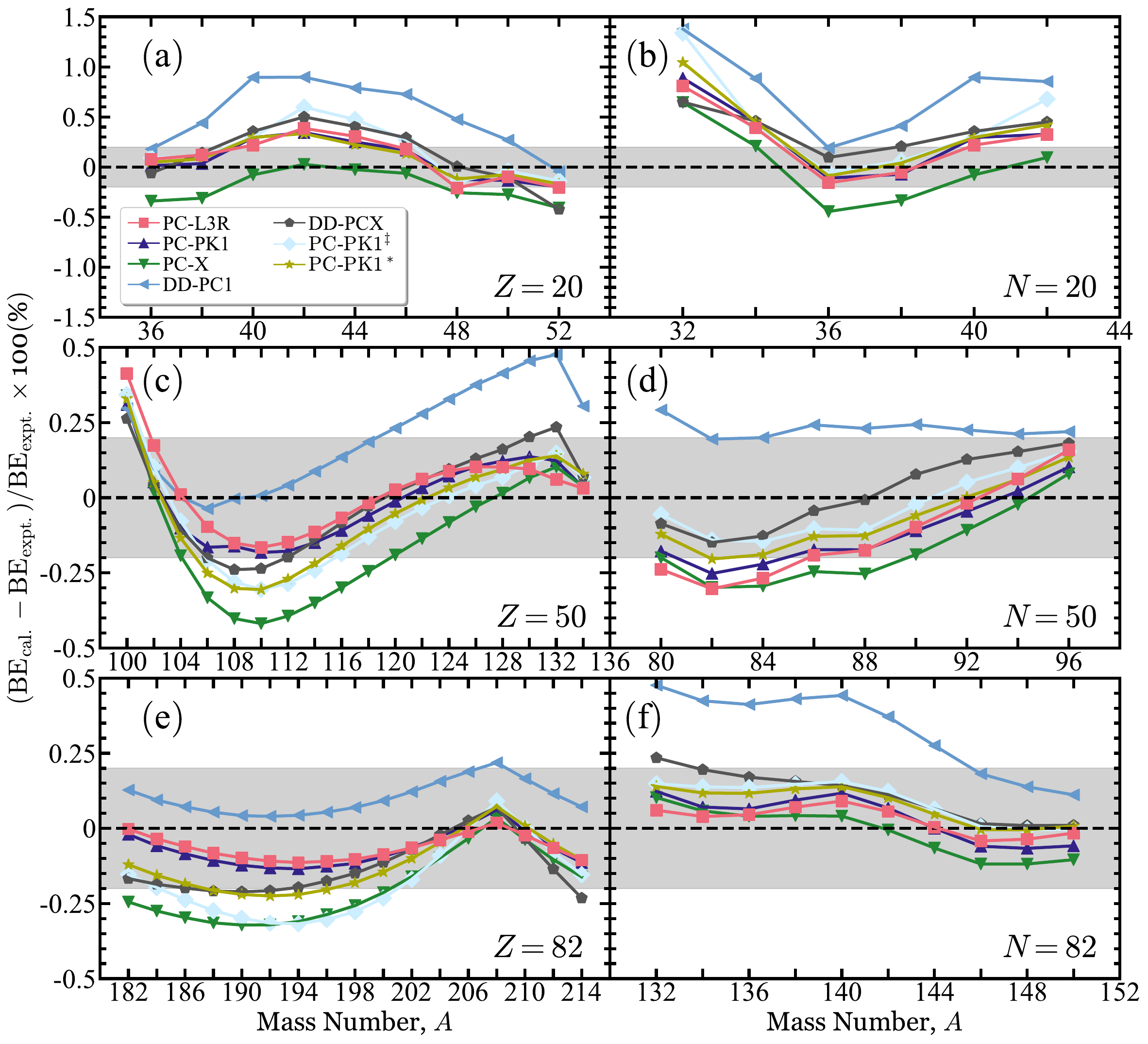}
%\vspace{-0.4cm}
\caption{\footnotesize The relative accuracy of the theoretical description of the AME2020 experimental data \cite{AME2020} for the binding energies of the $Z\!=\!20$, $50$, and $82$ isotopes, and for the $N\!=\!20$, $50$, and $82$ isotones. See the footnote of Tables~\ref{tab:RMS} and \ref{tab:BindingEnergy} for the theoretical frameworks in this figure. 
%The data of PC-PK1 (black triangles) and DD-PC1 (blue squares) are from Ref.~\cite{PRC2010Zhao}, whereas the PC-PK1$^\ddag$ data is from Ref.~\cite{ADNDT2018Xia}. 
The gray regions in all panels are the $\pm0.2~\!\%$ error band.}
%\cite{PLB2020Taninah}, \cite{PRC2008Niksic}
%The results of PC-L3R (red squares) (or PC-X (green up triangles) or PC-PK1 (cyan left triangles) with separable pairing force) interaction are calculated in the present work. 
\label{Fig2}
%(Color online) 
\vspace{-7mm}
\end{figure}

%\begin{figure}[pos=htb]
%\includegraphics[width=0.48\textwidth]{odd-even-staggering_Sn.pdf}
%%\vspace{-0.4cm}
%%\includegraphics[width=0.48\textwidth]{Delta.pdf}
%\caption{\footnotesize \textcolor{blue}{Single neutron separation energy $S_{n}$ obtained in the AME2020 experimental data \cite{AME2020} (black filled square) and the calculated binding energies (red filled circle) for the Ca (a), Ni (b), Sn (c), and Pb (d) isotopes}.}
%\label{fig:Single neutron}
%%(Color online)
%\vspace{-5mm}
%\end{figure}
%\begin{figure}[pos=htb]
%\includegraphics[width=0.48\textwidth]{odd-even-staggering_Sp.pdf}
%%\vspace{-0.4cm}
%%\includegraphics[width=0.48\textwidth]{Delta.pdf}
%\caption{\footnotesize \textcolor{blue}{Same as Fig.\ref{fig:Single neutron} but single proton separation energy $S_{p}$ for the N=20 (a), N=50 (b), N=82 (c), and N=126 (d) isotones.}}
%\label{fig:Single proton}
%%(Color online)
%\vspace{-5mm}
%\end{figure}
\begin{figure}[pos=htb]
\centering
%%trim={<left> <lower> <right> <upper>}
\includegraphics[width=0.45\textwidth,trim={0 0 0 0}]{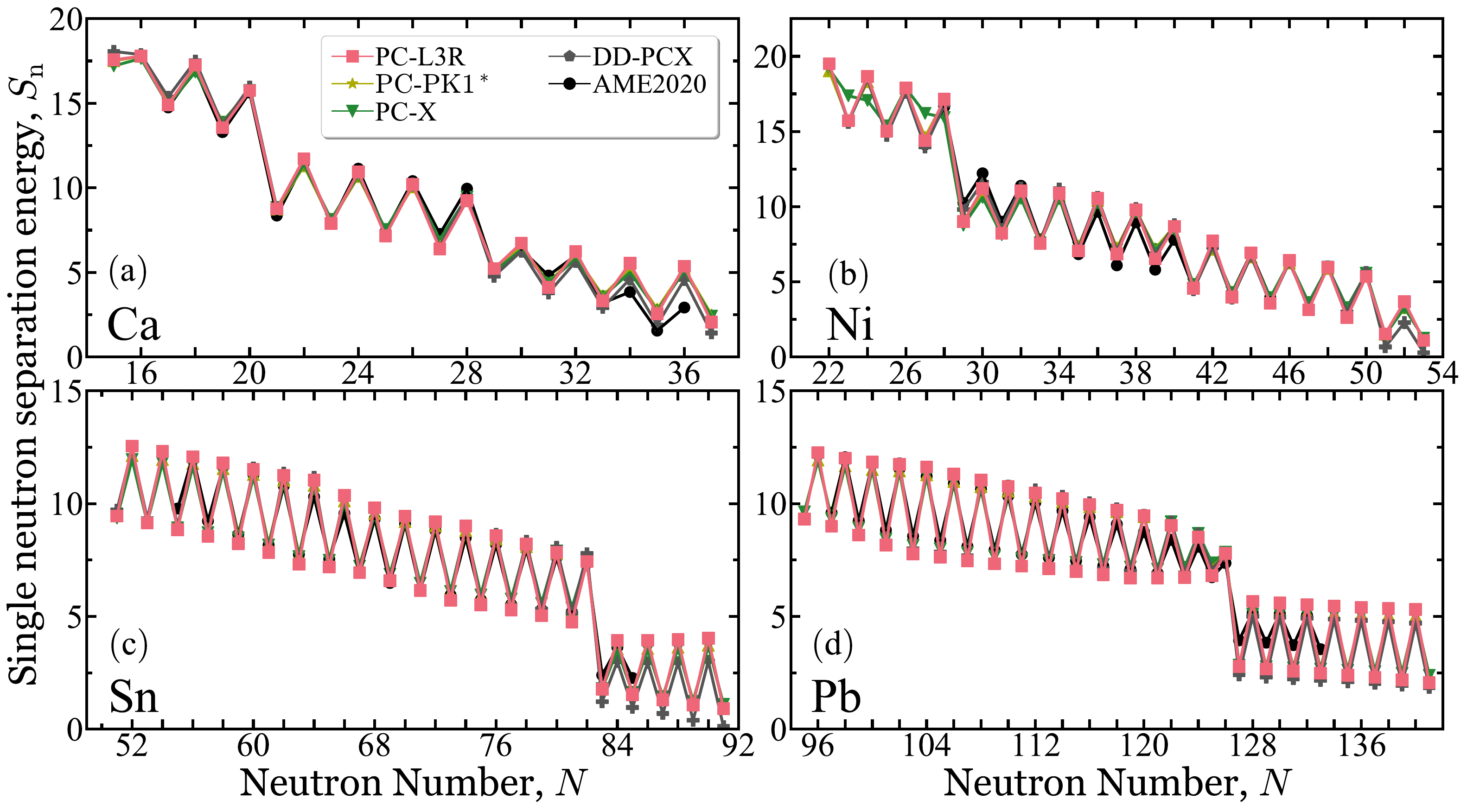}
\vspace{-4mm}
{\noindent}\rule[5mm]{0.48\textwidth}{0.05em}
\includegraphics[width=0.45\textwidth]{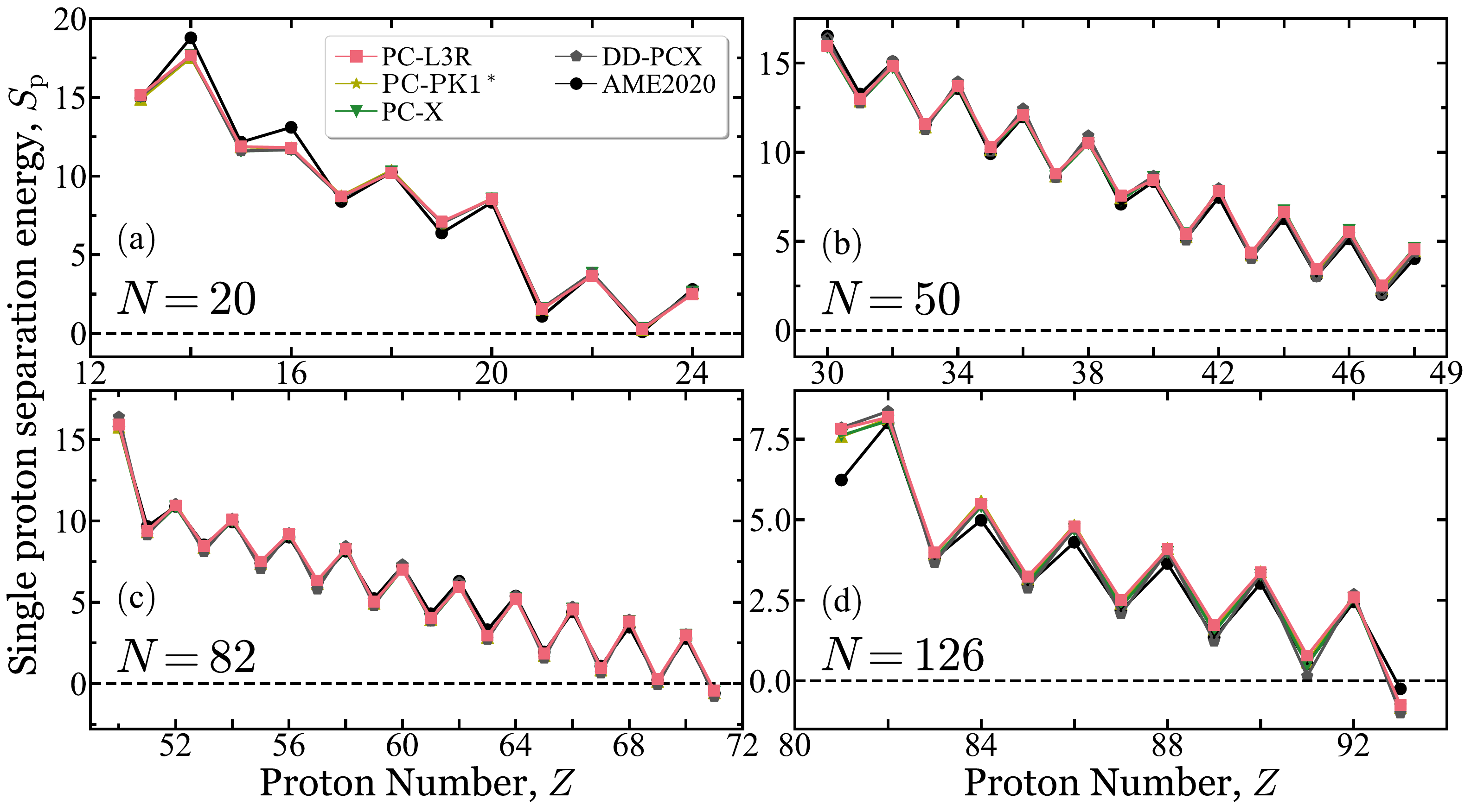}
%\vspace{-0.4cm}
%\includegraphics[width=0.48\textwidth]{Delta.pdf}
\caption{\footnotesize Single nucleon separation energy, $S_{i}$ ($i\!=\!$ n or p), obtained from the AME2020 experimental data~\cite{AME2020} (black dots) and from theoretical calculations. Top panel: $S_\mathrm{n}$ for the Ca, Ni, Sn, and Pb isotopes. Bottom panel: $S_\mathrm{p}$ for the $N\!=\!20$, $50$, $82$, and $126$ isotones. Experimental uncertainties are smaller than the symbol.}
\label{fig:Single_nucleon}
%(Color online)
\vspace{-5mm}
\end{figure}

\section{Binding energies of nuclei}
\label{Sec:BindingEnergies}
\subparagraph{\label{Sec:BEs}\normalsize\textbf{\emph{Binding energies of isotopic and isotonic chains}}}\!\!--
We present the comparison of the experimental and theoretical binding energies for some isotopic and isotonic chains based on the point-coupling interactions PC-L3R, PC-PK1, PC-X, DD-PC1, and DD-PCX used in the RHB, RMF, and RCHB frameworks in Fig.~\ref{Fig2}. For the $Z\!\!=\!\!20$ (Ca) isotopes, $N\!\!=\!\!20$ and $N\!\!=\!\!50$ isotones, the results from PC-L3R, PC-PK1, and PC-PK1$^*$ are rather similar, whereas for the $Z\!\!=\!\!50$~(Sn) and $Z\!\!=\!\!82$~(Pb) isotopes, especially for the even-even $^{122-134}$Sn, the PC-PK1$^*$ produces a set of binding energies more deviated from experiment compared to PC-L3R and PC-PK1, indicating the robustness %superiority 
of PC-L3R accustomed to the RHB model that yields an overall agreement with the prediction of PC-PK1 used in the RMF+BCS framework. We remark that a steep change at $^{208}$Pb is produced by PC-PK1, PC-PK1$^*$, PC-PK$^\ddag$, PC-X, DD-PC1, and DD-PCX, whereas PC-L3R results a much more reduced dip change (lower left panel in Fig.~\ref{Fig2}). A similar characteristic also happens in the $Z\!\!=\!\!50$ isotopic chain at $^{132}$Sn. For the $N\!\!=\!\!82$ isotones, PC-L3R reproduces the experimental binding energies with the average absolute deviation $0.535$~MeV, whereas PC-PK1, PC-PK1$^*$, PC-PK1$^\ddag$, PC-X, DD-PC1,  and DD-PCX produce the average absolute deviation of $0.839$ MeV, $0.931$~MeV, $1.07$~MeV, $0.818$~MeV, $4.278$~MeV, and 1.274 MeV, respectively. \textcolor{black}{This implies that the PC-L3R interaction is able to balance the Coulomb field and the isovector channel of the effective Lagrangian.} Thus, PC-L3R provides an appropriate prediction for not only the binding energy but also the isospin dependence.

We then evaluate the contribution of pairing energy, the experimental and theoretical single neutron and proton separation energies for some isotopic and isotonic chains based on the RHB frameworks using point-coupling interactions PC-L3R, PC-PK1, PC-X, and DD-PCX listed in Table~\ref{tab:RMS}. 
For the present calculations of odd-$N$ and/or odd-$Z$ nuclei, we consider the blocking effects of the unpaired nucleon(s)\noindent\footnote{\scriptsize\noindent The ground state of a nucleus with an odd neutron and/or proton number is one-quasiparticle state, $|\Phi_{1}\rangle = \beta_{i_\mathrm{b}}^{\dag}|\Phi_{0}\rangle$, which is constructed based on the ground state of an even-even nucleus $|\Phi_{0}\rangle$, c.f. Eq.~(\ref{eq:gs-wave}), where $\beta^{\dag}$ is the single-nucleon creation operator and $i_\mathrm{b}$ denotes the blocked quasiparticle state occupied by the unpaired nucleon(s). See Ref.~\cite{ManybodyProb1980} for the detail description of using the blocking effect. The same technique is employed for calculating the mirror displacement energies along the $N\!=\!Z$ line \cite{Lam2022}.}.
The PC-L3R adequately reproduces the experimental odd-even staggering trend of single nucleon separation energies (Fig.~\ref{fig:Single_nucleon}). 
%\textcolor{black}{The amplitudes of odd-even staggering effects happening in single nucleon separation energies for light (Ca and Ni) and heavy (Sn and Pb) nuclei indicates the effect imposed from the pairing force. One needs to consider a weaker pairing force for heavy nuclei, especially for $N\!>\!126$, e.g., the Pb isotope chain. This finding shows that the PC-L3R interaction exhibits a similar characteristic consistent with what was found by Afanasjev \emph{et al.}~\cite{Afanasjev2013} and by the Gogny group~\cite{Gogny}.}
The experimental single proton drip line at $^{152}$Yb ($N\!=\!82$) and $^{218}$U ($N\!=\!126$) are well reproduced by the RHB frameworks using point-coupling interactions PC-L3R, PC-PK1, and PC-X (subfigures (c) and (d) in the bottom panel of Fig.~\ref{fig:Single_nucleon}); nevertheless, the single proton drip line at $^{152}$Yb ($N\!=\!82$) generated from DD-PCX is not consistent with the experimental data.
 \textcolor{black}{Meanwhile, we find that the (absolute) magnitude of $G_\mathrm{p}$ is smaller than the $G_\mathrm{n}$ strength (Table~\ref{tab:PCL3M}). This is mainly due to the repulsive Coulomb force inducing a reduced pairing interaction for the protons~\cite{NPAAnguiano2001}.}
%本段文字表达：我们加入图3的动机是为了验证我们拟合得到的 Gn Gp a 的合理性, 并且通过单核子分离能比较直观的体现 pairing energy 贡献, 并且与实验比较，重现实验上所得到的单核子滴线。

\begin{figure}[pos=t]
%\vspace{-0.4cm}
\includegraphics[width=0.48\textwidth]{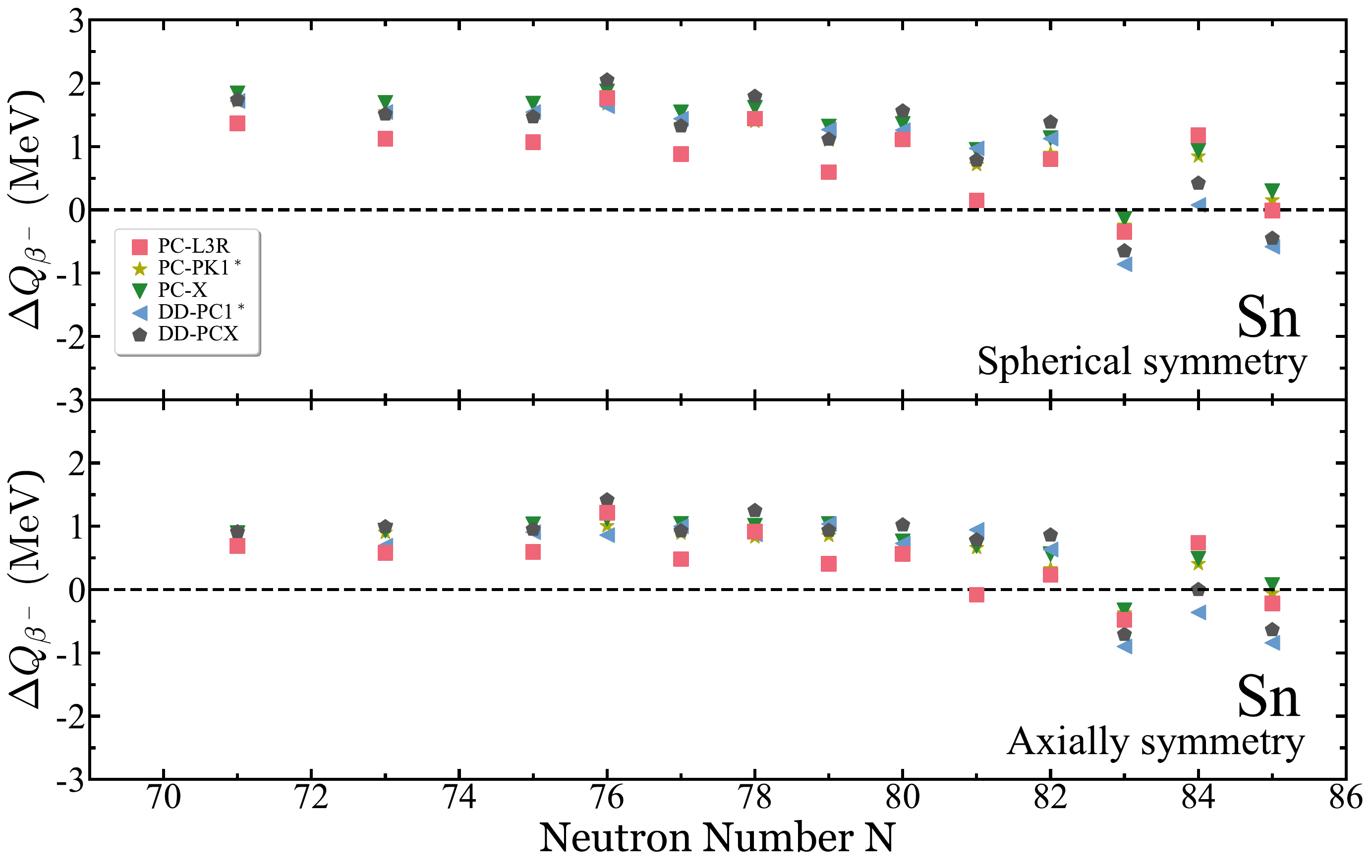}
\caption{\footnotesize \textcolor{black}{The deviations of the calculated $Q_{\beta^-}$ values from the AME2020 experimental data for the Sn isotopic chain. The top (bottom) panel shows the results of spherical (axial) symmetry}.}
%determined experimentally as well as for nuclides where the experimental value is seriously questioned
\label{fig:Q-beta}
%(Color online)
\vspace{-5mm}
\end{figure}

\textcolor{black}{
\subparagraph{\label{Sec:Q}\normalsize\textbf{\emph{$\beta$-decay $Q$ ($Q_{\beta^-}$) values}}}\!\!--
 Recent rapid-neutron capture (r-) process astrophysical models point out the decisive importance of $\beta$-decay half-lives of nuclei around magic neutron numbers $N = 50$ and $82$ \cite{Borzov2006}. We find that PC-L3R produces the lowest rms values for the $Q_{\beta^-}$ values based on spherical- or axial-symmetry calculations, i.e., $1.038$ MeV and $0.624$ MeV, respectively. The further improvement imposed from the axial-symmetry calculation is due to the existence of shape variation in the Sb isotopic chain. The deviations of the $Q_{\beta^-}$ values produced from the PC-L3R decreases with the increase of neutron number for the Sn isotopic chain (Fig.~\ref{fig:Q-beta}). This trend is very encouraging for us to use the PC-L3R interaction to predict the regions without experimentally determined $Q_{\beta^-}$ values.}

% 
%The deviations between AME2020 experimental data and the calculated $\beta^-$-decay $Q$ ($Q_{\beta^-}$) values for the Sn isotopic for the spherical (axially) symmetry calculation are considered in Fig.~\ref{fig:Q-beta} top (bottom) panel.

%红色标记:对于PC-L3R参数对于同中子素的结合能的结果, 说明了PC-L3R参数组比较好的平衡了拉氏量中的库伦场和同位旋矢量道的贡献。 
%It is noteworthy that a steep change of energy curves near the inflection point $^{208}$Pb by RHB\_PC-PK1, RHB\_PC-X and RMF$+$BCS\_PC-PK1. However, the trend of energy curve is smoother in RHB\_PC-L3R.

%These test cases are more sensitive to the newly constructed PC-L3R interaction.

\begin{figure}[pos=t]
\includegraphics[width=0.48\textwidth]{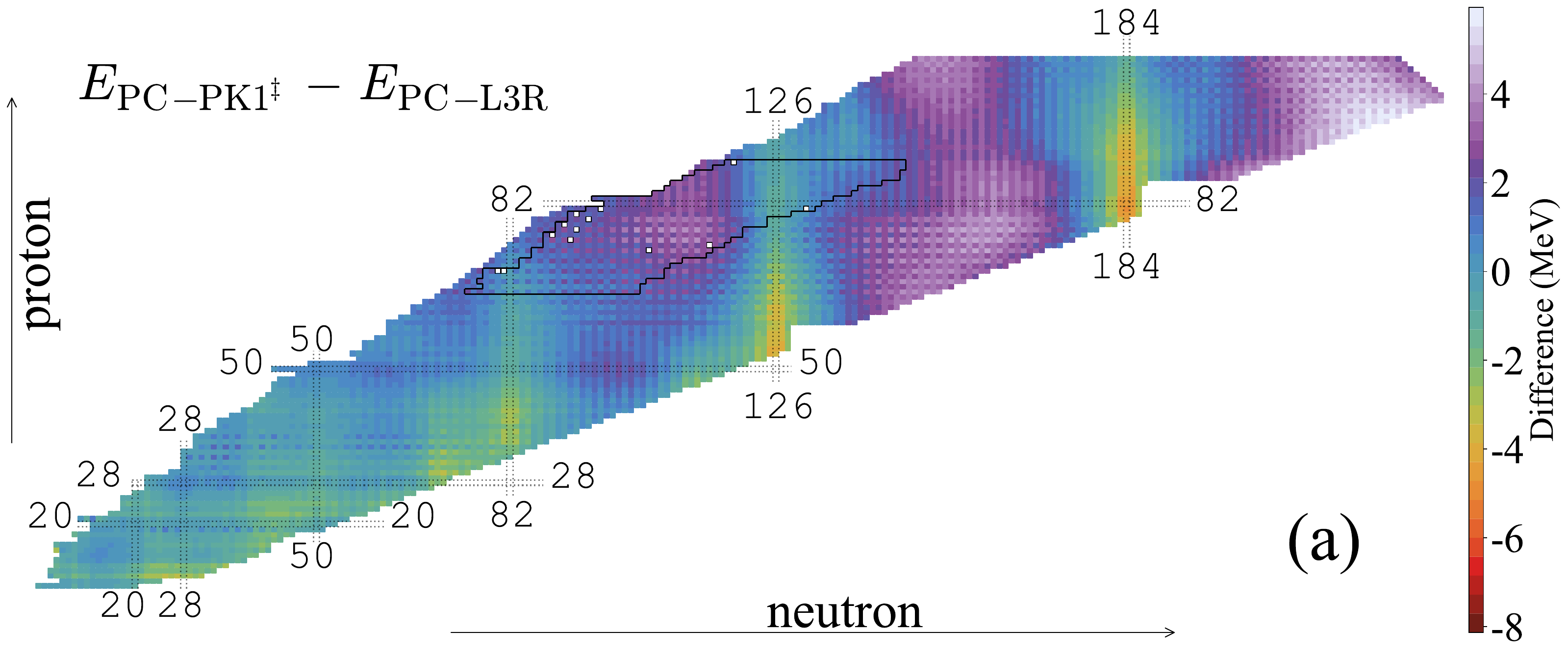}
%\vspace{-0.4cm}
\includegraphics[width=0.48\textwidth]{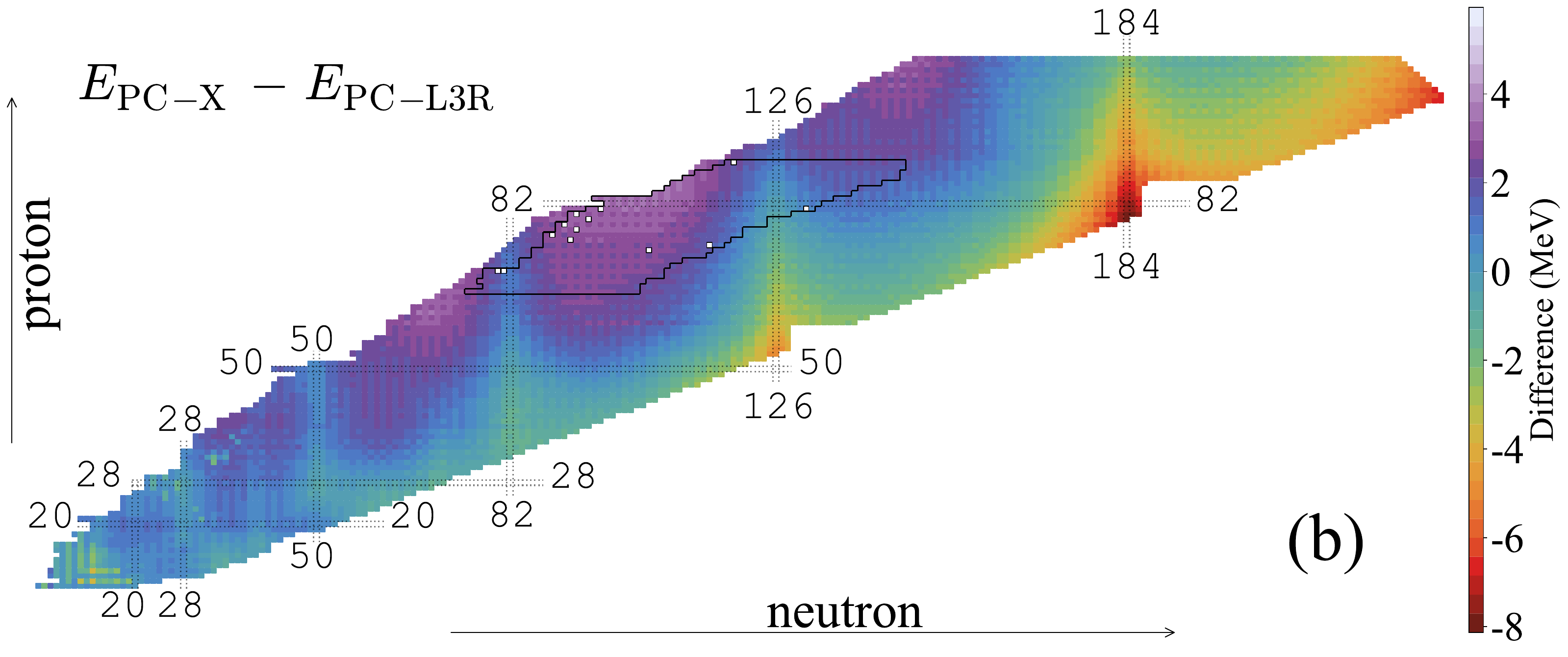}
\caption{\footnotesize The comparison of the binding energies of 7,373 nuclei calculated from the PC-L3R with the ones produced by the PC-PK1$^\ddag$ (panel a) or PC-X (panel b). See the footnote of Table~\ref{tab:BindingEnergy} for the theoretical frameworks of PC-PK1$^\ddag$, PC-X, and PC-L3R. The blank squared nuclei are excluded in the present comparison due to either unknown or ambiguous experimental data.}
%determined experimentally as well as for nuclides where the experimental value is seriously questioned
\label{fig:WeakNucChart}
%(Color online)
\vspace{-5mm}
\end{figure}

\subparagraph{\label{Sec:BE}\normalsize\textbf{\emph{Binding energies of 7,373 nuclei}}}\!\!--
We also compute the binding energies of $7,373$ nuclei using the RHB model with the PC-L3R interaction with an optimized separable pairing force. The comparison of both binding energies from PC-L3R and PC-PK1$^\ddag$ (PC-X) is given in Fig.~\ref{fig:WeakNucChart}a (\ref{fig:WeakNucChart}b). 
%We also compute and compare the binding energies of \textcolor{black}{$7373$} nuclei using the RHB model with the PC-L3R (or PC-PK1) interaction with a separable pairing force (Fig.~\ref{fig:WeakNucChart}A). The comparison of both binding energies from PC-L3R and PC-PK1$^\ddag$ is given in Fig.~\ref{fig:WeakNucChart}B. 
For the light to medium nuclei at the region $Z\!\!=\!8$-$50$ and $N\!\!=\!20$-$126$, the PC-L3R results are considerable close to the ones of PC-PK1$^\ddag$ and PC-X.
%, except at the region $Z\!\!=\!28$-$50$ and $N\!\!=\!82$-$126$ for PC-PK$^\ddag$. 
%\textcolor{black}{At the region $Z\!\!>\!\!50$ and $N\!\!>\!\!126$, the difference between the data of PC-L3R and PC-PK1$^*$ (PC-PK1$^\ddag$) is obvious.}
We compare the theoretical energies with currently available experimental data \cite{AME2020} at the region $Z\!\!=\!65$-$90$ and $N\!\!=\!75$-$150$, encircled with the black line in Figs.~\ref{fig:WeakNucChart}a and \ref{fig:WeakNucChart}b. 
The rrs for PC-L3R, PC-PK1$^\ddag$, and PC-X is $0.586\%$, $0.69\%$, and $0.723\%$, respectively. Besides, the rrs produced from the DD-PC1 and DD-PCX are $0.518\%$ and $0.692\%$, respectively.

For the heavy nuclei far from stability at regions $Z\!\!=\!60$-$110$ and $N\!\!=\!126$-$184$, PC-PK1$^\ddag$ (PC-X) estimates a set of binding energies, up to $4.59$~MeV ($8.133$~MeV) larger than the energies produced from PC-L3R. Meanwhile, for the region at $Z\!\!>\!\!82$ and $N\!\!>\!\!184$, the binding energies produced from PC-PK1$^\ddag$ and PC-X further deviate from the ones generated by PC-L3R; for PC-PK1$^\ddag$ and PC-X, the largest absolute deviation is $5.95$~MeV and $7.500$~MeV, respectively. 
%\textcolor{blue}{Pairing interactions plays important roles at these regions.}
%The deviations of binding energies is come from different point-coupling parameters and pairing strength.
We anticipate that the RHB with %the consideration of 
axial symmetry or triaxial deformation could further reduce the deviation between the theoretical and experiment binding energies for the region $Z\!\!=\!65$-$90$ and $N\!\!=\!75$-$150$. 
%the rms values of PC-L3R, PC-PK1$^*$, and PC-PK$^\ddag$, is \textcolor{black}{$1.599$~MeV}, \textcolor{black}{$1.918$~MeV}, and \textcolor{black}{$2.120$~MeV}, respectively. 

\section{Summary}
\label{Sec:Summary}
%and Eqs.~(\ref{eq:fermion}), (\ref{eq:HO}), (\ref{eq:derivative})
%The numerical detailed of PC-L3R coupling constants and pairing strength are given in Table.~\ref{tab:PCL3M}.
In this work, we propose a newly optimized nonlinear point-coupling parameterized interaction, PC-L3R, for the RHB framework with a further optimized separable pairing force \textcolor{black}{of $G_\mathrm{n}$ and $G_\mathrm{p}$ strengths (Table~\ref{tab:PCL3M}), which significantly improves the point-coupling interaction in describing the finite nuclear properties}. The PC-L3R interaction is fitted to the observables of $91$ spherical nuclei, i.e., the binding energies, charge radii, and another important constraint from pairing gaps. The lowest rms deviations, $1.176$~MeV (rms$_{60}$), $1.245$~MeV (rms$_{65}$), and $1.339$~MeV (rms$_{91}$), yielded by the implementation of the PC-L3R interaction in the RHB framework with the separable pairing force indicates the %capability and feasibility 
robustness 
of PC-L3R in the RHB framework unifying the treatment of pairing correlations and mean-field potentials, compared to other point-coupling interactions, i.e., PC-PK1, PC-X, DD-PC1, DD-PCX, PC-F1, and PC-LA, which are either used in the RHB framework with the separable pairing force or the RMF+BCS framework with the $\delta$ pairing force, or PC-PK1 used in the RCHB framework with the $\delta$ pairing force (Table~\ref{tab:BindingEnergy}). Meanwhile, the PC-L3R for the RHB framework provides the same description as other point-coupling interactions for the charge radii of the selected $63$ spherical nuclei, and produces a set of charge radii in a good agreement with experiment. A set of similar saturation properties is also obtained for the symmetric nuclear matter using the PC-L3R interaction, compared to the saturation properties generated from other point-coupling interactions.
%the saturation properties of the symmetric nuclear matter and bulk properties of the finite nuclei. 

We compare the experimental binding energies of the $Z\!\!=\!20$, $50$, and $82$ isotopic and $N\!\!=\!20$, $50$, and $82$ isotonic chains with the theoretical values resulted from the point-coupling interactions, PC-L3R, PC-PK1, PC-X, DD-PC1, and DD-PCX used in RHB, RMF+BCS, and RCHB frameworks. The PC-L3R provides an appropriate prediction for not only the binding energy but also for the isospin dependence. \textcolor{black}{The PC-L3R has the potential to predict a set of theoretical $Q_{\beta^{-}}$ values close to the future experimental $Q_{\beta^{-}}$ at the $Z\!\geq\!50$, $N\!>\!82$ regions}. The binding energies of $7,373$ nuclei computed from the RHB model with the PC-L3R interaction are compared with the results from the RHB model with the PC-X interaction and with the ones from RCHB model with the PC-PK1 interaction~\cite{ADNDT2018Xia}. The binding energies produced from PC-L3R, PC-PK1$^\ddag$, and PC-X are rather close at the relatively light region. For the heavy region at $Z\!\!=\!65$-$90$ and $N\!\!=\!75$-$150$, the PC-L3R provides the lowest rrs deviation, $0.586\%$, among PC-L3R, PC-PK1$^\ddag$, and PC-X, compared with the currently available $807$ experimental nuclear masses. The PC-L3R interaction could be a competently alternative point-coupling interaction for the RHB framework.\\
% satisfactorily

\noindent
{\bf Declaration of competing interest}\\

The authors declare that they have no known competing financial interests or personal relationships that could have appeared to influence the work reported in this paper.\\

\noindent
{\bf Acknowledgments}\\

%We are deeply grateful to P. Ring for reading our manuscript and for providing thoughtful and constructive suggestions to improve the manuscript. We are very thankful to M. Bender for suggestions and fruitful discussion. 
%We also thank the anonymous referee for the helpful comments and remarks on this manuscript.
This work was financially supported by 
the Strategic Priority Research Program of Chinese Academy of Sciences (CAS, Grant Nos. XDB34020100) 
and National Natural Science Foundation of China (No. 11775277). 
We are appreciative of the computing resource provided by the Institute of Physics (PHYS\_T3 cluster) and Academia Sinica Grid Computing Center (Grant No. AS-CFII-112-103; QDR4 and FDR5 clusters) of Academia Sinica, Taiwan.
Part of the numerical calculations were performed at the Gansu Advanced Computing Center. 
YHL gratefully acknowledges the financial supports from the Chinese Academy of Sciences President's International Fellowship Initiative (No. 2019FYM0002) and appreciates the laptop (Dell M4800) partially sponsored by Pin-Kok Lam and Fong-Har Tang during the pandemic of COVID-19. 
PR gratefully acknowledges financial support from the Deutsche Forschungsgemeinschaft (DFG, German Research Foundation) under Germany Excellence Strategy EXC-2094-390783311. 

%\newpage
\appendix
%\clearpage
\section{Supplementary material}
%Supplementary material related to this article can be found online at \href{https://doi.org/}{URL}.
Supplementary material related to this article is attached in this preprint.
\printcredits

%%%%%%%%%%%%%%%%%%%%%%%%%%%%%%%%%%%%%%%%%%%%%%%%%%%%%%%%%%%%%%%%%
%%% REFERENCES AND NOTES
%%%%%%%%%%%%%%%%%%%%%%%%%%%%%%%%%%%%%%%%%%%%%%%%%%%%%%%%%%%%%%%%%
%%% Loading bibliography style file

%%\bibliographystyle{model1-num-names}
%%\bibliographystyle{cas-model2-names}
\bibliographystyle{elsarticle-num-names}
%%\bibliographystyle{bibliography}

%%% Loading bibliography database
% \bibliography{bibliography}
% \bibliography{PCRHB-01}
%%%%%%%%%%%%%%%%%%%%%%%%%%%%%%%%%%%%%%%%%%%%%%%%%%%%%%%%%%%%%%%%%
%%%%%%%%%%%%%%%%%%%%%%%%%%%%%%%%%%%%%%%%%%%%%%%%%%%%%%%%%%%%%%%%%
%%%%%%%%%%%%%%%%%%%%%%%%%%%%%%%%%%%%%%%%%%%%%%%%%%%%%%%%%%%%%%%%%
%change bibliography.bib to PC-RHB.bib
%miktex/bin/bibtex PC-RHB
\providecommand{\noopsort}[1]{}\providecommand{\singleletter}[1]{#1}%

%%%%%%%%%%%%%%%%%%%%%%%%%%%%%%%%%%%%%%%%%%%%%%%%%%%%%%%%%%%%%%%%%
%%%%%%%%%%%%%%%%%%%%%%%%%%%%%%%%%%%%%%%%%%%%%%%%%%%%%%%%%%%%%%%%%
%%%%%%%%%%%%%%%%%%%%%%%%%%%%%%%%%%%%%%%%%%%%%%%%%%%%%%%%%%%%%%%%%

%\putbib
%\end{bibunit}
%%%%%%%%%%%%%%%%%%%%%%%%%%%%%%%%%%%%%%%%%%%%%%%%%%%%%%%%%%%%%%%%%
%%% Supplemental Material title
%%%%%%%%%%%%%%%%%%%%%%%%%%%%%%%%%%%%%%%%%%%%%%%%%%%%%%%%%%%%%%%%%
\newpage
\clearpage
%\onecolumn
%\begin{center}
%{\large\bf Supplementary material}
%\end{center}

\twocolumn[
\begin{@twocolumnfalse}
\begin{boxA}
{\begin{center}\noindent\large\bf Supplementary Material\end{center}}
\end{boxA}
\end{@twocolumnfalse}
]

%\vspace{\parskip}
%\begin{minipage}{\textwidth}
%\begin{boxA}
%{\noindent\large\bf Supplementary material}
%\end{boxA}
%\end{minipage}
%\vspace{\parskip}

%\begin{minipage}[t!]{2.\textwidth}
%{\noindent\large\bf Supplementary material}
%\end{minipage}
%\bigskip
%\twocolumn
%%%%%%%%%%%%%%%%%%%%%%%%%%%%%%%%%%%%%%%%%%%%%%%%%%%%%%%%%%%%%%%%%
%\vspace{-30mm}
\begin{figure}[pos=t!]
\renewcommand{\thefigure}{S1a}
%%%%%%%%%%%%%%%
%% 1st row
%%%%%%%%%%%%%%%
\centering
\includegraphics[width=0.9\columnwidth, angle=0]{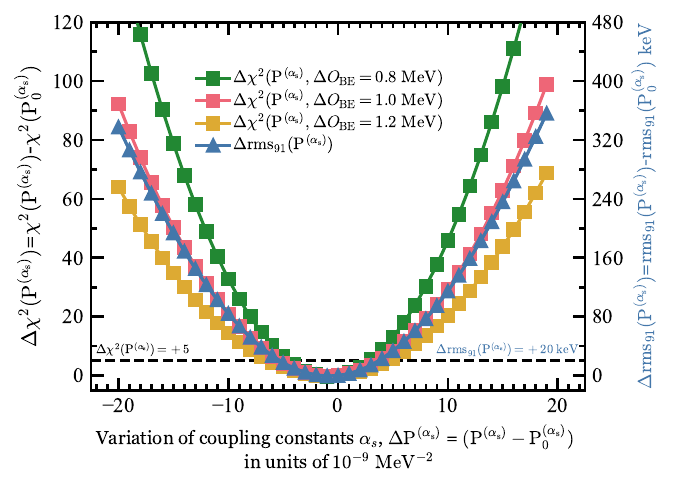}
%\caption{}
%\label{fig:as-x2-rms91}
\includegraphics[width=0.9\columnwidth, angle=0]{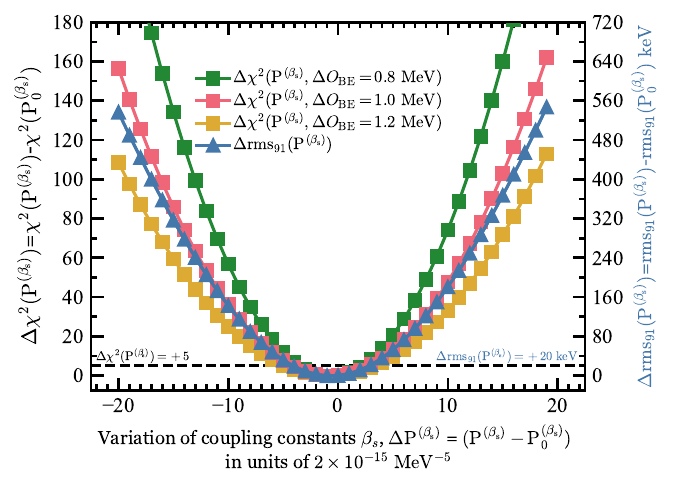}
%\caption{}
%\label{fig:bs-x2-rms91}
\includegraphics[width=0.9\columnwidth, angle=0]{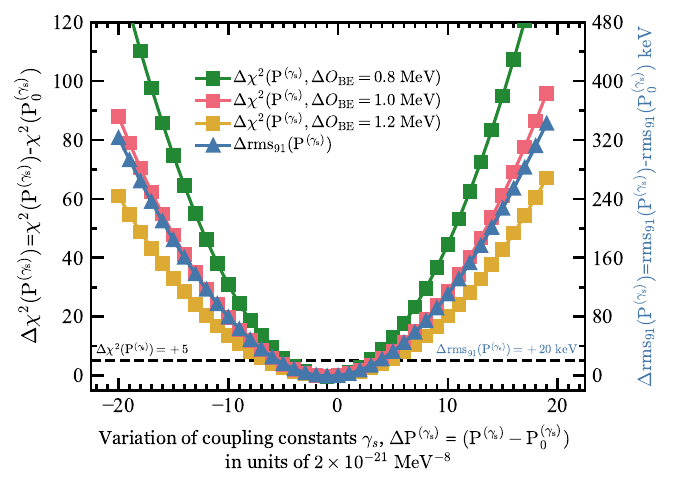}
%\caption{}
%\label{fig:cs-x2-rms91}
\caption{\label{fig:X2-rms91} 
\footnotesize
The variation of the coupling constants corresponding to the local minima obtained by the $\chi^2$-minimization (green, red, and yellow squares) and the rms of 91 spherical nuclei binding energies (blue triangles) for the relativistic density functional PC-L3R. The optimal parametrization is found for the parameter set $\rm P_{0}$ corresponding to the minimum of the penalty function $\chi^{2} (\rm P_{0})$. Each $x$-axis indicates the parameter variations, $\Delta {\rm P}$, from the ${\rm P_0}$ at the respective local minima. Each left $y$-axis shows the $\chi^{2}({\rm P})$ deviations, $\Delta \chi^{2}({\rm P})$, from the $\chi^{2}({\rm P}_0)$ at the respective local minima. Each right $y$-axis displays the rms deviations, $\Delta \rm rms_{91}({\rm P})$, due to the variations of the respective parameter from the optimized $\rm rms_{91}({\rm P}_0)$. 
Three sets of $\Delta O_{\rm BE}$ theoretical weights are studied to perceive the influence of the defined  weight on the optimization. The $\Delta \chi^{2}$ based on $\Delta O_{\rm BE}=$ $0.8$~MeV, $1$~MeV, and $1.2$~MeV are indicated as green, red, and yellow squares, respectively. The blue dash line is the constraint at $\Delta {\rm rms_{91}}\!\lessapprox\!20$~keV, for defining the parameter uncertainties. 
This figure set illustrates the optimizations of point-coupling constants, $\alpha_{S}$ (top panel), $\beta_{S}$ (middle panel), and $\gamma_{S}$ (bottom panel). See Figs.~\ref{fig:optimization2}, \ref{fig:optimization3}, and \ref{fig:optimization4} for optimizations of other parameters.
}
\vspace{-12mm}
\end{figure}%

\begin{figure}[pos=t!]
\renewcommand{\thefigure}{S1b}
%%%%%%%%%%%%%%%
%% 2nd row
%%%%%%%%%%%%%%%
\centering
\includegraphics[width=0.9\columnwidth, angle=0]{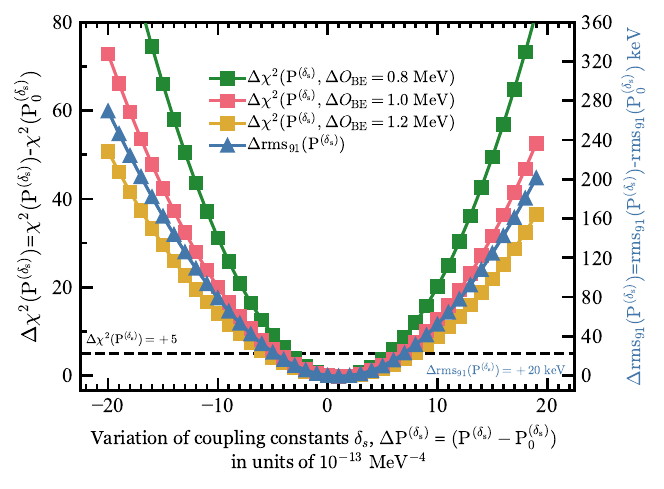}
%\caption{}
%\label{fig:ds-x2-rms91}
\includegraphics[width=0.9\columnwidth, angle=0]{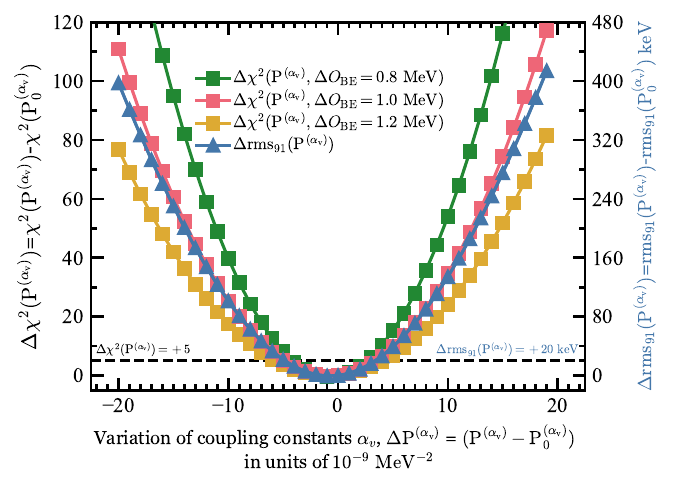}
%\caption{}
%\label{fig:av-x2-rms91}
\includegraphics[width=0.9\columnwidth, angle=0]{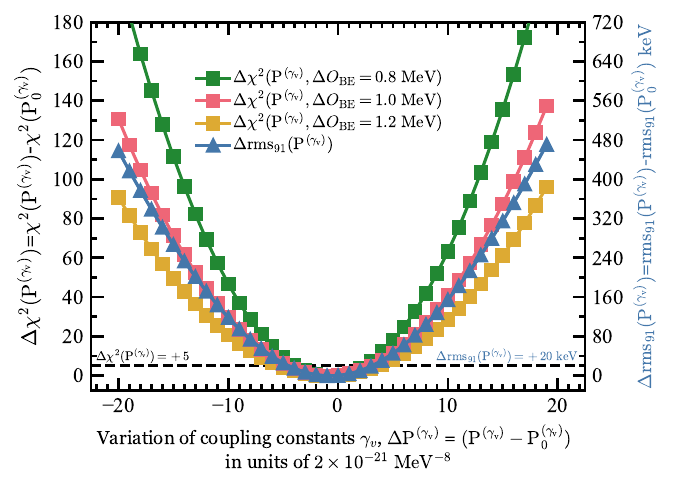}
%\caption{}
%\label{fig:cv-x2-rms91}
\caption{\label{fig:optimization2}
\footnotesize
The optimizations of point-coupling constants, $\delta_{S}$ (top panel), $\alpha_{V}$ (middle panel), and $\gamma_{V}$ (bottom panel). See Figs.~\ref{fig:X2-rms91}, \ref{fig:optimization3}, and \ref{fig:optimization4} for optimizations of other parameters. For further description, see Fig.~\ref{fig:X2-rms91}.
}
\end{figure}

\begin{figure}[pos=t!]
\renewcommand{\thefigure}{S1c}
%%%%%%%%%%%%%%%
%% 3rd row
%%%%%%%%%%%%%%%
\centering
\includegraphics[width=0.9\columnwidth, angle=0]{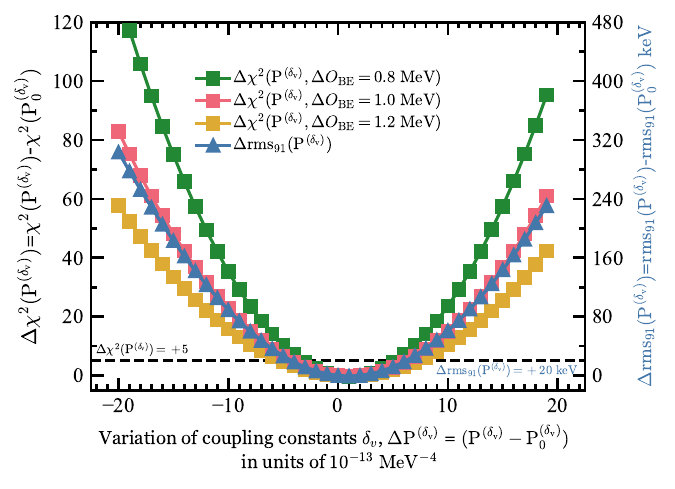}
%\caption{}
%\label{fig:dv-x2-rms91}
\includegraphics[width=0.9\columnwidth, angle=0]{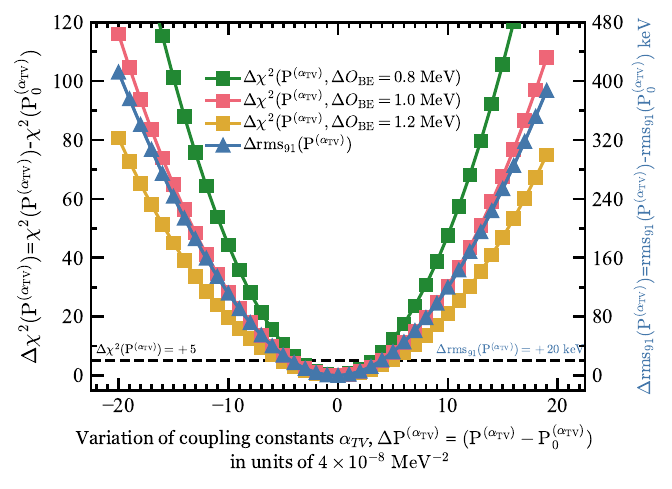}
%\caption{}
%\label{fig:atv-x2-rms91}
\includegraphics[width=0.9\columnwidth, angle=0]{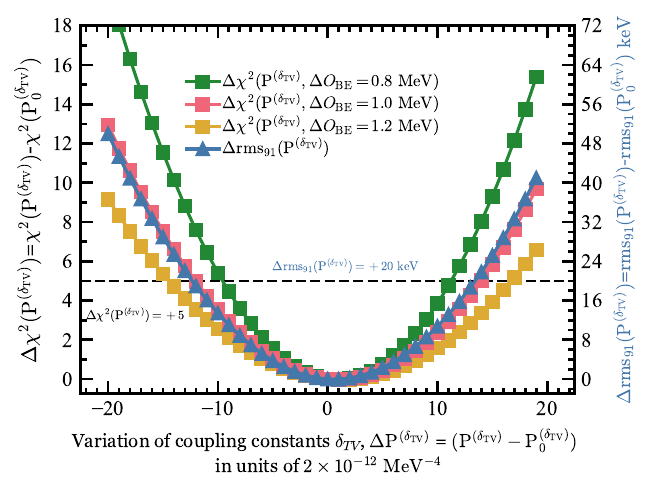}
%\caption{}
%\label{fig:dtv-x2-rms91}
\caption{\label{fig:optimization3}
\footnotesize
The optimizations of point-coupling constants, $\delta_{V}$ (top panel), $\alpha_{TV}$ (middle panel), and $\delta_{TV}$ (bottom panel). See Figs.~\ref{fig:X2-rms91}, \ref{fig:optimization2}, and \ref{fig:optimization4} for optimizations of other parameters. For further description, see Fig.~\ref{fig:X2-rms91}.
With referring to the results presented in Figs.~\ref{fig:X2-rms91}, \ref{fig:optimization2}, and \ref{fig:optimization3}, the variations of $\Delta O_{\rm BE}$ do not affect the local minima of point-coupling constants, i.e., $\Delta\chi^{2}({\rm P}_0)$ and $\Delta{\rm rms_{91}}({\rm P}_0)$. Therefore, assuming $\Delta O_{\rm BE}\!=\!1$~MeV is justified and in accord with the definition of previous works \cite{PRC2010Zhao,PRC2008Niksic,PRC2019Yuksel,PLB2020Taninah}. The influence of variations in $\Delta O_{\rm BE}$ on the local minima of pairing strengths and $a$ factor is given in Fig.~\ref{fig:optimization4}. 
}
\vspace{21.9mm}
\end{figure}

\begin{figure}[pos=t!]
\renewcommand{\thefigure}{S1d}
%%%%%%%%%%%%%%%
%% 4th row
%%%%%%%%%%%%%%%
\centering
\includegraphics[width=0.9\columnwidth, angle=0]{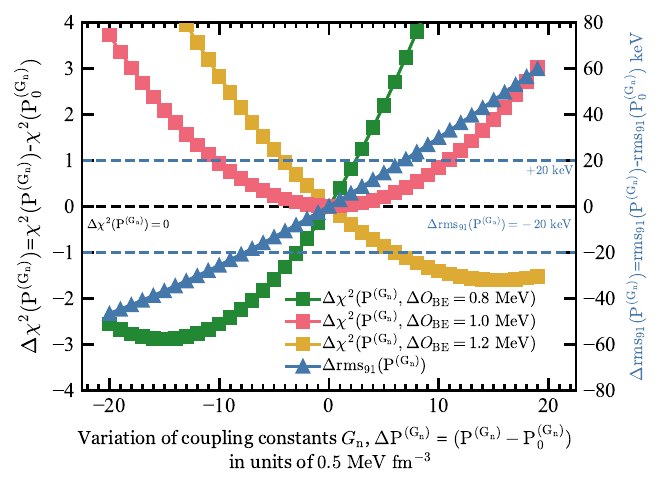}
%\caption{}
%\label{fig:Gn-x2-rms91}
\includegraphics[width=0.9\columnwidth, angle=0]{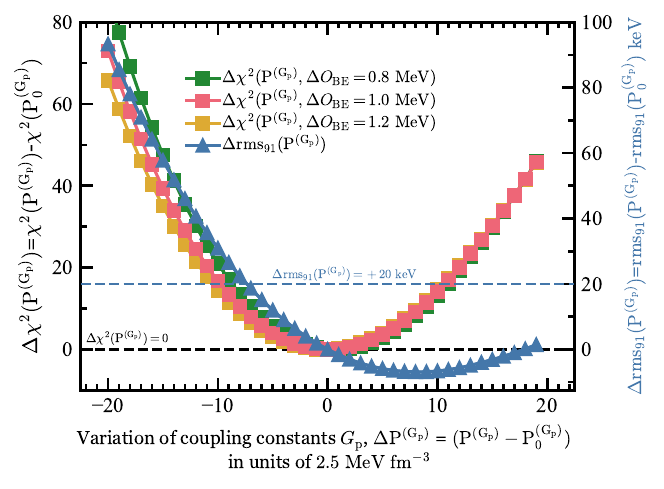}
%\caption{}
%\label{fig:Gp-x2-rms91}
\includegraphics[width=0.9\columnwidth, angle=0]{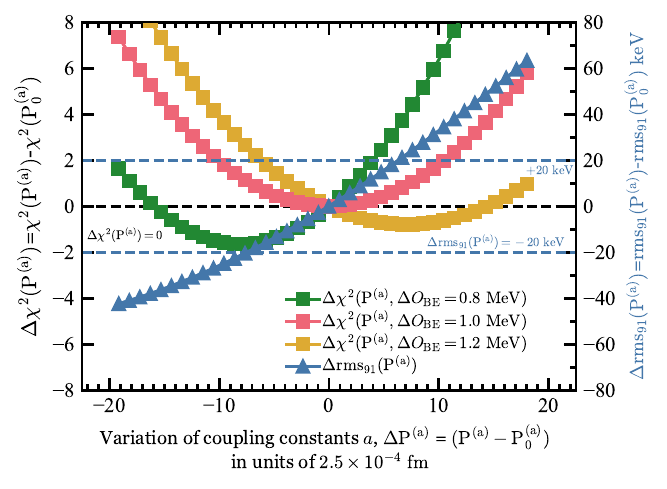}
%\caption{}
%\label{fig:a-x2-rms91}
\caption{
\label{fig:optimization4}
\footnotesize
The optimizations of $G_{\rm n}$ (top panel) and $G_{\rm p}$ (middle panel) pairing strengths, and $a$ factor (bottom panel). See Figs.~\ref{fig:X2-rms91}, \ref{fig:optimization2}, and \ref{fig:optimization3} for optimizations of other parameters. For further description, see Fig.~\ref{fig:X2-rms91}. 
The local minima of the $G_{\rm n}$ pairing strength, $\Delta\chi^{2}({\rm P}^{(G_{\rm n})}_0)$, and 
$a$ factor, $\Delta\chi^{2}({\rm P}^{(a)}_0)$, and the corresponding $\Delta {\rm rms_{91}^{(G_{\rm n})}}$ and $\Delta {\rm rms_{91}^{(a)}}$, are varied accordingly with variations in $\Delta O_{\rm BE}$. Nevertheless, the influence of $\Delta O_{\rm BE}$ variations on the $\Delta\chi^{2}({\rm P}^{(G_{\rm p})})$ corresponding to the $G_{\rm p}$ pairing strength is negligible. Meanwhile, the rms values of mean pairing gaps, rms$_{\rm PG}$, are inversely varied accordingly with variations in $\Delta O_{\rm BE}$ as well, i.e., for the pairing strength $G_{\rm n}$, rms$_{\rm PG}\!=\!0.237$~MeV ($\Delta O_{\rm BE}\!=\!0.8$~MeV), rms$_{\rm PG}\!=\!0.217$~MeV ($\Delta O_{\rm BE}\!=\!1$~MeV), rms$_{\rm PG}\!=\!0.196$~MeV ($\Delta O_{\rm BE}\!=\!1.2$~MeV), and for the $a$ factor, rms$_{\rm PG}\!=\!0.228$~MeV ($\Delta O_{\rm BE}\!=\!0.8$~MeV), rms$_{\rm PG}\!=\!0.217$~MeV ($\Delta O_{\rm BE}\!=\!1$~MeV), rms$_{\rm PG}\!=\!0.208$~MeV ($\Delta O_{\rm BE}\!=\!1.2$~MeV). Therefore, it is still appropriate to define $\Delta O_{\rm BE}\!=\!1$~MeV for a balance weight to constrain the point-coupling interaction. 
%${\rm P}^{(G_{n})}$, ${\rm P}^{(G_{n})}$, ${\rm P}^{(G_{n})}$, 
%${\rm P}^{(a)}$, ${\rm P}^{(a)}$, ${\rm P}^{(a)}$, 
%$\rm rms_{91}({\rm P}_0)$, and factor $a$, $\Delta {\rm rms_{91} }$ decrease with the $\Delta O_{\rm BE}$.
}
\end{figure}

%\newpage
\clearpage

\begin{table*}[pos=b,width=\textwidth]
\renewcommand\arraystretch{1.3}
\renewcommand{\thetable}{S1}
\centering
\small
%\footnotesize
\caption{\footnotesize The coupling constants and penalty function $\chi^2$ of the non-linear point-coupling interaction PC-L3R, PC-F1, PC-PK1, and PC-X. The separable pairing force (modeled D1S) is implemented in the PC-F1, PC-PK1, and PC-X interactions.} 
\label{tab:PC SET}
%\scriptsize
\footnotesize
%\begin{tabular}{lcc}
%\begin{tabular}[width=\linewidth]{@{\extracolsep{\fill}}lcc}
%\begin{tabular}{@{\extracolsep{\fill}}lr}
\begin{tabular}{@{}@{\hspace{7mm}}l@{\hspace{7mm}}r@{\hspace{7mm}}r@{\hspace{7mm}}r@{\hspace{7mm}}r@{\hspace{7mm}}r@{}}
\toprule
\midrule
%\hline
%\hlinehttps://www.overleaf.com/project/6078f9e923c7c00d2fd483b7
%$\rm {Coupling~~constant}$ & $\rm{Quantity}$ and $\rm {physical~unit(s)}$\\
   & PC-L3R &  PC-X & PC-PK1 & PC-F1\\

\hline
 $\alpha_{S} $($\rm MeV^{-2}$)  & $-3.99289\times10^{-04}$ & $-4.0242\times10^{-04}$ & $-3.96291\times10^{-04}$ &  $-3.83577\times10^{-04}$   \\
 $\beta_{S}  $($\rm MeV^{-5}$)  & $8.65504 \times10^{-11}$ & $8.5958 \times10^{-11}$ & $8.6653  \times10^{-11}$ &  $7.68567 \times10^{-11}$  \\
 $\gamma_{S} $($\rm MeV^{-8}$)  & $-3.83950\times10^{-17}$ & $-3.7304\times10^{-17}$ & $-3.80724\times10^{-17}$ &  $-2.90443\times10^{-17}$   \\
 $\delta_{S} $($\rm MeV^{-4}$)  & $-1.20749\times10^{-10}$ & $-0.52181\times10^{-10}$ & $-1.09108\times10^{-10}$ &  $-4.1853 \times10^{-10}$   \\
 $\alpha_{V} $($\rm MeV^{-2}$)  & $2.71991 \times10^{-04}$ & $2.7534 \times10^{-04}$ & $2.6904  \times10^{-04}$ &  $2.59333 \times10^{-04}$  \\
 $\gamma_{V} $($\rm MeV^{-8}$)  & $-3.72107\times10^{-18}$ & $-4.9012\times10^{-18}$ & $-3.64219\times10^{-18}$ &  $-3.879  \times10^{-18}$   \\
 $\delta_{V} $($\rm MeV^{-4}$)  & $-4.26653\times10^{-10}$ & $-5.9440\times10^{-10}$ & $-4.32619\times10^{-10}$ &  $-1.1921 \times10^{-10}$   \\
 $\alpha_{TV}$($\rm MeV^{-2}$)  & $2.96688 \times10^{-05}$ & $2.8188 \times10^{-05}$ & $2.95018 \times10^{-05}$ &  $3.4677  \times10^{-05}$  \\
 $\delta_{TV}$($\rm MeV^{-4}$)  & $-4.65682\times10^{-10}$ & $-5.2838\times10^{-10}$ & $-4.11112\times10^{-10}$ &  $-4.2    \times10^{-11}$   \\ 
\hline
 $\chi^{2}$            & $274.478$     & $494.875$  &  $364.449$  & $751.225$ \\
\bottomrule
\end{tabular}
\end{table*}
\renewcommand\arraystretch{1.0}

{\bf The Justification of the Point-Coupling Interaction, PC-L3R}\,--\,
Table~\ref{tab:PC SET} shows the coupling constants and penalty functions of the point-coupling interactions PC-L3R, PC-F1, PC-PK1, and PC-X, obtained from Eq.~(\ref{eq:minimization}). Referring to the work of Manohar and Georgi~\cite{Manohar1984}, we can scale the newly optimized coupling constants in the point-coupling interaction to a generic QCD-based Lagrangian term using the expression, %a generic Lagrangian term of the physical series as
\begin{equation}\label{eq:QCD-cln}
\mathcal{L}\sim -c_{ln}[\frac{\bar{\psi}\psi}{f_{\pi}^{2}\Lambda}]^{l}[\frac{\partial^{\mu}}{\Lambda}]^{n}f_{\pi}^{2}\Lambda^{2},
\end{equation}
where $\psi$ is the nucleon field, $f_{\pi}$ is the pion decay constants, $\Lambda\!=\! 770$~MeV is a generic QCD large-mass scale. The chiral symmetry in weakening the N-body forces demands 
%Taking into account the role of chiral symmetry in weakening N-body forces demands 
$\Delta\!=\!l+n-2 \geq 0$. 
By scaling the coupling constants in accordance with the QCD-based Lagrangian, the naturalness in effective theories can be justified. 
If the theory is natural, the Lagrangian should lead to dimensionless coefficients $c_{ln}$ within numerical factors of the order of unity \cite{PRC2010Zhao,PRC2002Burvenich}. The corresponding QCD-scaled coupling constants $c_{ln}$ for each coupling constants of the relativistic point-coupling interactions PC-L3R, PC-F1, PC-PK1, and PC-X are presented in Table~\ref{tab:PC-cln}. The numerical values of these corresponding QCD-scaled coupling constants $c_{ln}$ indicate that the nine coupling constants of the PC-L3R, PC-F1, PC-PK1, and PC-X interactions are natural.

\begin{table}[pos=h,width=\columnwidth]
\renewcommand\arraystretch{1.3}
\renewcommand{\thetable}{S2}
\centering
\footnotesize
\caption{\footnotesize 
The corresponding QCD-scaled coupling constants $c_{ln}$ to each coupling constants of the point-coupling interactions PC-L3R, PC-F1, PC-PK1, and PC-X.} 
\label{tab:PC-cln}
\scriptsize
\begin{tabular}{@{}@{\hspace{2mm}}l@{\hspace{2mm}}r@{\hspace{2mm}}r@{\hspace{2mm}}r@{\hspace{2mm}}r@{\hspace{2mm}}r@{}}
\toprule
\midrule
   & $c_{ln}$ (PC-L3R) & $c_{ln}$ (PC-X) & $c_{ln}$ (PC-PK1) & $c_{ln}$ (PC-F1)\\
\hline
 $\alpha_{S} $& $-1.708$ & $-1.722$ & $-1.695$ &  $-1.641$ \\
 $\beta_{S}  $& $1.626 $ & $1.615 $ & $1.628 $ &  $1.443 $\\
 $\gamma_{S} $& $-3.565$ & $-3.464$ & $-3.535$ &  $-2.695$ \\
 $\delta_{S} $& $-0.306$ & $-0.132$ & $-0.277$ &  $-1.061$ \\
 $\alpha_{V} $& $1.164 $ & $1.178 $ & $1.151 $ &  $1.109 $\\
 $\gamma_{V} $& $-0.345$ & $-0.455$ & $-0.338$ &  $-0.360$ \\
 $\delta_{V} $& $-1.082$ & $-1.508$ & $-1.097$ &  $-0.302$ \\
 $\alpha_{TV}$& $0.508 $ & $0.482 $ & $0.505 $ &  $0.593 $\\
 $\delta_{TV}$& $-4.725$ & $-5.361$ & $-4.171$ &  $-0.426$ \\ 
\bottomrule
\vspace{-5mm}
\end{tabular}
\end{table}
\renewcommand\arraystretch{1.0}

{\bf The Constraint of Pairing Interaction \,--\,}
We further optimize the separable pairing force, of which the $G$ pairing strength is separated to the neutron ($G_n$) and proton ($G_p$) pairing strengths to improve the description of the finite nuclear properties. An empirical pairing gap, based on the five-point formula~\cite{Bender2000}, is expressed as  
\begin{eqnarray}\label{eq:five-point}
\Delta_{q}^{(5)}(N_{0}) = \!\!\!\!\!\! &-& \!\!\!\!\!\! \frac{\pi_{N_{0}}}{8}[E(N_{0}+2)-4E(N_{0}+1)+6E(N_{0}) \nonumber \\ 
 &-&\!\!\!\!\!\!4E(N_{0}-1)+E(N_{0}-2)],
\end{eqnarray}
where $N_{0}$ is either the proton (or neutron) number of a selected nucleus, $\pi_{N_{0}}\!=\!-1^{N_{0}}$ is the number parity, and $E(N_{0})$ is the binding energy of the selected nucleus. For the present work, we consider twelve sets of mean pairing gaps consisting of 54 nuclei in total, of which the $N_0$ nuclei are: $^{84}$Se, $^{88}$Sr, $^{92}$Mo, $^{112}$Sn, $^{118}$Sn, $^{124}$Sn, $^{138}$Ba, $^{142}$Nd, $^{146}$Gd, $^{204}$Pb, $^{212}$Rn, and $^{214}$Ra.

The rms value of comparing experimental and theoretical pairing gaps produced from the PC-L3R interaction used in RHB model with a further optimized separable pairing interaction is $0.217$~MeV, whereas the rms values yielded from the PC-X, PC-PK1*, DD-PCX, and DD-PC1* are $0.109$~MeV, $0.124$~MeV, $0.273$~MeV, and $0.115$~MeV, respectively. These mean pairing gaps are exhibited in Fig.~\ref{fig:Mean-PG}. We remark that the $\Delta O_{\rm PG}$ weight can be defined as either $0.1$~MeV or $0.05$~MeV for Eq.~(\ref{eq:minimization}). Defining $\Delta O_{\rm PG}\!=\!0.05$~MeV can somehow impose a reduction of $\sim\!\!70$~keV on the rms of pairing gap. Nevertheless, such reduction of rms does not significantly alter  the theoretical mean pairing gaps. Therefore, defining $\Delta O_{\rm PG}\!=\!0.1$~MeV is rather appropriate to reduce both rms values of mean pairing gaps and binding energies. 
%, the corresponding new pairing strengths $G_n$ and $G_p$ are $-718.01$~MeV and $-683.08$~MeV, respectively. 
%The rms value of pairing gaps is then reduced to $0.143$~MeV, however, the rms values of binding energies are increased, i.e., $1.234$~MeV (60 nuclei), $1.383$~MeV (65 nuclei), and $1.591$~MeV (91 nuclei). 

\begin{table*}[pos=t,width=\textwidth]
\renewcommand{\thetable}{S3}
\centering
\caption{\footnotesize The binding energies (in MeV) of the selected 91 spherical nuclei.}
%$\scriptstyle {\rm RMS} = \scriptstyle \sqrt{\Sigma_{i}^{N}(E^{i}_{expt.}-E^{i}_{calc})^{2}/N}$
%The experimental values of the corresponding nuclei are used in the parametrization fitting.
\label{tab:BindingEnergy}
\scriptsize
%\tiny
%\resizebox{.90\columnwidth}{!}{
%\begin{tabular}{lcccccccccccc}
\begin{threeparttable}
\begin{tabular}{@{}l@{\hspace{2mm}}c@{\hspace{2mm}}c@{\hspace{2mm}}c@{\hspace{2mm}}c@{\hspace{2mm}}c@{\hspace{2mm}}c@{\hspace{2mm}}c@{\hspace{2mm}}c@{\hspace{2mm}}c@{\hspace{2mm}}c@{\hspace{2mm}}c@{\hspace{2mm}}c@{\hspace{2mm}}c@{}}
%\hline\hline
\toprule
\midrule
$\rm {Nuclei}$ & AME2020 &  PC-L3R$^a$ & PC-PK1$^b$ & PC-PK1$^{\ast c}$ & PC-PK1$^{\ddag d}$ & PC-X$^e$ & DD-PC1$^f$ & DD-PC1$^{\ast g}$ & DD-PCX$^{h}$ & PC-F1$^i$ & PC-F1$^{\ast j}$ & PC-LA$^k$ & PC-LA$^{\ast l}$  \\
%\hline
\midrule
$\rm\mathbf{ ~^{16}O}$     &      $-127.619  $    &     $ -127.139  $   &  $ -127.280  $ & $ -127.283    $ &  $ -127.29  $   & $ -126.827  $   &  $ -128.527   $  &  $ -128.540   $  &  $ -127.300    $ &  $ -127.691    $  &  $ -127.694     $  &  $ -127.407    $   &  $ -127.656     $       \\
$\rm\mathbf{~^{18}O}$      &      $-139.513  $    &     $ -140.714  $   &  $ -140.223  $ & $ -140.496    $ &  $ -141.63  $   & $ -140.127  $   &  $ -141.145   $  &  $ -141.337   $  &  $ -140.526    $ &  $ -140.028    $  &  $ -140.697     $  &  $ -140.356    $   &  $ -140.638     $       \\
$\rm\mathbf{~^{20}O}$      &      $-151.371  $    &     $ -152.502  $   &  $ -151.962  $ & $ -152.278    $ &  $ -153.16  $   & $ -151.940  $   &  $ -152.790   $  &  $ -152.972   $  &  $ -151.806    $ &  $ -151.606    $  &  $ -152.394     $  &  $ -152.228    $   &  $ -152.490     $       \\
$\rm\mathbf{~^{22}O}$      &      $-162.028  $    &     $ -162.570  $   &  $ -162.285  $ & $ -162.586    $ &  $ -162.91  $   & $ -162.194  $   &  $ -163.141   $  &  $ -163.293   $  &  $ -161.332    $ &  $ -162.054    $  &  $ -162.659     $  &  $ -162.665    $   &  $ -162.884     $       \\
$\rm\mathbf{ ^{18}Ne}$     &      $-132.143  $    &     $ -132.406  $   &  $ -132.088  $ & $ -132.668    $ &  $ -133.74  $   & $ -132.282  $   &  $ -132.923   $  &  $ -133.427   $  &  $ -132.383    $ &  $ -132.216    $  &  $ -132.922     $  &  $ -132.317    $   &  $ -132.815     $       \\
$\rm\mathbf{ ^{20}Mg}$     &      $-134.561  $    &     $ -134.990  $   &  $ -134.563  $ & $ -135.352    $ &  $ -136.23  $   & $ -134.966  $   &  $ -135.141   $  &  $ -135.785   $  &  $ -134.290    $ &  $ -134.613    $  &  $ -135.563     $  &  $ -134.992    $   &  $ -135.503     $       \\
$\rm ^{32}Mg $    &      $-249.723  $    &     $ -251.742  $   &  $ -251.940  $ & $ -252.330    $ &  $ -253.06  $   & $ -251.335  $   &  $ -253.164   $  &  $ -253.511   $ &   $ -251.341    $ &  $ -252.756    $  &  $ -253.241     $  &  $ -251.370    $   &  $ -251.562     $      \\
$\rm\mathbf{ ^{34}Si}$     &      $-283.464  $    &     $ -284.568  $   &  $ -284.727  $ & $ -284.739    $ &  $ -284.63  $   & $ -284.056  $   &  $ -285.967   $  &  $ -286.000   $  &  $ -284.765    $ &  $ -285.067    $  &  $ -285.036     $  &  $ -283.989    $   &  $ -284.281     $       \\
$\rm\mathbf{~^{36}S}$      &      $-308.714  $    &     $ -308.230  $   &  $ -308.374  $ & $ -308.444    $ &  $ -308.55  $   & $ -307.345  $   &  $ -309.305   $  &  $ -309.648   $  &  $ -309.011    $ &  $ -308.973    $  &  $ -309.402     $  &  $ -307.221    $   &  $ -307.748     $       \\
$\rm\mathbf{ ^{38}Ar}$     &      $-327.343  $    &     $ -327.167  $   &  $ -327.107  $ & $ -327.477    $ &  $ -327.54  $   & $ -326.247  $   &  $ -328.691   $  &  $ -329.030   $  &  $ -328.017    $ &  $ -328.540    $  &  $ -329.034     $  &  $ -326.755    $   &  $ -327.238     $       \\
$\rm\mathbf{ ^{36}Ca}$     &      $-281.372  $    &     $ -281.592  $   &  $ -281.412  $ & $ -281.505    $ &  $ -281.63  $   & $ -280.418  $   &  $ -281.878   $  &  $ -281.996   $  &  $ -281.202    $ &  $ -282.001    $  &  $ -282.471     $  &  $ -280.454    $   &  $ -280.576     $       \\
$\rm\mathbf{ ^{38}Ca}$     &      $-313.122  $    &     $ -313.499  $   &  $ -313.230  $ & $ -313.381    $ &  $ -313.46  $   & $ -312.150  $   &  $ -314.501   $  &  $ -314.563   $  &  $ -313.567    $ &  $ -314.415    $  &  $ -314.946     $  &  $ -312.901    $   &  $ -313.013     $       \\
$\rm\mathbf{ ^{40}Ca}$     &      $-342.052  $    &     $ -342.803  $   &  $ -343.060  $ & $ -343.065    $ &  $ -343.07  $   & $ -341.788  $   &  $ -345.113   $  &  $ -345.152   $  &  $ -343.272    $ &  $ -345.041    $  &  $ -345.054     $  &  $ -343.202    $   &  $ -343.460     $       \\
$\rm\mathbf{ ^{42}Ca}$     &      $-361.896  $    &     $ -363.294  $   &  $ -363.142  $ & $ -363.104    $ &  $ -364.06  $   & $ -361.993  $   &  $ -365.143   $  &  $ -365.021   $  &  $ -363.706    $ &  $ -364.411    $  &  $ -364.813     $  &  $ -363.685    $   &  $ -363.747     $       \\
$\rm\mathbf{ ^{44}Ca}$     &      $-380.960  $    &     $ -382.129  $   &  $ -381.915  $ & $ -381.814    $ &  $ -382.77  $   & $ -380.861  $   &  $ -383.967   $  &  $ -383.740   $  &  $ -382.503    $ &  $ -382.748    $  &  $ -383.237     $  &  $ -382.789    $   &  $ -382.803     $       \\
$\rm\mathbf{ ^{46}Ca}$     &      $-398.773  $    &     $ -399.495  $   &  $ -399.451  $ & $ -399.307    $ &  $ -399.86  $   & $ -398.523  $   &  $ -401.668   $  &  $ -401.422   $  &  $ -399.941    $ &  $ -400.060    $  &  $ -400.394     $  &  $ -400.627    $   &  $ -400.714     $       \\
$\rm\mathbf{ ^{48}Ca}$     &      $-416.001  $    &     $ -415.134  $   &  $ -415.492  $ & $ -415.501    $ &  $ -415.39  $   & $ -414.935  $   &  $ -417.973   $  &  $ -418.038   $  &  $ -416.022    $ &  $ -416.085    $  &  $ -416.060     $  &  $ -416.969    $   &  $ -417.433     $       \\
$\rm\mathbf{ ^{50}Ca}$     &      $-427.508  $    &     $ -427.098  $   &  $ -426.937  $ & $ -427.193    $ &  $ -427.34  $   & $ -426.335  $   &  $ -428.660   $  &  $ -428.827   $  &  $ -427.054    $ &  $ -427.302    $  &  $ -427.827     $  &  $ -426.883    $   &  $ -427.242     $      \\
$\rm ^{52}Ca $    &      $-438.328  $    &     $ -437.428  $   &  $ -437.455  $ & $ -437.586    $ &  $ -437.73  $   & $ -436.542  $   &  $ -438.129   $  &  $ -438.337   $ &   $ -436.479    $ &  $ -437.571    $  &  $ -438.088     $  &  $ -435.512    $   &  $ -435.772     $      \\
$\rm\mathbf{ ^{42}Ti}$     &      $-346.888  $    &     $ -348.017  $   &  $ -348.024  $ & $ -348.352    $ &  $ -349.24  $   & $ -347.219  $   &  $ -349.848   $  &  $ -350.097   $  &  $ -348.448    $ &  $ -349.701    $  &  $ -350.141     $  &  $ -348.626    $   &  $ -349.013     $       \\
$\rm\mathbf{ ^{50}Ti}$     &      $-437.786  $    &     $ -436.424  $   &  $ -436.445  $ & $ -436.536    $ &  $ -437.27  $   & $ -436.326  $   &  $ -437.761   $  &  $ -437.942   $  &  $ -437.037    $ &  $ -436.171    $  &  $ -436.454     $  &  $ -437.223    $   &  $ -437.821     $      \\
$\rm ^{54}Ni $    &      $-453.224  $    &     $ -452.788  $   &  $ -452.362  $ & $ -452.000    $ &  $ -452.31  $   & $ -452.480  $   &  $ -451.409   $  &  $ -451.110   $ &   $ -451.154    $ &  $ -450.310    $  &  $ -450.476     $  &  $ -451.965    $   &  $ -452.196     $         \\
$\rm\mathbf{ ^{56}Ni}$     &      $-483.996  $    &     $ -484.316  $   &  $ -483.669  $ & $ -483.663    $ &  $ -483.68  $   & $ -484.610  $   &  $ -481.447   $  &  $ -481.552   $  &  $ -481.885    $ &  $ -480.758    $  &  $ -480.812     $  &  $ -481.826    $   &  $ -482.571     $      \\
$\rm\mathbf{ \textcolor{blue}{^{58}Ni}}$     &      $-506.460  $    &     $ -504.525  $   &  $ -503.636  $ & $ -503.653    $ &  $ -504.08  $   & $ -503.989  $   &  $ -502.587   $  &  $ -502.475   $  &  $ -503.308    $ &  $ -501.646    $  &  $ -502.117     $  &  $ -502.623    $   &  $ -502.964     $      \\
$\rm ^{60}Ni $    &      $-526.847  $    &     $ -523.786  $   &  $ -523.004  $ & $ -522.814    $ &  $ -523.31  $   & $ -522.682  $   &  $ -522.872   $  &  $ -522.415   $ &   $ -523.280    $ &  $ -521.653    $  &  $ -522.307     $  &  $ -522.567    $   &  $ -522.432     $         \\
$\rm ^{62}Ni $    &      $-545.262  $    &     $ -542.218  $   &  $ -541.712  $ & $ -541.288    $ &  $ -541.63  $   & $ -540.791  $   &  $ -542.216   $  &  $ -541.609   $ &   $ -542.099    $ &  $ -540.891    $  &  $ -541.571     $  &  $ -541.527    $   &  $ -541.143     $         \\
$\rm ^{64}Ni $    &      $-561.758  $    &     $ -559.821  $   &  $ -559.374  $ & $ -559.088    $ &  $ -559.22  $   & $ -558.303  $   &  $ -560.592   $  &  $ -560.109   $ &   $ -559.971    $ &  $ -559.308    $  &  $ -559.950     $  &  $ -559.281    $   &  $ -559.084     $         \\
$\rm ^{66}Ni $    &      $-576.808  $    &     $ -576.474  $   &  $ -576.185  $ & $ -576.105    $ &  $ -576.10  $   & $ -575.079  $   &  $ -578.000   $  &  $ -577.817   $ &   $ -576.908    $ &  $ -576.808    $  &  $ -577.367     $  &  $ -576.079    $   &  $ -576.068     $         \\
$\rm ^{68}Ni $    &      $-590.408  $    &     $ -591.692  $   &  $ -591.958  $ & $ -591.998    $ &  $ -592.00  $   & $ -590.706  $   &  $ -594.209   $  &  $ -594.338   $ &   $ -592.534    $ &  $ -593.481    $  &  $ -593.538     $  &  $ -591.242    $   &  $ -591.404     $         \\
$\rm ^{70}Ni $    &      $-602.300  $    &     $ -603.986  $   &  $ -603.998  $ & $ -603.863    $ &  $ -604.42  $   & $ -602.798  $   &  $ -606.157   $  &  $ -605.928   $ &   $ -604.303    $ &  $ -604.435    $  &  $ -604.862     $  &  $ -603.468    $   &  $ -603.345     $         \\
$\rm\mathbf{ ^{72}Ni}$     &      $-613.455  $    &     $ -614.904  $   &  $ -614.875  $ & $ -614.710    $ &  $ -615.29  $   & $ -613.822  $   &  $ -617.071   $  &  $ -616.757   $  &  $ -614.926    $ &  $ -614.646    $  &  $ -615.195     $  &  $ -614.486    $   &  $ -614.296     $      \\
$\rm ^{80}Zn $    &      $-673.884  $    &     $ -672.278  $   &  $ -672.686  $ & $ -673.066    $ &  $ -673.51  $   & $ -672.541  $   &  $ -675.851   $  &  $ -676.293   $ &   $ -673.303    $ &  $ -672.234    $  &  $ -672.783     $  &  $ -673.537    $   &  $ -674.110     $      \\
$\rm ^{82}Ge $    &      $-702.228  $    &     $ -700.099  $   &  $ -700.458  $ & $ -700.797    $ &  $ -701.24  $   & $ -700.133  $   &  $ -703.597   $  &  $ -704.002   $ &   $ -701.183    $ &  $ -700.724    $  &  $ -701.302     $  &  $ -701.867    $   &  $ -702.434     $       \\
$\rm\mathbf{ ^{84}Se}$     &      $-727.339  $    &     $ -725.394  $   &  $ -725.732  $ & $ -725.961    $ &  $ -726.28  $   & $ -725.200  $   &  $ -728.792   $  &  $ -729.131   $  &  $ -726.413    $ &  $ -726.609    $  &  $ -727.097     $  &  $ -727.605    $   &  $ -728.141     $      \\
$\rm\mathbf{ ^{86}Kr}$     &      $-749.235  $    &     $ -747.799  $   &  $ -747.939  $ & $ -748.271    $ &  $ -748.46  $   & $ -747.393  $   &  $ -751.050   $  &  $ -751.461   $  &  $ -748.911    $ &  $ -749.427    $  &  $ -749.969     $  &  $ -750.313    $   &  $ -750.949     $      \\
$\rm\mathbf{ ^{88}Sr}$     &      $-768.468  $    &     $ -767.120  $   &  $ -767.138  $ & $ -767.501    $ &  $ -767.64  $   & $ -766.522  $   &  $ -770.240   $  &  $ -770.649   $  &  $ -768.412    $ &  $ -769.143    $  &  $ -769.652     $  &  $ -769.742    $   &  $ -770.397     $      \\
$\rm\mathbf{ ^{90}Zr}$     &      $-783.897  $    &     $ -783.135  $   &  $ -783.033  $ & $ -783.430    $ &  $ -783.72  $   & $ -782.416  $   &  $ -785.806   $  &  $ -786.219   $  &  $ -784.505    $ &  $ -785.348    $  &  $ -785.811     $  &  $ -785.565    $   &  $ -786.228     $      \\
$\rm\mathbf{ ^{92}Mo}$     &      $-796.511  $    &     $ -796.365  $   &  $ -796.148  $ & $ -796.530    $ &  $ -796.92  $   & $ -795.653  $   &  $ -798.308   $  &  $ -798.717   $  &  $ -797.525    $ &  $ -798.191    $  &  $ -798.746     $  &  $ -798.719    $   &  $ -799.372     $      \\
$\rm\mathbf{ ^{94}Ru}$     &      $-806.864  $    &     $ -807.363  $   &  $ -807.034  $ & $ -807.359    $ &  $ -807.67  $   & $ -806.684  $   &  $ -808.575   $  &  $ -808.950   $  &  $ -808.100    $ &  $ -808.731    $  &  $ -809.281     $  &  $ -809.695    $   &  $ -810.343     $      \\
$\rm ^{96}Pd $    &      $-815.042  $    &     $ -816.342  $   &  $ -815.875  $ & $ -816.141    $ &  $ -816.30  $   & $ -815.695  $   &  $ -816.833   $  &  $ -817.179   $ &   $ -816.518    $ &  $ -817.205    $  &  $ -817.687     $  &  $ -818.641    $   &  $ -819.296     $       \\
$\rm\mathbf{ ^{98}Cd}$     &      $-821.073  $    &     $ -823.394  $   &  $ -822.765  $ & $ -822.970    $ &  $ -823.02  $   & $ -822.775  $   &  $ -823.162   $  &  $ -823.501   $  &  $ -822.908    $ &  $ -823.668    $  &  $ -824.051     $  &  $ -825.580    $   &  $ -826.292     $      \\
$\rm\mathbf{ \textcolor{blue}{^{100}Sn}}$    &      $-825.160  $    &     $ -828.563  $   &  $ -827.715  $ & $ -827.882    $ &  $ -827.99  $   & $ -827.963  $   &  $ -827.609   $  &  $ -827.953   $  &  $ -827.331    $ &  $ -828.156    $  &  $ -828.387     $  &  $ -830.582    $   &  $ -831.348     $      \\
$\rm ^{102}Sn$    &      $-849.089  $    &     $ -850.561  $   &  $ -849.531  $ & $ -849.526    $ &  $ -849.96  $   & $ -849.284  $   &  $ -849.910   $  &  $ -849.899   $ &   $ -849.459    $ &  $ -850.289    $  &  $ -850.820     $  &  $ -852.657    $   &  $ -853.005     $         \\
$\rm ^{104}Sn$    &      $-871.927  $    &     $ -872.018  $   &  $ -871.042  $ & $ -870.759    $ &  $ -871.25  $   & $ -870.246  $   &  $ -871.957   $  &  $ -871.435   $ &   $ -871.005    $ &  $ -871.989    $  &  $ -872.677     $  &  $ -874.200    $   &  $ -874.175     $         \\
$\rm\mathbf{ ^{106}Sn}$    &      $-893.796  $    &     $ -892.931  $   &  $ -892.323  $ & $ -891.559    $ &  $ -891.94  $   & $ -890.823  $   &  $ -893.469   $  &  $ -892.551   $  &  $ -892.005    $ &  $ -893.370    $  &  $ -893.958     $  &  $ -895.447    $   &  $ -894.851     $      \\
$\rm\mathbf{ ^{108}Sn}$    &      $-914.655  $    &     $ -913.279  $   &  $ -913.179  $ & $ -911.895    $ &  $ -912.09  $   & $ -910.979  $   &  $ -914.627   $  &  $ -913.215   $  &  $ -912.466    $ &  $ -914.236    $  &  $ -914.642     $  &  $ -916.165    $   &  $ -914.997     $      \\
$\rm ^{110}Sn$    &      $-934.570  $    &     $ -933.026  $   &  $ -932.866  $ & $ -931.716    $ &  $ -931.72  $   & $ -930.658  $   &  $ -934.633   $  &  $ -933.368   $ &   $ -932.365    $ &  $ -934.129    $  &  $ -934.690     $  &  $ -935.613    $   &  $ -934.549     $         \\
$\rm\mathbf{ \textcolor{blue}{^{112}Sn}}$    &      $-953.525  $    &     $ -952.113  $   &  $ -951.831  $ & $ -950.942    $ &  $ -950.80  $   & $ -949.771  $   &  $ -953.922   $  &  $ -952.907   $  &  $ -951.640    $ &  $ -953.367    $  &  $ -954.029     $  &  $ -954.258    $   &  $ -953.385     $      \\
$\rm ^{114}Sn$    &      $-971.573  $    &     $ -970.464  $   &  $ -970.121  $ & $ -969.442    $ &  $ -969.23  $   & $ -968.172  $   &  $ -972.427   $  &  $ -971.622   $ &   $ -970.176    $ &  $ -971.826    $  &  $ -972.546     $  &  $ -971.985    $   &  $ -971.310     $         \\
$\rm\mathbf{ \textcolor{blue}{^{116}Sn}}$    &      $-988.682  $    &     $ -988.017  $   &  $ -987.601  $ & $ -987.091    $ &  $ -986.84  $   & $ -985.726  $   &  $ -990.019   $  &  $ -989.271   $  &  $ -987.853    $ &  $ -989.326    $  &  $ -990.125     $  &  $ -989.016    $   &  $ -988.215     $      \\
$\rm ^{118}Sn$    &      $-1004.951 $    &     $ -1004.782 $   &  $ -1004.354 $ & $ -1003.910   $ &  $ -1003.64 $   & $ -1002.492 $   &  $ -1006.817  $  &  $ -1006.004  $ &   $ -1004.671   $ &  $ -1005.902   $  &  $ -1006.760    $  &  $ -1005.207   $   &  $ -1004.332    $         \\
$\rm\mathbf{ \textcolor{blue}{^{120}Sn}}$    &      $-1020.539 $    &     $ -1020.814 $   &  $ -1020.415 $ & $ -1020.001   $ &  $ -1019.73 $   & $ -1018.596 $   &  $ -1022.902  $  &  $ -1022.045  $ &   $ -1020.728   $ &  $ -1021.704   $  &  $ -1022.534    $  &  $ -1020.767   $   &  $ -1019.865    $       \\
$\rm\mathbf{ ^{122}Sn}$    &      $-1035.523 $    &     $ -1036.168 $   &  $ -1035.860 $ & $ -1035.445   $ &  $ -1035.18 $   & $ -1034.115 $   &  $ -1038.417  $  &  $ -1037.525  $ &   $ -1036.128   $ &  $ -1036.755   $  &  $ -1037.547    $  &  $ -1035.794   $   &  $ -1034.908    $       \\
$\rm\mathbf{ ^{124}Sn}$    &      $-1049.958 $    &     $ -1050.884 $   &  $ -1050.715 $ & $ -1050.303   $ &  $ -1050.02 $   & $ -1049.096 $   &  $ -1053.402  $  &  $ -1052.522  $ &   $ -1050.954   $ &  $ -1051.160   $  &  $ -1051.888    $  &  $ -1050.327   $   &  $ -1049.509    $         \\
$\rm\mathbf{ ^{126}Sn}$    &      $-1063.884 $    &     $ -1064.983 $   &  $ -1064.993 $ & $ -1064.612   $ &  $ -1064.31 $   & $ -1063.565 $   &  $ -1067.877  $  &  $ -1067.084  $ &   $ -1065.266   $ &  $ -1064.978   $  &  $ -1065.612    $  &  $ -1064.381   $   &  $ -1063.691    $         \\
$\rm\mathbf{ ^{128}Sn}$    &      $-1077.372 $    &     $ -1078.470 $   &  $ -1078.688 $ & $ -1078.395   $ &  $ -1078.11 $   & $ -1077.536 $   &  $ -1081.835  $  &  $ -1081.240  $ &   $ -1079.104   $ &  $ -1078.234   $  &  $ -1078.748    $  &  $ -1077.945   $   &  $ -1077.462    $         \\
$\rm\mathbf{ ^{130}Sn}$    &      $-1090.286 $    &     $ -1091.328 $   &  $ -1091.774 $ & $ -1091.657   $ &  $ -1091.48 $   & $ -1091.011 $   &  $ -1095.253  $  &  $ -1095.002  $ &   $ -1092.491   $ &  $ -1090.930   $  &  $ -1091.302    $  &  $ -1090.993   $   &  $ -1090.817    $         \\
$\rm\mathbf{ ^{132}Sn}$    &      $-1102.843 $    &     $ -1103.507 $   &  $ -1104.202 $ & $ -1104.382   $ &  $ -1104.48 $   & $ -1103.975 $   &  $ -1108.096  $  &  $ -1108.370  $ &   $ -1105.432   $ &  $ -1103.057   $  &  $ -1103.250    $  &  $ -1103.484   $   &  $ -1103.740    $         \\
$\rm\mathbf{ ^{134}Sn}$    &      $-1108.873 $    &     $ -1109.218 $   &  $ -1109.253 $ & $ -1109.749   $ &  $ -1109.62 $   & $ -1109.288 $   &  $ -1112.253  $  &  $ -1112.507  $ &   $ -1109.652   $ &  $ -1107.330   $  &  $ -1108.139    $  &  $ -1106.707   $   &  $ -1106.988    $         \\
$\rm\mathbf{ ^{134}Te}$    &      $-1123.408 $    &     $ -1123.853 $   &  $ -1124.205 $ & $ -1124.731   $ &  $ -1124.96 $   & $ -1124.061 $   &  $ -1128.176  $  &  $ -1128.711  $ &   $ -1125.604   $ &  $ -1124.193   $  &  $ -1124.834    $  &  $ -1124.613   $   &  $ -1125.105    $         \\
$\rm\mathbf{ ^{136}Xe}$    &      $-1141.882 $    &     $ -1142.397 $   &  $ -1142.621 $ & $ -1143.213   $ &  $ -1143.43 $   & $ -1142.341 $   &  $ -1146.587  $  &  $ -1147.165  $ &   $ -1143.825   $ &  $ -1143.601   $  &  $ -1144.363    $  &  $ -1143.997   $   &  $ -1144.540    $         \\
$\rm\mathbf{ ^{138}Ba}$    &      $-1158.292 $    &     $ -1159.109 $   &  $ -1159.381 $ & $ -1159.811   $ &  $ -1159.99 $   & $ -1158.784 $   &  $ -1163.283  $  &  $ -1163.731  $ &   $ -1160.104   $ &  $ -1161.245   $  &  $ -1161.862    $  &  $ -1161.575   $   &  $ -1162.038    $         \\
$\rm\mathbf{ ^{140}Ce}$    &      $-1172.683 $    &     $ -1173.748 $   &  $ -1174.054 $ & $ -1174.309   $ &  $ -1174.51 $   & $ -1173.159 $   &  $ -1177.868  $  &  $ -1178.187  $ &   $ -1174.352   $ &  $ -1176.722   $  &  $ -1177.129    $  &  $ -1176.953   $   &  $ -1177.356    $         \\
$\rm\mathbf{ ^{142}Nd}$    &      $-1185.136 $    &     $ -1185.801 $   &  $ -1185.938 $ & $ -1186.345   $ &  $ -1186.60 $   & $ -1185.075 $   &  $ -1189.537  $  &  $ -1189.953  $ &   $ -1186.465   $ &  $ -1189.138   $  &  $ -1189.781    $  &  $ -1189.292   $   &  $ -1189.723    $         \\
$\rm\mathbf{ ^{144}Sm}$    &      $-1195.730 $    &     $ -1195.780 $   &  $ -1195.736 $ & $ -1196.293   $ &  $ -1196.50 $   & $ -1194.948 $   &  $ -1199.024  $  &  $ -1199.536  $ &   $ -1196.511   $ &  $ -1199.353   $  &  $ -1200.206    $  &  $ -1199.420   $   &  $ -1199.896    $         \\
$\rm\mathbf{ ^{146}Gd}$    &      $-1204.428 $    &     $ -1203.924 $   &  $ -1203.712 $ & $ -1204.380   $ &  $ -1204.50 $   & $ -1203.003 $   &  $ -1206.614  $  &  $ -1207.198  $ &   $ -1204.615   $ &  $ -1207.635   $  &  $ -1208.652    $  &  $ -1207.687   $   &  $ -1208.201    $         \\
$\rm\mathbf{ ^{148}Dy}$    &      $-1210.779 $    &     $ -1210.343 $   &  $ -1209.974 $ & $ -1210.722   $ &  $ -1210.73 $   & $ -1209.344 $   &  $ -1212.454  $  &  $ -1213.090  $ &   $ -1210.891   $ &  $ -1214.117   $  &  $ -1215.265    $  &  $ -1214.258   $   &  $ -1214.803    $         \\
$\rm\mathbf{ ^{150}Er}$    &      $-1215.330 $    &     $ -1215.139 $   &  $ -1214.624 $ & $ -1215.424   $ &  $ -1215.29 $   & $ -1214.064 $   &  $ -1216.686  $  &  $ -1217.356  $ &   $ -1215.454   $ &  $ -1218.943   $  &  $ -1220.177    $  &  $ -1219.236   $   &  $ -1219.810    $         \\
$\rm\mathbf{ ^{206}Hg}$    &      $-1621.049 $    &     $ -1620.734 $   &  $ -1621.321 $ & $ -1621.914   $ &  $ -1621.79 $   & $ -1621.317 $   &  $ -1623.820  $  &  $ -1624.392  $ &   $ -1621.358   $ &  $ -1620.353   $  &  $ -1621.121    $  &  $ -1616.956   $   &  $ -1617.304    $         \\
$\rm ^{182}Pb$    &      $-1411.652 $    &     $ -1411.621 $   &  $ -1411.370 $ & $ -1409.951   $ &  $ -1409.49 $   & $ -1408.196 $   &  $ -1413.454  $  &  $ -1411.829  $ &   $ -1409.295   $ &  $ -1415.601   $  &  $ -1416.297    $  &  $ -1416.580   $   &  $ -1414.841    $         \\
$\rm ^{184}Pb$    &      $-1432.022 $    &     $ -1431.522 $   &  $ -1431.194 $ & $ -1429.787   $ &  $ -1429.20 $   & $ -1428.087 $   &  $ -1433.373  $  &  $ -1431.627  $ &   $ -1429.360   $ &  $ -1435.047   $  &  $ -1435.821    $  &  $ -1436.089   $   &  $ -1434.255    $         \\
$\rm ^{186}Pb$    &      $-1451.793 $    &     $ -1450.919 $   &  $ -1450.561 $ & $ -1449.122   $ &  $ -1448.35 $   & $ -1447.481 $   &  $ -1452.835  $  &  $ -1450.943  $ &   $ -1448.909   $ &  $ -1453.972   $  &  $ -1454.820    $  &  $ -1455.077   $   &  $ -1453.224    $         \\
$\rm ^{188}Pb$    &      $-1471.066 $    &     $ -1469.851 $   &  $ -1469.491 $ & $ -1468.024   $ &  $ -1467.04 $   & $ -1466.443 $   &  $ -1471.859  $  &  $ -1469.851  $ &   $ -1467.998   $ &  $ -1472.428   $  &  $ -1473.358    $  &  $ -1473.628   $   &  $ -1471.774    $         \\
$\rm ^{190}Pb$    &      $-1489.815 $    &     $ -1488.354 $   &  $ -1487.983 $ & $ -1486.540   $ &  $ -1485.36 $   & $ -1485.023 $   &  $ -1490.447  $  &  $ -1488.403  $ &   $ -1486.673   $ &  $ -1490.541   $  &  $ -1491.481    $  &  $ -1491.709   $   &  $ -1489.922    $         \\
$\rm ^{192}Pb$    &      $-1508.092 $    &     $ -1506.450 $   &  $ -1506.108 $ & $ -1504.701   $ &  $ -1503.34 $   & $ -1503.255 $   &  $ -1508.697  $  &  $ -1506.630  $ &   $ -1504.965   $ &  $ -1508.223   $  &  $ -1509.217    $  &  $ -1509.391   $   &  $ -1507.681    $         \\
$\rm ^{194}Pb$    &      $-1525.891 $    &     $ -1524.157 $   &  $ -1523.838 $ & $ -1522.528   $ &  $ -1521.05 $   & $ -1521.160 $   &  $ -1526.554  $  &  $ -1524.551  $ &   $ -1522.894   $ &  $ -1525.607   $  &  $ -1526.584    $  &  $ -1526.586   $   &  $ -1525.056    $         \\
$\rm ^{196}Pb$    &      $-1543.175 $    &     $ -1541.484 $   &  $ -1541.228 $ & $ -1540.031   $ &  $ -1538.50 $   & $ -1538.749 $   &  $ -1544.007  $  &  $ -1542.174  $ &   $ -1540.473   $ &  $ -1542.618   $  &  $ -1543.591    $  &  $ -1543.439   $   &  $ -1542.049    $         \\
\hline
%\bottomrule
\end{tabular}
\begin{tablenotes}
\item ~~~~~~~~~~~~~~~~~~~~~~~~~~~~~~~~~~~~~~~~~~~~~~~~~~~~~~~~~~~~~~~~~~~~~~~~~~~~~~~~~~~~~~~~~~~~~~~~~~~~~~~~~
~~~~~~~~~~~~~~~~~~~~~~~~~~~~~~~~~~~~~~~~~~~~~~
(continued on next page)
\end{tablenotes}
\end{threeparttable}
\end{table*}

\begin{table*}[pos=h!,width=\textwidth]
%\nonumber
\renewcommand{\thetable}{S3}
%\toprule
\centering
\caption{\footnotesize \centering (continued)}
%$\scriptstyle {\rm RMS} = \scriptstyle \sqrt{\Sigma_{i}^{N}(E^{i}_{expt.}-E^{i}_{calc})^{2}/N}$
%The experimental values of the corresponding nuclei are used in the parametrization fitting.
%\label{tab:BindingEnergy}
\scriptsize
%\tiny
%\resizebox{.90\columnwidth}{!}{
%\begin{tabular}{lcccccccccccc}
\begin{tabular}{@{}l@{\hspace{2mm}}c@{\hspace{2mm}}c@{\hspace{2mm}}c@{\hspace{2mm}}c@{\hspace{2mm}}c@{\hspace{2mm}}c@{\hspace{2mm}}c@{\hspace{2mm}}c@{\hspace{2mm}}c@{\hspace{2mm}}c@{\hspace{2mm}}c@{\hspace{2mm}}c@{\hspace{2mm}}c@{}}
%\hline\hline
\toprule
\midrule
$\rm {Nuclei}$ & AME2020 &  PC-L3R$^a$ & PC-PK1$^b$ & PC-PK1$^{\ast c}$ & PC-PK1$^{\ddag d}$ & PC-X$^e$ & DD-PC1$^f$ & DD-PC1$^{\ast g}$ & DD-PCX$^{h}$ & PC-F1$^i$ & PC-F1$^{\ast j}$ & PC-LA$^k$ & PC-LA$^{\ast l}$  \\
%\hline
\midrule
$\rm ^{198}Pb$    &      $-1560.036 $    &     $ -1558.433 $   &  $ -1558.218 $ & $ -1557.214   $ &  $ -1555.71 $   & $ -1556.029 $   &  $ -1561.123  $  &  $ -1559.500  $ &   $ -1557.703   $ &  $ -1559.292   $  &  $ -1560.242    $  &  $ -1559.829   $   &  $ -1558.650    $         \\
$\rm\mathbf{ ^{200}Pb}$    &      $-1576.362 $    &     $ -1574.998 $   &  $ -1574.885 $ & $ -1574.075   $ &  $ -1572.71 $   & $ -1572.996 $   &  $ -1577.817  $  &  $ -1576.520  $ &   $ -1574.578   $ &  $ -1575.666   $  &  $ -1576.532    $  &  $ -1575.769   $   &  $ -1574.841    $         \\
$\rm\mathbf{ ^{202}Pb}$    &      $-1592.195 $    &     $ -1591.163 $   &  $ -1591.172 $ & $ -1590.601   $ &  $ -1589.49 $   & $ -1589.639 $   &  $ -1594.139  $  &  $ -1593.212  $ &   $ -1591.079   $ &  $ -1591.675   $  &  $ -1592.448    $  &  $ -1591.240   $   &  $ -1590.592    $         \\
$\rm\mathbf{ ^{204}Pb}$    &      $-1607.506 $    &     $ -1606.896 $   &  $ -1607.068 $ & $ -1606.765   $ &  $ -1606.05 $   & $ -1605.928 $   &  $ -1610.026  $  &  $ -1609.526  $ &   $ -1607.165   $ &  $ -1607.325   $  &  $ -1607.963    $  &  $ -1606.187   $   &  $ -1605.845    $         \\
$\rm\mathbf{ ^{206}Pb}$    &      $-1622.325 $    &     $ -1622.133 $   &  $ -1622.525 $ & $ -1622.498   $ &  $ -1622.32 $   & $ -1621.784 $   &  $ -1625.385  $  &  $ -1625.335  $ &   $ -1622.743   $ &  $ -1622.563   $  &  $ -1623.014    $  &  $ -1620.490   $   &  $ -1620.454    $         \\
$\rm\mathbf{ ^{208}Pb}$    &      $-1636.430 $    &     $ -1636.735 $   &  $ -1637.438 $ & $ -1637.626   $ &  $ -1637.92 $   & $ -1637.004 $   &  $ -1640.008  $  &  $ -1640.281  $ &   $ -1637.578   $ &  $ -1637.241   $  &  $ -1637.442    $  &  $ -1633.865   $   &  $ -1633.990    $         \\
$\rm\mathbf{ ^{210}Pb}$    &      $-1645.553 $    &     $ -1645.180 $   &  $ -1645.449 $ & $ -1645.684   $ &  $ -1645.59 $   & $ -1644.936 $   &  $ -1648.272  $  &  $ -1648.032  $ &   $ -1644.993   $ &  $ -1644.793   $  &  $ -1645.397    $  &  $ -1641.484   $   &  $ -1641.006    $         \\
$\rm\mathbf{ ^{212}Pb}$    &      $-1654.516 $    &     $ -1653.439 $   &  $ -1653.425 $ & $ -1653.658   $ &  $ -1653.19 $   & $ -1652.803 $   &  $ -1656.428  $  &  $ -1655.716  $ &   $ -1652.275   $ &  $ -1652.275   $  &  $ -1653.182    $  &  $ -1648.887   $   &  $ -1647.897    $         \\
$\rm\mathbf{ ^{214}Pb}$    &      $-1663.293 $    &     $ -1661.535 $   &  $ -1661.397 $ & $ -1661.552   $ &  $ -1660.73 $   & $ -1660.609 $   &  $ -1664.481  $  &  $ -1663.324  $ &   $ -1659.437   $ &  $ -1659.697   $  &  $ -1660.804    $  &  $ -1656.073   $   &  $ -1654.660    $         \\
$\rm\mathbf{ ^{210}Po}$    &      $-1645.213 $    &     $ -1646.217 $   &  $ -1646.703 $ & $ -1647.137   $ &  $ -1647.25 $   & $ -1646.206 $   &  $ -1649.441  $  &  $ -1649.855  $ &   $ -1646.732   $ &  $ -1647.760   $  &  $ -1648.346    $  &  $ -1644.643   $   &  $ -1644.878    $         \\
$\rm\mathbf{ ^{212}Rn}$    &      $-1652.497 $    &     $ -1654.242 $   &  $ -1654.632 $ & $ -1655.131   $ &  $ -1655.11 $   & $ -1653.940 $   &  $ -1657.476  $  &  $ -1657.930  $ &   $ -1654.342   $ &  $ -1656.863   $  &  $ -1657.576    $  &  $ -1653.921   $   &  $ -1654.215    $         \\
$\rm\mathbf{ ^{214}Ra}$    &      $-1658.323 $    &     $ -1660.831 $   &  $ -1661.172 $ & $ -1661.643   $ &  $ -1661.55 $   & $ -1660.229 $   &  $ -1664.092  $  &  $ -1664.540  $ &   $ -1660.438   $ &  $ -1664.512   $  &  $ -1665.213    $  &  $ -1661.709   $   &  $ -1662.043    $         \\
$\rm\mathbf{ ^{216}Th}$    &      $-1662.695 $    &     $ -1665.941 $   &  $ -1666.248 $ & $ -1666.633   $ &  $ -1666.59 $   & $ -1665.027 $   &  $ -1669.244  $  &  $ -1669.659  $ &   $ -1664.982   $ &  $ -1670.649   $  &  $ -1671.232    $  &  $ -1667.967   $   &  $ -1668.335    $         \\
$\rm\mathbf{ ^{218}~U}$    &      $-1665.677 $    &     $ -1669.317 $   &  $ -1669.602 $ & $ -1669.847   $ &  $ -1670.05 $   & $ -1668.086 $   &  $ -1672.733  $  &  $ -1673.060  $ &   $ -1667.816   $ &  $ -1675.109   $  &  $ -1675.372    $  &  $ -1672.491   $   &  $ -1672.882    $         \\
\hline
$\rm\mathbf{ rms_{60}}$$^m$     &    &     $ 1.176 $   &  $ 1.248  $  &  $ 1.417   $  &  $ 1.64   $ &  $ 1.378   $ &  $ 3.139  $  &  $ 3.269  $ &  $ 1.413  $ &  $ 2.589  $ &  $ 2.913  $ &  $ 2.581  $&  $ 2.890  $\\  

$\rm\textcolor{blue}{\mathbf{rms_{65}}}$$^{n}$    &   &     $ 1.245 $   &  $ 1.312  $  &  $ 1.495   $  &  $ 1.69   $ &  $ 1.542   $ &  $ 3.088  $  &  $ 3.206  $ &  $ 1.461  $ &  $ 2.590  $ &  $ 2.895  $ &  $ 2.615  $&  $ 2.914  $\\                                                     
$\rm rms_{91 }$$^{o}$     &        &     $ 1.339 $   &  $ 1.453  $  &  $ 1.796   $  &  $ 2.11   $ &  $ 2.168   $ &  $ 2.805  $  &  $ 2.925  $ &  $ 1.712  $ &  $ 2.493  $ &  $ 2.755  $ &  $ 2.555  $&  $ 2.714  $\\  
\hline
$\rm\mathbf{rrs_{60}}$ &   &  $ 0.22 \% $   &  $ 0.18 \%  $  &  $ 0.22 \%   $  &  $ 0.39 \%   $ &  $ 0.20 \%   $ &  $ 0.47 \%  $  &  $ 0.50 \% $ &  $ 0.22 \%  $ &  $ 0.30 \%  $ &  $ 0.39 \%  $ &  $ 0.29 \%  $&  $ 0.34 \% $\\  

$\rm\textcolor{blue}{\mathbf{rrs_{65}}}$   &  &     $ 0.23 \% $   &  $ 0.19 \%  $  &  $ 0.23 \%   $  &  $ 0.39\%   $ &  $ 0.22 \%   $ &  $ 0.45 \%  $  &  $ 0.49 \%  $ &  $0.23 \% $ &  $ 0.32 \%  $ &  $ 0.40 \%  $ &  $ 0.30 \%  $&  $ 0.35 \%  $\\                                                     
$\rm rrs_{91 }$  &  &   $ 0.24\% $   &  $ 0.23 \%  $  &  $ 0.27 \%   $  &  $ 0.39 \%   $ &  $ 0.27 \%   $ &  $ 0.44 \%  $  &  $ 0.48 \%  $ &  $ 0.26 \%  $ &  $ 0.35 \%  $ &  $ 0.40 \%  $ &  $ 0.32 \%  $&  $ 0.35 \%  $ \\ 
%\hline\hline
\bottomrule
\end{tabular}
%}
%\begin{minipage}{\columnwidth}
\begin{minipage}{\textwidth}
%\vskip5pt
%{\sc Note}---\\
% \textbf{Note.}\\
\tiny
The experimental binding energies compiled in AME2020 \cite{AME2020}. 
$^a$ PC-L3R: calculated from the RHB model with the PC-L3R interaction.
$^b$ PC-PK1: quoted from the RMF model with the Bardeen–Cooper–Schrieffer (BCS) theory using the PC-PK1 interaction \cite{PRC2010Zhao} and $\delta$ pairing force.
$^c$ PC-PK1$^*$: calculated from the RHB model using the PC-PK1 interaction \cite{PRC2010Zhao} and separable pairing force \cite{PLB2009Tian_676_44}.
$^d$ PC-PK1$^\ddag$: quoted from the relativistic continuum Hartree-Bogoliubov (RCHB) model \cite{ADNDT2018Xia} using the PC-PK1 interaction \cite{PRC2010Zhao} and $\delta$ pairing force.
$^e$ PC-X: calculated from the RHB model using the PC-X interaction \cite{PLB2020Taninah} and separable pairing force \cite{PLB2009Tian_676_44}.
$^f$ DD-PC1: quoted from the RMF model with the BCS theory using the DD-PC1 interaction \cite{PRC2008Niksic} and $\delta$ pairing force.
$^g$ DD-PC1$^*$: calculated from the RHB model using the DD-PC1 interaction \cite{PRC2008Niksic} and separable pairing force \cite{PLB2009Tian_676_44}.
$^h$ DD-PCX: calculated from the RHB model using the DD-PCX interaction \cite{PRC2019Yuksel} and separable pairing force \cite{PRC2019Yuksel}.
$^i$ PC-F1: quoted from the RMF model with the BCS theory using the PC-F1 interaction \cite{PRC2002Burvenich} and $\delta$ pairing force.
$^j$ PC-F1$^*$: calculated from the RHB model using the PC-F1 interaction \cite{PRC2002Burvenich} and separable pairing force \cite{PLB2009Tian_676_44}.
$^k$ PC-LA: quoted from the RMF model with the BCS theory using the PC-LA interaction \cite{PRC1992Nikolaus} and $\delta$ pairing force.
$^l$ PC-LA$^*$: calculated from the RHB model using the PC-LA interaction \cite{PRC1992Nikolaus} and separable pairing force \cite{PLB2009Tian_676_44}.
The root mean square (rms) and the root of relative square (rrs) deviations of comparing the theoretical and experimental binding energies are listed in the last rows. 
$^m$ rms$_{60}$: calculated from the binding energies of $60$ spherical nuclei (black bold texts), selected for fitting the PC-PK1~\cite{PRC2010Zhao}. 
$^n$ rms$_{65}$: calculated from the binding energies of $65$ spherical nuclei (blue bold texts), referred by previous works in fitting point coupling interactions~\cite{PRC2010Zhao,PRC2019Yuksel,PLB2020Taninah}. 
$^o$ rms$_{91}$: calculated from the binding energies of $91$ spherical nuclei (all nuclei in this table), selected for the present work fitting the PC-L3R interaction. 
The rrs have a similar label as rms.
%\vspace{10mm}
\end{minipage}
\end{table*}
%%%

%%%%

%{\bf Supplementary material}\\
%\vspace{-10mm}
\begin{table*}[pos=h!,width=\textwidth]
\renewcommand{\thetable}{S4}
\centering
\caption{\footnotesize The comparison of charge radii (in fm) of the selected \textcolor{blue}{$63$} spherical nuclei.}
%$\scriptstyle {\rm RMS} = \scriptstyle \sqrt{\Sigma_{i}^{N}(E^{i}_{expt.}-E^{i}_{calc})^{2}/N}$
%The experimental values of the corresponding nuclei are used in the parametrization fitting.
\label{tab:ChargeRadii}
\scriptsize
%\resizebox{.90\columnwidth}{!}{
%\begin{tabular}{lcccccccccccc}
\begin{tabular}{@{}l@{\hspace{2mm}}c@{\hspace{2mm}}c@{\hspace{2mm}}c@{\hspace{2mm}}c@{\hspace{2mm}}c@{\hspace{2mm}}c@{\hspace{2mm}}c@{\hspace{2mm}}c@{\hspace{2mm}}c@{\hspace{2mm}}c@{\hspace{2mm}}c@{\hspace{2mm}}c@{\hspace{2mm}}c@{}}
%\hline\hline
\toprule
\midrule
$\rm {Nuclei}$ & Expt. & PC-L3R$^a$ & PC-PK1$^b$ & PC-PK1$^{\ast c}$ & PC-PK1$^{\ddag d}$ & PC-X$^e$ & DD-PC1$^f$ & DD-PC1$^{\ast g}$ & DD-PCX$^{h}$ & PC-F1$^i$ & PC-F1$^{\ast j}$ & PC-LA$^k$ & PC-LA$^{\ast l}$  \\
%\hline
\midrule
$\rm\mathbf{ ^{16 }O} $     &      $2.6991 $    &     $ 2.7676  $   &  $ 2.7677  $   & $ 2.7677  $   &  $ 2.768  $   & $ 2.7698  $   &  $ 2.7472  $  &  $ 2.7471  $ &  $ 2.7594   $   &  $ 2.7633$    &  $ 2.7632  $     &  $ 2.7528  $     &  $ 2.7528  $       \\
$\rm ^{18 }O $     &      $2.7726 $    &     $ 2.7599  $   &  $ ~       $   & $ 2.7594  $   &  $ 2.763  $   & $ 2.7603  $   &  $ ~       $  &  $ 2.7475  $ &  $ 2.7622   $   &  $ ~     $    &  $ 2.7563  $     &  $ ~       $     &  $ 2.7505  $       \\   
$\rm ^{18 }Ne$     &      $2.9714 $    &     $ 2.9694  $   &  $ ~       $   & $ 2.9700  $   &  $ 2.960  $   & $ 2.9691  $   &  $ ~       $  &  $ 2.9456  $ &  $ 2.9564   $   &  $ ~     $    &  $ 2.9693  $     &  $ ~       $     &  $ 2.9508  $       \\   
$\rm ^{36 }S$      &      $3.2985 $    &     $ 3.2852  $   &  $ ~       $   & $ 3.2872  $   &  $ 3.289  $   & $ 3.2904  $   &  $ ~       $  &  $ 3.2898  $ &  $ 3.2926   $   &  $ ~     $    &  $ 3.2912  $     &  $ ~       $     &  $ 3.3016  $       \\   
$\rm ^{38 }Ar$     &      $3.4028 $    &     $ 3.3918  $   &  $ ~       $   & $ 3.3916  $   &  $ 3.391  $   & $ 3.3942  $   &  $ ~       $  &  $ 3.3785  $ &  $ 3.3837   $   &  $ ~     $    &  $ 3.3903  $     &  $ ~       $     &  $ 3.3891  $       \\   
$\rm\mathbf{ ^{40 }Ca}$     &      $3.4776 $    &     $ 3.4826  $   &  $ 3.4815  $   & $ 3.4814  $   &  $ 3.481  $   & $ 3.4832  $   &  $ 3.4566  $  &  $ 3.4561  $ &  $ 3.4617   $   &  $ 3.4777$    &  $ 3.4774  $     &  $ 3.4678  $     &  $ 3.4674  $      \\    
$\rm\mathbf{ ^{42 }Ca}$     &      $3.5081 $    &     $ 3.4821  $   &  $ 3.4805  $   & $ 3.4808  $   &  $ 3.482  $   & $ 3.4820  $   &  $ 3.4626  $  &  $ 3.4622  $ &  $ 3.4689   $   &  $ 3.4778$    &  $ 3.4782  $     &  $ 3.4729  $     &  $ 3.4725  $      \\   
$\rm \mathbf{^{44 }Ca}$     &      $3.5179 $    &     $ 3.4848  $   &  $ 3.4826  $   & $ 3.4832  $   &  $ 3.486  $   & $ 3.4838  $   &  $ 3.4709  $  &  $ 3.4705  $ &  $ 3.4777   $   &  $ 3.4809$    &  $ 3.4817  $     &  $ 3.4810  $     &  $ 3.4806  $      \\   
$\rm\mathbf{ ^{46 }Ca}$     &      $3.4953 $    &     $ 3.4890  $   &  $ 3.4865  $   & $ 3.4869  $   &  $ 3.490  $   & $ 3.4871  $   &  $ 3.4806  $  &  $ 3.4799  $ &  $ 3.4867   $   &  $ 3.4860$    &  $ 3.4868  $     &  $ 3.4912  $     &  $ 3.4908  $      \\    
$\rm\mathbf{ ^{48 }Ca}$     &      $3.4771 $    &     $ 3.4935  $   &  $ 3.4890  $   & $ 3.4889  $   &  $ 3.494  $   & $ 3.4890  $   &  $ 3.4895  $  &  $ 3.4891  $ &  $ 3.4935   $   &  $ 3.4906$    &  $ 3.4916  $     &  $ 3.5023  $     &  $ 3.5020  $       \\  
$\rm^{50 }Ca$     &      $3.5168 $    &     $ 3.5157  $   &  $ ~       $   & $ 3.5122  $   &  $ 3.515  $   & $ 3.5130  $   &  $ ~       $  &  $ 3.5077  $ &  $ 3.5155   $   &  $ ~     $    &  $ 3.5112  $     &  $ ~       $     &  $ 3.5180  $      \\    
$\rm\mathbf{ ^{50 }Ti}$     &      $3.5704 $    &     $ 3.5537  $   &  $ 3.5558  $   & $ 3.5543  $   &  $ 3.555  $   & $ 3.5522  $   &  $ 3.5696  $  &  $ 3.5673  $ &  $ 3.5680   $   &  $ 3.5664$    &  $ 3.5648  $     &  $ 3.5868  $     &  $ 3.5846  $      \\    
$\rm\mathbf{ ^{58 }Ni}$     &      $3.7757 $    &     $ 3.7358  $   &  $ 3.7372  $   & $ 3.7362  $   &  $ 3.737  $   & $ 3.7308  $   &  $ 3.7761  $  &  $ 3.7744  $ &  $ 3.7621   $   &  $ 3.7645$    &  $ 3.7652  $     &  $ 3.8065  $     &  $ 3.8055  $      \\    
$\rm ^{60 }Ni$     &      $3.8118 $    &     $ 3.7676  $   &  $ ~       $   & $ 3.7673    $ &  $ 3.768  $   & $ 3.7637    $ &  $ ~        $ &  $ 3.7972   $&  $ 3.7865    $  &  $ ~       $  &  $ 3.7892     $  &  $ ~         $   &  $ 3.8246     $       \\  
$\rm ^{62 }Ni$     &      $3.8399 $    &     $ 3.7983  $   &  $ ~       $   & $ 3.7982    $ &  $ 3.797  $   & $ 3.7961    $ &  $ ~        $ &  $ 3.8211   $&  $ 3.8136    $  &  $ ~       $  &  $ 3.8135     $  &  $ ~         $   &  $ 3.8445     $         \\
$\rm ^{64 }Ni$     &      $3.8572 $    &     $ 3.8275  $   &  $ ~       $   & $ 3.8278    $ &  $ 3.827  $   & $ 3.8270    $ &  $ ~        $ &  $ 3.8449   $&  $ 3.8408    $  &  $ ~       $  &  $ 3.8372     $  &  $ ~         $   &  $ 3.8644     $         \\
$\rm ^{86 }Kr$     &      $4.1835 $    &     $ 4.1818  $   &  $ ~       $   & $ 4.1813    $ &  $ 4.180  $   & $ 4.1801    $ &  $ ~        $ &  $ 4.1823   $&  $ 4.1847    $  &  $ ~       $  &  $ 4.1821     $  &  $ ~         $   &  $ 4.1965     $         \\
$\rm\mathbf{ ^{88 }Sr}$     &      $4.2240 $    &     $ 4.2245  $   &  $ 4.2247  $   & $ 4.2240    $ &  $ 4.223  $   & $ 4.2222    $ &  $ 4.2231   $ &  $ 4.2218   $&  $ 4.2250    $  &  $ 4.2269  $  &  $ 4.2261     $  &  $ 4.2379    $   &  $ 4.2365     $         \\
$\rm\mathbf{ ^{90 }Zr}$     &      $4.2694 $    &     $ 4.2688  $   &  $ 4.2695  $   & $ 4.2681    $ &  $ 4.267  $   & $ 4.2662    $ &  $ 4.2664   $ &  $ 4.2654   $&  $ 4.2673    $  &  $ 4.2724  $  &  $ 4.2719     $  &  $ 4.2847    $   &  $ 4.2828     $         \\
$\rm\mathbf{ ^{92 }Mo}$     &      $4.3151 $    &     $ 4.3111  $   &  $ 4.3125  $   & $ 4.3107    $ &  $ 4.310  $   & $ 4.3081    $ &  $ 4.3140   $ &  $ 4.3120   $&  $ 4.3112    $  &  $ 4.3192  $  &  $ 4.3178     $  &  $ 4.3333    $   &  $ 4.3308     $         \\
$\rm ^{108}Sn$     &      $4.5605 $    &     $ 4.5455  $   &  $ ~       $   & $ 4.5456    $ &  $ 4.544  $   & $ 4.5417    $ &  $ ~        $ &  $ 4.5572   $&  $ 4.5486    $  &  $ ~       $  &  $ 4.5562     $  &  $ ~         $   &  $ 4.5774     $         \\
$\rm ^{110}Sn$     &      $4.5785 $    &     $ 4.5645  $   &  $ ~       $   & $ 4.5647    $ &  $ 4.563  $   & $ 4.5613    $ &  $ ~        $ &  $ 4.5738   $&  $ 4.5673    $  &  $ ~       $  &  $ 4.5728     $  &  $ ~         $   &  $ 4.5922     $         \\
$\rm\mathbf{ ^{112}Sn}$     &      $4.5948 $    &     $ 4.5825  $   &  $ 4.5801  $   & $ 4.5826    $ &  $ 4.582  $   & $ 4.5796    $ &  $ 4.5894   $ &  $ 4.5891   $&  $ 4.5849    $  &  $ 4.5870  $  &  $ 4.5883     $  &  $ 4.6044    $   &  $ 4.6059     $         \\
$\rm ^{114}Sn$     &      $4.6099 $    &     $ 4.5993  $   &  $ ~       $   & $ 4.5993  $   &  $ 4.599  $   & $ 4.5967  $   &  $ ~       $  &  $ 4.6030  $ &  $ 4.6014   $   &  $ ~     $    &  $ 4.6028  $     &  $ ~       $     &  $ 4.6186  $       \\
$\rm\mathbf{ ^{116}Sn}$     &      $4.6250 $    &     $ 4.6150  $   &  $ 4.6121  $   & $ 4.6148  $   &  $ 4.614  $   & $ 4.6126  $   &  $ 4.6174  $  &  $ 4.6165  $ &  $ 4.6168   $   &  $ 4.6168$    &  $ 4.6164  $     &  $ 4.6307  $     &  $ 4.6314  $       \\   
$\rm ^{118}Sn$     &      $4.6393 $    &     $ 4.6299  $   &  $ ~       $   & $ 4.6293  $   &  $ 4.629  $   & $ 4.6275  $   &  $ ~       $  &  $ 4.6300  $ &  $ 4.6314   $   &  $ ~     $    &  $ 4.6296  $     &  $ ~       $     &  $ 4.6447  $       \\   
$\rm ^{120}Sn$     &      $4.6519 $    &     $ 4.6441  $   &  $ ~       $   & $ 4.6431  $   &  $ 4.643  $   & $ 4.6417  $   &  $ ~       $  &  $ 4.6435  $ &  $ 4.6455   $   &  $ ~     $    &  $ 4.6424  $     &  $ ~       $     &  $ 4.6582  $       \\   
$\rm\mathbf{ ^{122}Sn}$     &      $4.6634 $    &     $ 4.6579  $   &  $ 4.6561  $   & $ 4.6566  $   &  $ 4.657  $   & $ 4.6553  $   &  $ 4.6579  $  &  $ 4.6571  $ &  $ 4.6593   $   &  $ 4.6549$    &  $ 4.6552  $     &  $ 4.6728  $     &  $ 4.6718  $       \\   
$\rm\mathbf{ ^{124}Sn}$     &      $4.6735 $    &     $ 4.6714  $   &  $ 4.6694  $   & $ 4.6698  $   &  $ 4.670  $   & $ 4.6687  $   &  $ 4.6714  $  &  $ 4.6708  $ &  $ 4.6727   $   &  $ 4.6677$    &  $ 4.6681  $     &  $ 4.6864  $     &  $ 4.6856  $       \\   
$\rm ^{126}Sn$     &      $4.6833 $    &     $ 4.6847  $   &  $ ~       $   & $ 4.6830  $   &  $ 4.683  $   & $ 4.6819  $   &  $ ~       $  &  $ 4.6844  $ &  $ 4.6859   $   &  $ ~     $    &  $ 4.6812  $     &  $ ~       $     &  $ 4.6994  $      \\    
$\rm ^{128}Sn$     &      $4.6921 $    &     $ 4.6979  $   &  $ ~       $   & $ 4.6961  $   &  $ 4.696  $   & $ 4.6949  $   &  $ ~       $  &  $ 4.6981  $ &  $ 4.6987   $   &  $ ~     $    &  $ 4.6943  $     &  $ ~       $     &  $ 4.7134  $      \\    
$\rm ^{130}Sn$     &      $4.7019 $    &     $ 4.7108  $   &  $ ~       $   & $ 4.7091  $   &  $ 4.709  $   & $ 4.7078  $   &  $ ~       $  &  $ 4.7116  $ &  $ 4.7111   $   &  $ ~     $    &  $ 4.7074  $     &  $ ~       $     &  $ 4.7274  $      \\    
$\rm ^{132}Sn$     &      $4.7093 $    &     $ 4.7232  $   &  $ ~       $   & $ 4.7218  $   &  $ 4.722  $   & $ 4.7205  $   &  $ ~       $  &  $ 4.7250  $ &  $ 4.7229   $   &  $ ~     $    &  $ 4.7202  $     &  $ ~       $     &  $ 4.7415  $      \\    
$\rm ^{134}Te$     &      $4.7569 $    &     $ 4.7691  $   &  $ ~       $   & $ 4.7682  $   &  $ 4.767  $   & $ 4.7666  $   &  $ ~       $  &  $ 4.7693  $ &  $ 4.7683   $   &  $ ~     $    &  $ 4.7663  $     &  $ ~       $     &  $ 4.7850  $      \\    
$\rm ^{136}Xe$     &      $4.7964 $    &     $ 4.8116  $   &  $ ~       $   & $ 4.8107  $   &  $ 4.809  $   & $ 4.8089  $   &  $ ~       $  &  $ 4.8108  $ &  $ 4.8104   $   &  $ ~     $    &  $ 4.8088  $     &  $ ~       $     &  $ 4.8261  $      \\    
$\rm\mathbf{ ^{138}Ba}$     &      $4.8378 $    &     $ 4.8513  $   &  $ 4.8508  $   & $ 4.8503  $   &  $ 4.848  $   & $ 4.8483  $   &  $ 4.8511  $  &  $ 4.8501  $ &  $ 4.8497   $   &  $ 4.8494$    &  $ 4.8487  $     &  $ 4.8667  $     &  $ 4.8654  $      \\    
$\rm\mathbf{ ^{140}Ce}$     &      $4.8771 $    &     $ 4.8883  $   &  $ 4.8879  $   & $ 4.8872    $ &  $ 4.885  $   & $ 4.8850    $ &  $ 4.8879   $ &  $ 4.8867   $&  $ 4.8861    $  &  $ 4.8871  $  &  $ 4.8861     $  &  $ 4.9037    $   &  $ 4.9024     $       \\  
$\rm ^{142}Nd$     &      $4.9123 $    &     $ 4.9217  $   &  $ ~       $   & $ 4.9209    $ &  $ 4.919  $   & $ 4.9183    $ &  $ ~        $ &  $ 4.9188   $&  $ 4.9199    $  &  $ ~       $  &  $ 4.9204     $  &  $ ~         $   &  $ 4.9337     $         \\
$\rm\mathbf{ ^{144}Sm}$     &      $4.9524 $    &     $ 4.9541  $   &  $ 4.9544  $   & $ 4.9534    $ &  $ 4.951  $   & $ 4.9504    $ &  $ 4.9521   $ &  $ 4.9505   $&  $ 4.9523    $  &  $ 4.9547  $  &  $ 4.9539     $  &  $ 4.9676    $   &  $ 4.9657     $         \\
$\rm ^{146}Gd$     &      $4.9801 $    &     $ 4.9862  $   &  $ ~       $   & $ 4.9857    $ &  $ 4.983  $   & $ 4.9822    $ &  $ ~        $ &  $ 4.9829   $&  $ 4.9846    $  &  $ ~       $  &  $ 4.9875     $  &  $ ~         $   &  $ 4.9989     $         \\
$\rm ^{148}Dy$     &      $5.0455 $    &     $ 5.0183  $   &  $ ~       $   & $ 5.0179    $ &  $ 5.015  $   & $ 5.0139    $ &  $ ~        $ &  $ 5.0161   $&  $ 5.0169    $  &  $ ~       $  &  $ 5.0212     $  &  $ ~         $   &  $ 5.0328     $         \\
$\rm ^{150}Er$     &      $5.0548 $    &     $ 5.0501  $   &  $ ~       $   & $ 5.0499    $ &  $ 5.047  $   & $ 5.0454    $ &  $ ~        $ &  $ 5.0493   $&  $ 5.0490    $  &  $ ~       $  &  $ 5.0546     $  &  $ ~         $   &  $ 5.0666     $         \\
$\rm ^{206}Hg$     &      $5.4837 $    &     $ 5.5053  $   &  $ ~       $   & $ 5.5036  $   &  $ 5.503  $   & $ 5.5006  $   &  $ ~       $  &  $ 5.4967  $ &  $ 5.4949   $   &  $ ~     $    &  $ 5.4995  $     &  $ ~       $     &  $ 5.5121  $      \\   
$\rm ^{182}Pb$     &      $5.3788 $    &     $ 5.3856  $   &  $ ~       $   & $ 5.3846    $ &  $ 5.384  $   & $ 5.3793    $ &  $ ~        $ &  $ 5.3813   $&  $ 5.3811    $  &  $ ~       $  &  $ 5.3877     $  &  $ ~         $   &  $ 5.3955     $         \\
$\rm ^{184}Pb$     &      $5.3930 $    &     $ 5.3966  $   &  $ ~       $   & $ 5.3952    $ &  $ 5.394  $   & $ 5.3900    $ &  $ ~        $ &  $ 5.3916   $&  $ 5.3919    $  &  $ ~       $  &  $ 5.3977     $  &  $ ~         $   &  $ 5.4056     $         \\
$\rm ^{186}Pb$     &      $5.4027 $    &     $ 5.4076  $   &  $ ~       $   & $ 5.4059    $ &  $ 5.405  $   & $ 5.4008    $ &  $ ~        $ &  $ 5.4023   $&  $ 5.4028    $  &  $ ~       $  &  $ 5.4080     $  &  $ ~         $   &  $ 5.4160     $         \\
$\rm ^{188}Pb$     &      $5.4139 $    &     $ 5.4185  $   &  $ ~       $   & $ 5.4167    $ &  $ 5.416  $   & $ 5.4116    $ &  $ ~        $ &  $ 5.4131   $&  $ 5.4136    $  &  $ ~       $  &  $ 5.4183     $  &  $ ~         $   &  $ 5.4267     $         \\
$\rm ^{190}Pb$     &      $5.4222 $    &     $ 5.4295  $   &  $ ~       $   & $ 5.4275  $   &  $ 5.427  $   & $ 5.4225  $   &  $ ~       $  &  $ 5.4239  $ &  $ 5.4243   $   &  $ ~     $    &  $ 5.4287  $     &  $ ~       $     &  $ 5.4375  $       \\
$\rm ^{192}Pb$     &      $5.4300 $    &     $ 5.4403  $   &  $ ~       $   & $ 5.4383  $   &  $ 5.438  $   & $ 5.4334  $   &  $ ~       $  &  $ 5.4347  $ &  $ 5.4349   $   &  $ ~     $    &  $ 5.4390  $     &  $ ~       $     &  $ 5.4484  $       \\   
$\rm ^{194}Pb$     &      $5.4372 $    &     $ 5.4511  $   &  $ ~       $   & $ 5.4490  $   &  $ 5.448  $   & $ 5.4443  $   &  $ ~       $  &  $ 5.4454  $ &  $ 5.4453   $   &  $ ~     $    &  $ 5.4493  $     &  $ ~       $     &  $ 5.4593  $       \\   
$\rm ^{196}Pb$     &      $5.4444 $    &     $ 5.4618  $   &  $ ~       $   & $ 5.4596  $   &  $ 5.459  $   & $ 5.4550  $   &  $ ~       $  &  $ 5.4560  $ &  $ 5.4556   $   &  $ ~     $    &  $ 5.4595  $     &  $ ~       $     &  $ 5.4702  $       \\   
$\rm ^{198}Pb$     &      $5.4524 $    &     $ 5.4723  $   &  $ ~       $   & $ 5.4701  $   &  $ 5.470  $   & $ 5.4657  $   &  $ ~       $  &  $ 5.4664  $ &  $ 5.4656   $   &  $ ~     $    &  $ 5.4696  $     &  $ ~       $     &  $ 5.4808  $       \\   
$\rm ^{200}Pb$     &      $5.4611 $    &     $ 5.4826  $   &  $ ~       $   & $ 5.4804  $   &  $ 5.480  $   & $ 5.4762  $   &  $ ~       $  &  $ 5.4766  $ &  $ 5.4754   $   &  $ ~     $    &  $ 5.4796  $     &  $ ~       $     &  $ 5.4913  $       \\   
$\rm\mathbf{ ^{202}Pb}$     &      $5.4705 $    &     $ 5.4928  $   &  $ 5.4908  $   & $ 5.4905  $   &  $ 5.490  $   & $ 5.4865  $   &  $ 5.4869  $  &  $ 5.4865  $ &  $ 5.4849   $   &  $ 5.4892$    &  $ 5.4893  $     &  $ 5.4996  $     &  $ 5.5014  $      \\    
$\rm\mathbf{ ^{204}Pb}$     &      $5.4803 $    &     $ 5.5026  $   &  $ 5.5005  $   & $ 5.5003  $   &  $ 5.500  $   & $ 5.4966  $   &  $ 5.4962  $  &  $ 5.4960  $ &  $ 5.4940   $   &  $ 5.4987$    &  $ 5.4988  $     &  $ 5.5112  $     &  $ 5.5110  $      \\    
$\rm\mathbf{ ^{206}Pb}$     &      $5.4902 $    &     $ 5.5119  $   &  $ 5.5098  $   & $ 5.5098  $   &  $ 5.509  $   & $ 5.5063  $   &  $ 5.5049  $  &  $ 5.5047  $ &  $ 5.5027   $   &  $ 5.5078$    &  $ 5.5079  $     &  $ 5.5200  $     &  $ 5.5197  $      \\    
$\rm\mathbf{ ^{208}Pb}$     &      $5.5012 $    &     $ 5.5204  $   &  $ 5.5185  $   & $ 5.5186  $   &  $ 5.518  $   & $ 5.5154  $   &  $ 5.5129  $  &  $ 5.5128  $ &  $ 5.5109   $   &  $ 5.5162$    &  $ 5.5162  $     &  $ 5.5279  $     &  $ 5.5275  $      \\    
$\rm ^{210}Pb$     &      $5.5208 $    &     $ 5.5405  $   &  $ ~       $   & $ 5.5385  $   &  $ 5.538  $   & $ 5.5346  $   &  $ ~       $  &  $ 5.5327  $ &  $ 5.5292   $   &  $ ~     $    &  $ 5.5363  $     &  $ ~       $     &  $ 5.5464  $      \\    
$\rm ^{212}Pb$     &      $5.5396 $    &     $ 5.5599  $   &  $ ~       $   & $ 5.5580  $   &  $ 5.558  $   & $ 5.5539  $   &  $ ~       $  &  $ 5.5520  $ &  $ 5.5472   $   &  $ ~     $    &  $ 5.5556  $     &  $ ~       $     &  $ 5.5649  $      \\    
$\rm\mathbf{ ^{214}Pb}$     &      $5.5577 $    &     $ 5.5789  $   &  $ 5.5798  $   & $ 5.5773    $ &  $ 5.578  $   & $ 5.5730    $ &  $ 5.5711   $ &  $ 5.5707   $&  $ 5.5650    $  &  $ 5.5762  $  &  $ 5.5743     $  &  $ 5.5813    $   &  $ 5.5828     $       \\  
$\rm ^{210}Po$     &      $5.5704 $    &     $ 5.5554  $   &  $ ~       $   & $ 5.5540    $ &  $ 5.552  $   & $ 5.5346    $ &  $ ~        $ &  $ 5.5468   $&  $ 5.5459    $  &  $ ~       $  &  $ 5.5513     $  &  $ ~         $   &  $ 5.5606     $         \\
$\rm ^{212}Rn$     &      $5.5915 $    &     $ 5.5884  $   &  $ ~       $   & $ 5.5871    $ &  $ 5.585  $   & $ 5.5836    $ &  $ ~        $ &  $ 5.5792   $&  $ 5.5791    $  &  $ ~       $  &  $ 5.5840     $  &  $ ~         $   &  $ 5.5922     $         \\
$\rm ^{214}Ra$     &      $5.6079 $    &     $ 5.6200  $   &  $ ~       $   & $ 5.6188    $ &  $ 5.617  $   & $ 5.6151    $ &  $ ~        $ &  $ 5.6103   $&  $ 5.6108    $  &  $ ~       $  &  $ 5.6153     $  &  $ ~         $   &  $ 5.6229     $         \\
\hline                                                                                              
$\rm\mathbf{rms_{23}}$$^n$ &~  &     $ 0.0224 $   &  $ 0.0222 
$  &  $ 0.0220    $  &  $ 0.0217   $ &  $ 0.0222   $ &  $ 0.0197  $  &  $ 0.0197 $&  $ 0.0195 $ &  $ 0.0197 $  &  $ 0.0194 $ &  $ 0.0250  $ &  $ 0.0247  $\\  

$\rm rms_{63}$$^{o}$ &~   &     $ 0.0187 $   &  $ ~  $  &  $ 0.0183   $  &  $ 0.0182   $ &  $ 0.0187   $ &  $ ~  $  &  $ 0.0157  $ &  $ 0.0153 $&  ~ &  $ 0.0153  $ &  $ ~  $ &  $ 0.0211  $ \\
\hline
$\rm\mathbf{rrs_{23}}$   & &    $ 0.68 \% $   &  $ 0.68 \%  $  &  $ 0.67 \%   $  &  $ 0.67 \%   $ &  $ 0.69 \%   $ &  $ 0.58 \%  $  &  $ 0.59 \% $ &  $ 0.61 \% $&  $ 0.61 \%  $ &  $ 0.61\%  $ &  $ 0.67 \%  $ &  $ 0.66 \%  $\\  

$\rm rrs_{63}$     &    & $ 0.51 \% $   &  $~  $  &  $ 0.50 \%   $  &  $ 0.50\%   $ &  $ 0.52 \%   $ &  $ ~  $  &  $ 0.44 \%  $ &  $ 0.43 \% $&  ~ &  $ 0.43\%  $ &  $ ~  $ &  $ 0.51\%  $ \\
%\hline\hline                                                                                               
\bottomrule
\end{tabular}
%}
%\begin{minipage}{\columnwidth}
\begin{minipage}{\textwidth}
%\vskip5pt
%{\sc Note}---\\
% \textbf{Note.}\\
\tiny
The experimental data is quoted from the compilation by Angeli \& Marinova \cite{ADNDT2013Angeli}. \\
See the footnote of Tables~\ref{tab:RMS} and ~\ref{tab:BindingEnergy} for further description of the theoretical data. \\
%\vspace{10mm}
\end{minipage}
\end{table*}
%\end{minipage}

%\begin{figure*}[pos=thb!]
%\renewcommand{\thefigure}{S1}
%\centering
%\footnotesize
%\includegraphics[width=1.0\textwidth]{PC-L3R.pdf}
%\caption{\footnotesize\textcolor{blue}{ The variation of the parameters corresponding to local minima obtained by minimization and the rms of 91 spherical nuclei binding energies for the functional PC-L3R. The left y-axis indicates the $\Delta \chi ^{2}({\rm P}) $ value of the $\chi ^{2} ({\rm P})$ for the functionals in these local minima where the former is expressed as $\Delta \chi ^{2} ({\rm P}) = \chi ^{2} (\rm P) - \chi ^{2} (\rm P_{0}) $. The right y-axis indicates the $\Delta \rm rms_{91}$ have same as $\Delta \chi ^{2}$. The $\Delta \chi ^{2}$ of $\Delta O_{\rm BE}=$ $0.8$~MeV (green), $1$~MeV (red), $1.2$~MeV (yellow) are calculated, respectively.}}
%\vspace{-5mm}
%\label{fig:X2-rms91}
%\end{figure*}
\textcolor{black}{
{\bf The Symmetric Nuclear Matter Energy \,--\,}
Figure~\ref{fig:Eos_EA} displays the binding energy curves for symmetric nuclear matter produced from the PC-L3R, PC-PK1, PC-X, DD-PC1, DD-PCX, PC-F1, and PC-LA interactions, and the \emph{ab initio} variational calculations (black dots)~\cite{PRC1998Akmal}. For the baryon density $\rho_\mathrm{B}\!\leq\!0.20$, the $E\!/\!A$ predicted by PC-L3R is consistent with all point-coupling interactions and the \emph{ab initio} variational calculation owing to the constraints from the properties of finite nuclei. For the high density region with $\rho_\mathrm{B}\!\geq\!0.30$, the DD-PCX binding energy curve is closer to the prediction from the \emph{ab initio} variational calculation, while softer than those from the PC-L3R, PC-PK1, PC-X, and PC-F1 interactions, and stiffer than the one from the DD-PC1, and PC-LA interactions. Note that the inconsistent description at the high-density nuclear matter region, $\rho_\mathrm{B}\!\geq\!0.30$, is not sensitive on the description of the low-energy bulk nuclear properties.}

\begin{figure}[pos=hb!]
\renewcommand{\thefigure}{S2}
\includegraphics[width=0.47\textwidth]{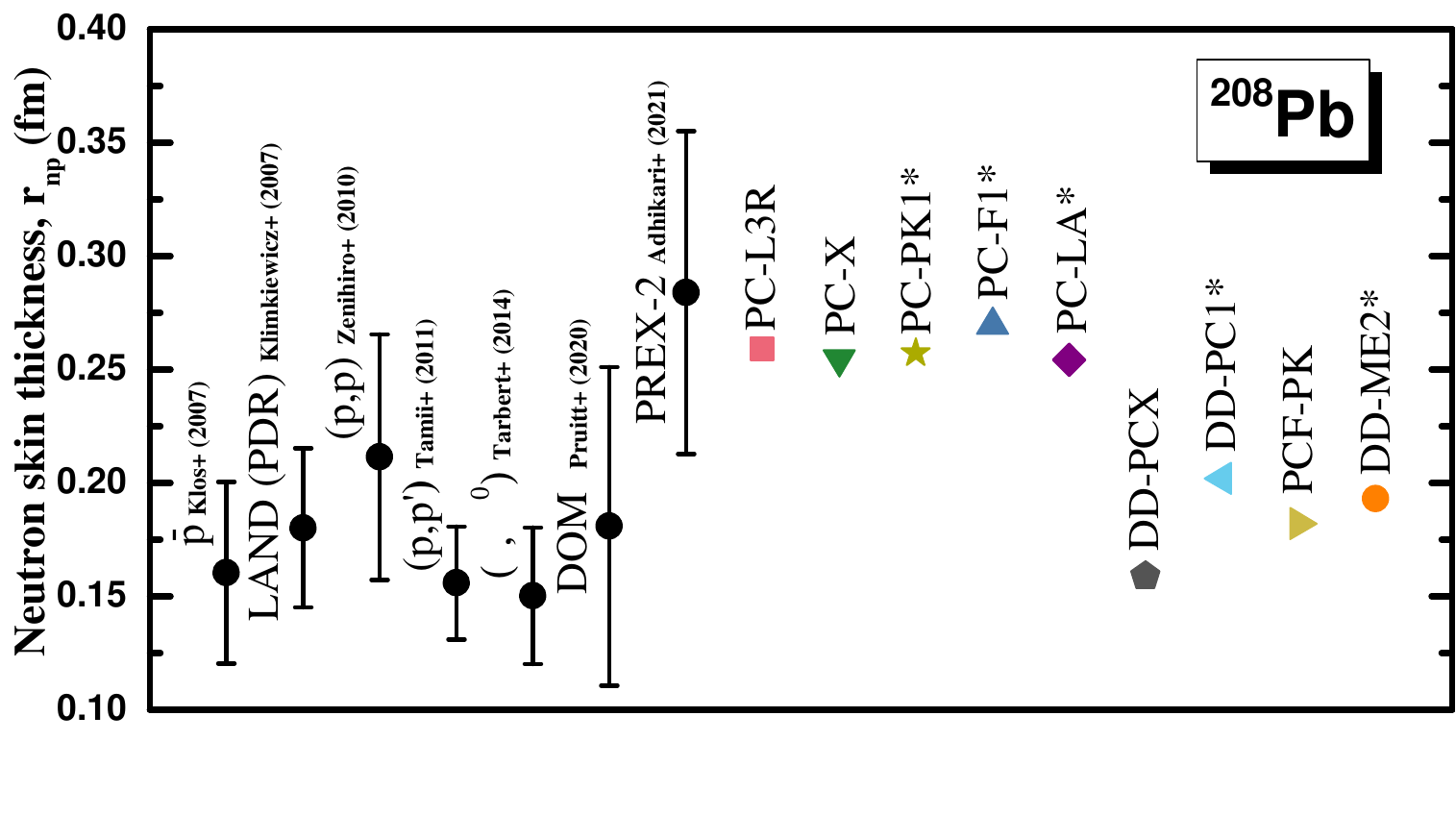}
%\vspace{-0.4cm}
\caption{\footnotesize The neutron skin thickness for $^{208}$Pb determined by experiments and predicted by nuclear energy density functionals. The experimental data are taken from $\bar{ p}$~\cite{PRCKlos2007}, LAND(PDR)~\cite{PRCKlimkiewicz2007}, $(p,p)$~\cite{PRCZenihiro2010}, $(p,p')$~\cite{PRLTamii2011}, ($\gamma, \pi^{0}$)~\cite{PRLTarbert2014}, DOM~\cite{PRLPruitt2020}, and PREX-2~\cite{PRLAdhikari2021}. The calculations are based on the frameworks PC-L3R, PC-X, PC-PK1$^\ast$, PC-F1$^\ast$, PC-LA$^\ast$, DD-PCX, DD-PC1$^\ast$, and DD-ME2$^\ast$, described in the footnotes of Tables~\ref{tab:RMS} and \ref{tab:BindingEnergy}. The result of PCF-PK1 is quoted from with Zhao et al.~\cite{PRCZhao2022}.}
\label{fig:rnp-Pb208}
\vspace{-5mm}
\end{figure}

\begin{figure}[pos=t]
\renewcommand{\thefigure}{S3}
\includegraphics[width=0.45\textwidth]{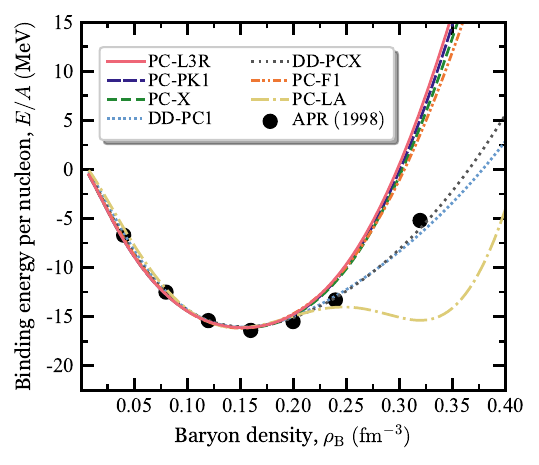}
%\vspace{-0.4cm}
\caption{\footnotesize The binding energy per nucleon, $E\!/\!A$, for the nuclear matter as a function of the baryon density, $\rho_\mathrm{B}$, see texts.}
%given by the PC-L3R, PC-PK1 \cite{PRC2010Zhao}, PC-X \cite{PLB2020Taninah}, DD-PC1 \cite{PRC2008Niksic}, PC-F1 \cite{PRC2002Burvenich}, and PC-LA \cite{PRC1992Nikolaus} interactions, as well as the \emph{ab initio} variational calculation \cite{PRC1998Akmal} (black dots).
\label{fig:Eos_EA}
%(Color online)
\vspace{-5mm}
\end{figure}

%The $\ast$ indicates the calculations was obtained by using the RHB model with separable pairing force (modeled D1S), see Table~\ref{tab:BindingEnergy} note.

\begin{figure}[pos=hb!]
\renewcommand{\thefigure}{S4}
\includegraphics[width=0.47\textwidth]{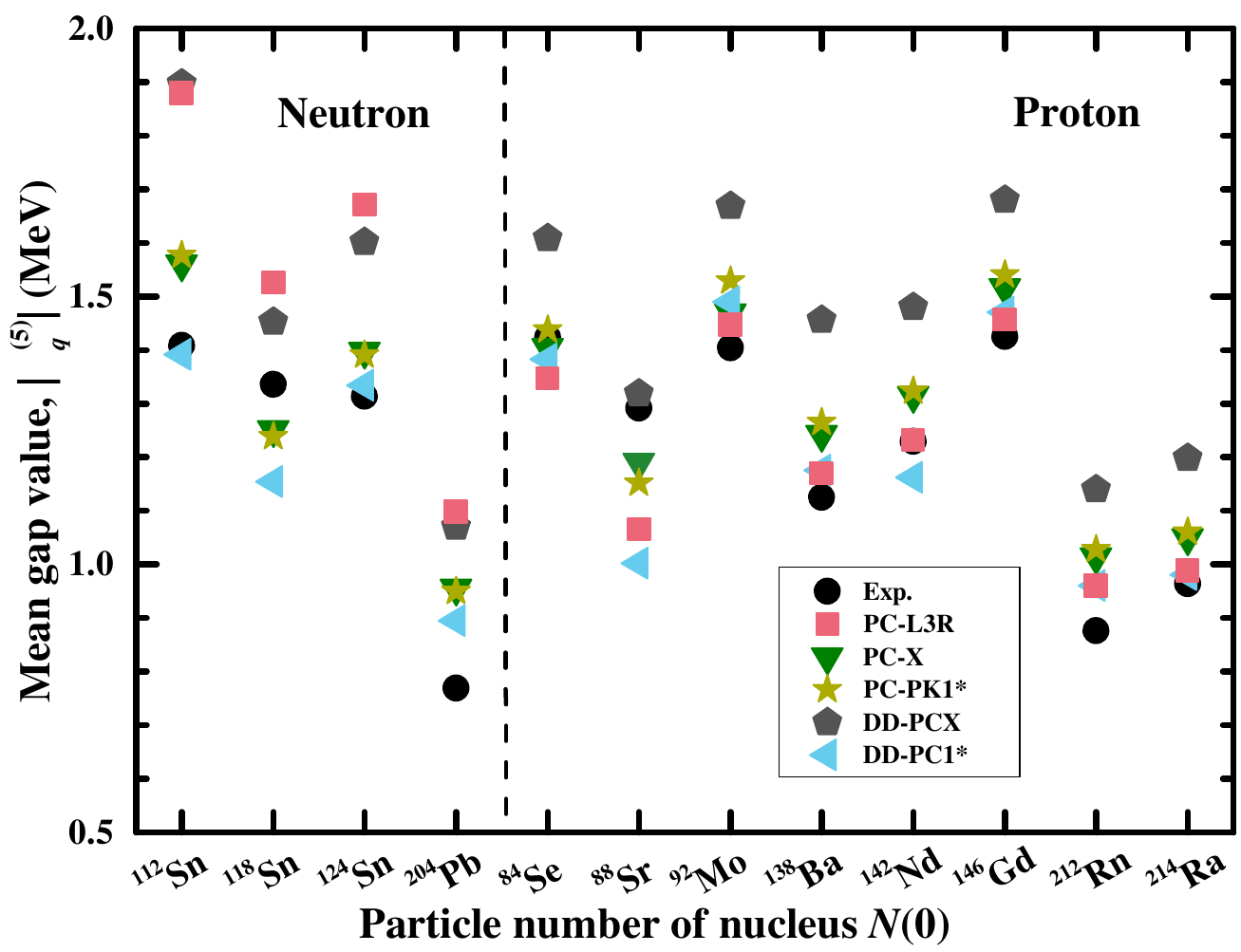}
%\vspace{-0.4cm}
\caption{\footnotesize The five-point mean pairing gap values from AME2020 and present calculations using the frameworks of PC-L3R, PC-X, PC-PK1$^\ast$, DD-PCX, and DD-PC1$^\ast$. See the footnotes of Tables~\ref{tab:RMS} and \ref{tab:BindingEnergy} for the description of these frameworks. The left panel shows the neutron mean gap values whereas the right panel illustrates the proton mean gap values.}
\label{fig:Mean-PG}
\vspace{-5mm}
\end{figure}
%The PC-X, PC-PK1$^\ast$ and DD-PC1$^\ast$ used separable pairing force (modeled D1S). 

\begin{figure}[pos=hb!]
\renewcommand{\thefigure}{S5}
\includegraphics[width=0.47\textwidth]{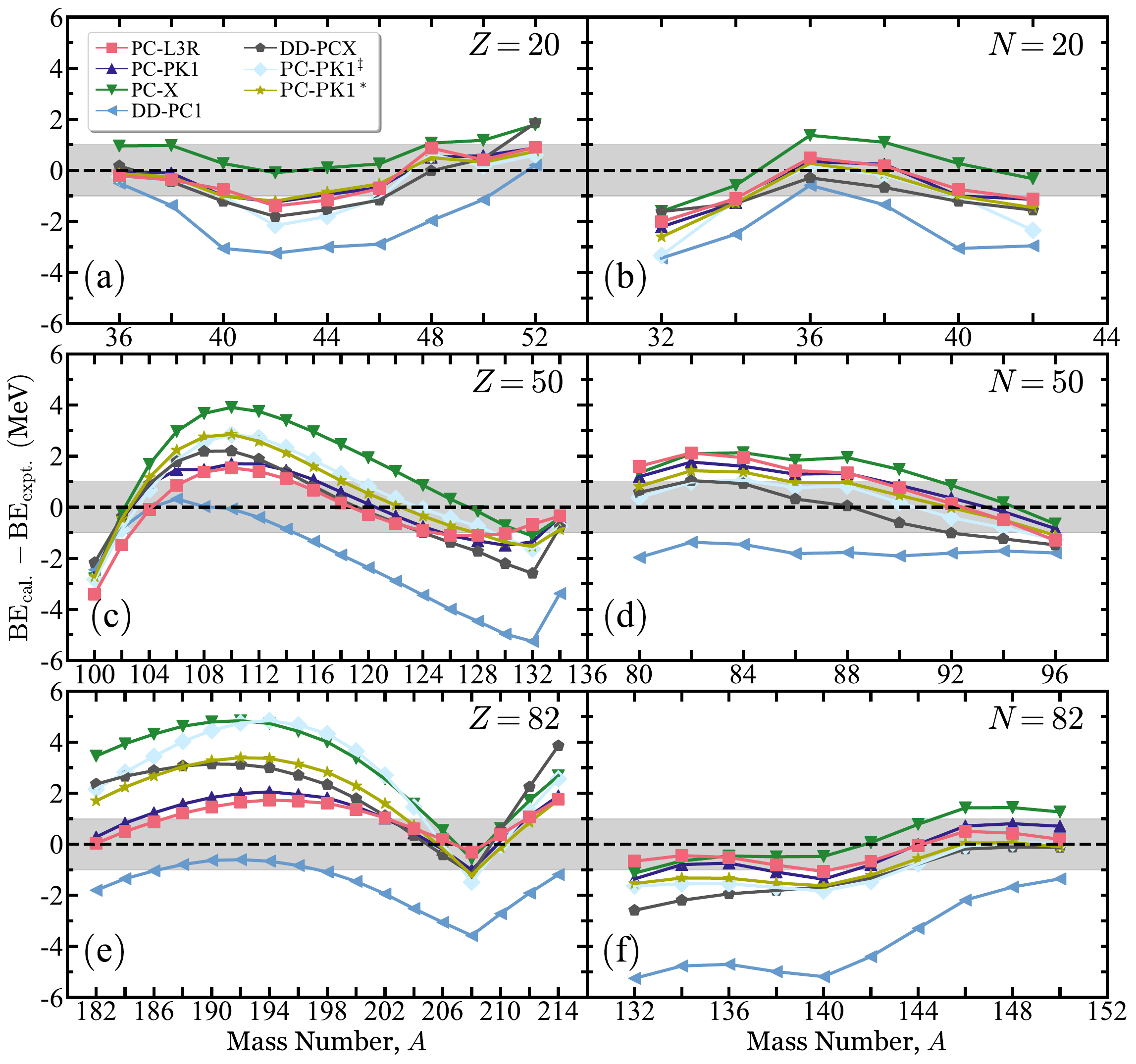}
%\vspace{-0.4cm}
\caption{\footnotesize The deviations between the AME2020 experimental data \cite{AME2020} and the calculated binding energies for the $Z=20$, $50$, and $82$ isotopes, and for the $N=20$, $50$, and $82$ isotones. See the footnote of Tables~\ref{tab:RMS} and ~\ref{tab:BindingEnergy} for the theoretical frameworks in this figure. The data of PC-PK1 (black triangles) and DD-PC1 (blue squares) are from Ref.~\cite{PRC2010Zhao}, whereas the PC-PK1$^\ddag$ data is from Ref.~\cite{ADNDT2018Xia}.}
%\cite{PLB2020Taninah}, \cite{PRC2008Niksic}
%The results of PC-L3R (red squares) (or PC-X (green up triangles) or PC-PK1 (cyan left triangles) with separable pairing force) interaction are calculated in the present work. 
%\label{Fig2}
%(Color online) 
\vspace{-5mm}
\end{figure}

\end{document}